# On the computation of finite bottom-quark mass effects in Higgs boson production

**Romain Mueller**

*Institute for Theoretical Physics, ETH Zürich, 8093 Zürich, Switzerland*
*E-mail:* muellrom@phys.ethz.ch

**Deniz Gizem Öztürk**

*Physik Institut, University of Zürich, 8057, Zürich, Switzerland*
*E-mail:* odeniz@physik.uzh.ch

Abstract: We present analytic results for the partonic cross-sections contributing to the top-bottom interference in Higgs production via gluon fusion at hadron colliders at NLO accuracy in QCD. We develop a method of expansion in small bottom-mass for master integrals and combine it with the usual infinite top-mass effective theory. Our method of expansion admits a simple algorithmic description and can be easily generalized to any small parameter. These results for the integrated cross-sections will be needed in the computation of the renormalization counter-terms entering the computation of finite bottom-quark mass effects at NNLO.

Keywords: Higgs Physics, QCD.

## 1. Introduction

There have been recent improvements in the determination of the inclusive Higgs boson cross section with the computation of the three-loop corrections to the production of a Higgs boson via gluon fusion in the infinite top-mass limit [1]. The uncertainty due to scale variation has been shown to be less than 3% in this case. With such a level of precision arises the question of the uncertainty coming from other contributions. In particular, corrections coming from a finite bottom-quark mass account for around 6% of the total Higgs boson production cross section at leading order and around 4% at NLO [2, 3]. This means that one can expect the uncertainty coming from the unknown finite bottom-quark mass effects at N$^2$LO to contribute substantially to the total uncertainty on the inclusive Higgs cross section at the LHC and the computation of such effects is desirable.

The dominant production mechanism of the Higgs boson within the Standard Model at the LHC is gluon fusion through a heavy-quark loop. Because of the large value of the Yukawa coupling $y_t$ of the top quark, the corresponding cross-section is dominated by contributions involving a top-quark loop [4]. Since the Higgs boson is light compared to the top quark, it is justified to work in the limit of an infinite top quark mass. This induces an effective interaction of the Higgs boson with gluons described an effective scalar operator

$$\mathcal{L}_{\text{eff.}} = -\frac{1}{4} c_H \, F^a_{\mu\nu} F^{\mu\nu;a} H/v, \tag{1.1}$$

where $c_H$ is the Wilson coefficient of the effective interaction, $F^a_{\mu\nu}$ is the gluon field-strength tensor, $H$ is the physical Higgs field, and $v$ is the vacuum expectation value of the Higgs field. The Wilson coefficient can be found in the literature, see for example [5–7].

Despite its much smaller mass, the bottom-quark also contributes substantially to the inclusive Higgs cross-section. In particular, the top-bottom interference, i.e. contributions where diagrams containing a top-quark loop are interfered with diagrams containing a bottom-quark loop, has to be taken into account. These contributions are suppressed by a factor of $y_b/y_t = m_b/m_t \simeq 0.025$ compared to the ones with two effective vertices but become important well above the bottom-quark threshold, where $m_b^2 \ll s \sim m_H^2$, due to large logarithms. Note that contributions where two effective operators are replaced by $b$-quark loops are further suppressed by a factor of $y_b/y_t$ and will not be considered here. These large logarithms can be resummed to all orders in the strong coupling constant and have been found to have only a mild impact on exclusive observables [8]. Here we concentrate on the fixed-order inclusive cross-section with complete $m_b$ dependence whose computation presents two main difficulties: Firstly because the bottom-quark mass is much smaller than the physical Higgs mass it is impossible to simplify the calculations using an infinite-mass effective theory for these contributions; secondly the integrals required at N$^2$LO with full dependence on $m_b^2$ are unknown in the literature and very difficult to compute because of the many mass scales involved. We suggest that a more amenable strategy is to compute these finite bottom-mass effects as an expansion in *small* $m_b^2$.

We explore the difficulties associated with such a double expansion – large top mass and small bottom mass – and obtain analytic expressions for the first order of the small $m_b^2$ expansion of the top-bottom interference contributions to the inclusive Higgs cross section via gluon fusion at NLO QCD. This shall serve both as a proof for the feasibility of such an approach and as a preparation for the computation of the corresponding $\mathrm{N^2LO}$ QCD corrections. In particular we develop tools and understanding that will be valuable for the $\mathrm{N^2LO}$ computation: (i) we introduce a systematic method of expansion of Feynman integrals based on differential equations and (ii) we explore how the phase-space integration create non-trivial analytic behaviour of the inclusive integrals as $m_b \to 0$ and we were able to obtain the correct small $m_b^2$ expansion. Note that the two-loop amplitudes needed at NLO along with the corresponding master integrals are already known in the literature with full $m_b$ dependence [4, 9–12] and that the real radiation matrix elements have been obtained in references [13–15]. While the known two-loop virtual master integrals will merely serve as a check of our method of expansion, we present analytic results for the cut two-loop real master integrals for the first time. We also present analytic expressions for the contribution of the top-bottom interference to the cross section for the production of the Higgs boson via gluon fusion at NLO at order $\epsilon$. This will serve for the computation of counter terms for the renormalization at NNLO.

This article is organized as follows. In Section 2 we present our notation and describe the calculation. In Section 3 we present our analytic results for the first term in the small $m_b^2$ expansion of the Higgs production cross section via gluon fusion at NLO QCD. In Section 4 we describe our method of expansion of the master integrals; and in section 5 we present the analytic results for the master integrals appearing in the virtual and real radiation contributions to the cross section. In Section 6 we give analytic results for the squared matrix elements in terms of master integrals.

## 2. Description of the calculation

We consider the hadronic production of a Higgs boson, denoted symbolically as

$$P(P_1) + P(P_2) \to H(p_H) + R(p_3, \ldots),$$

where $P$ denotes a proton and $R$ is any additional QCD radiation. Let us denote by $S = 2P_1 \cdot P_2$ the center-of-mass frame energy of the two colliding protons. Assuming the usual factorization, the total hadronic cross section can be written as

$$\sigma^{\text{tot.}}_{P_1 P_2 \to H} = \sum_{i,j} \sum_{R} \int_0^1 \mathrm{d}x_1 \mathrm{d}x_2\, f_i(x_1) f_j(x_2) \sigma_{ij \to RH}, \tag{2.1}$$

where $f_i$ is the parton density function for partons of type $i$. The partonic cross section for the process $i(p_1) + j(p_2) \to R(p_3, \ldots) + H(p_H)$ is given by

$$\sigma_{ij \to RH} = \frac{1}{2s} \int \mathrm{d}\Phi(p_1, p_2; p_3, \ldots, p_H)\, |M_{ij \to RH}|^2 ,$$

where $p_1 = x_1 P_1$ and $p_2 = x_2 P_2$ are the momenta of the incoming partons, $s = 2p_1 \cdot p_2$ denotes the center-of-mass frame energy of the colliding partons and $|M_{ij \to RH}|^2$ is the corresponding squared matrix element. We can rewrite the production phase-space as

$$\mathrm{d}\Phi(p_1, p_2; p_3, \ldots, p_H) = \mathrm{d}\Phi(p_1, p_2; p_3, \ldots, p_H)\, s\, \mathrm{d}z\, \delta(zs - m_H^2),$$

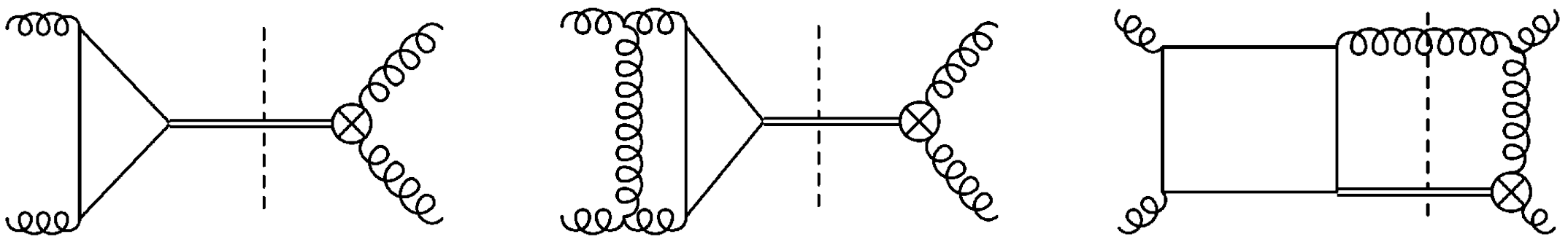

**Figure 1:** Typical contributions of order $y_t y_b$ to the $gg$-initiated cross-section: Born (left), virtual (middle) and real (right). Curly lines indicate a gluon, plain lines a $b$-quark, and doubled lines a Higgs boson. The $Ht\bar{t}$ effective vertex is indicated by a crossed circle.

where the four vector $p_H$ has the mass-shell condition $p_H^2 = sz$ on the right-hand side and we have $z = m_H^2/s \in [0,1]$. We can then use the mass-shell delta function $\delta(zs - m_H^2)$ to put constraint on $s$ rather than on $z$ and use the latter as an integration variable parametrizing the soft limit of the integral. This amounts to rewriting the total hadronic cross section (2.1) as

$$\sigma^{\text{tot.}}_{P_1 P_2 \to H} = \sum_{i,j} \sum_{R} \int_\tau^1 \mathrm{d}z\, \mathcal{L}_{ij}(z)\, \sigma_{ij\to RH}(z), \tag{2.2}$$

where $\tau = m_H^2/S$ is the production threshold, $\sigma_{ij\to RH}(z)$ is the partonic cross section with $p_H^2 = zs$, and we defined $\mathcal{L}_{ij}(z)$ as

$$\mathcal{L}_{ij}(z) = \int_0^1 \mathrm{d}x_1 \mathrm{d}x_2\, \left(x_1 f_i(x_1)\right) \left(x_2 f_j(x_2)\right)\, \delta\left(z x_1 x_2 - \tau\right). \tag{2.3}$$

This representation allows one to consider the inclusive cross-section as a distribution with respect to the integration variable $z$ and allows the implementation of the soft subtraction without any explicit reference to the luminosity. Note that such a representation requires that we choose $z$ and $m_H^2$ as independent variables such that we must set $s = m_H^2/z$.

We can now define more precisely the different pieces of this computation. The partonic cross-section can be decomposed as

$$\sigma_{ij\to RH} = \sigma^{\text{eff.}}_{ij\to RH} + \sigma^{\text{int.}}_{ij\to RH} + \mathcal{O}(y_b^2),$$

where ‘int.’ stands for interference and ‘eff.’ for effective theory. The cross section $\sigma^{\text{eff.}}_{ij\to H}$ is the cross section for the partonic process $ij \to RH$ in the infinite top-mass effective theory alone (with $y_b = 0$ and the effective interaction given by (1.1)) and starts at order $y_t^2 \propto m_t^2$. Such contributions can be found in the literature up to $\alpha_s^5$ [16]. The cross-section $\sigma^{\text{int.}}_{gg\to H}$ is the top-bottom interference and contains the first finite bottom-mass effects. It can be further decomposed in Born, virtual, and real contributions as

$$\sigma^{\text{int.}}_{ij\to H} = \sigma^{\text{B; int.}}_{ij\to H} + \sigma^{\text{V; int.}}_{ij\to H} + \sum_k \sigma^{\text{R; int.}}_{ij\to kH} + \mathcal{O}(\alpha_s^4),$$

where

$$\sigma^{B;\text{int.}}_{ij\to H} = \frac{n_{ij}}{2s}\int \mathrm{d}\Phi(p_1,p_2;p_H)\, 2\Re\,\mathcal{A}^{(0)}_{ij\to H}\left(\mathcal{B}^{(0)}_{ij\to H}\right)^*,$$

$$\sigma^{\text{R};\text{int.}}_{ij\to kH} = \frac{n_{ij}}{2s}\int \mathrm{d}\Phi(p_1,p_2;p_3,p_H)\, 2\Re\,\mathcal{A}^{(0)}_{ij\to kH}\left(\mathcal{B}^{(0)}_{ij\to kH}\right)^*,$$

$$\sigma^{\text{V};\text{int.}}_{ij\to H} = \frac{n_{ij}}{2s}\int \mathrm{d}\Phi(p_1,p_2;p_H)\, 2\Re\left\{\mathcal{A}^{(1)}_{ij\to H}\left(\mathcal{B}^{(0)}_{ij\to H}\right)^* + \mathcal{A}^{(0)}_{ij\to H}\left(\mathcal{B}^{(1)}_{ij\to H}\right)^*\right\},$$

where $n_{ij}$ is the averaging factor and we made the sum over external polarizations and spins implicit. We denote by $\mathcal{A}^{(n)}_{ij\to RH}$ the $n$-th order QCD correction to the matrix element for the production of $RH$ mediated by a $b$-quark loop (with $y_t = 0$), and by $\mathcal{B}^{(n)}_{ij\to RH}$ the $n$-th order QCD correction to the matrix element for the production of $RH$ in the infinite top-mass effective theory (with $y_b = 0$). Note that there is no single diagram involving both a top and a bottom loop at this order and that the Born matrix element $\mathcal{A}^{(0)}_{ij\to RH}$ contains one $b$-quark loop. Typical diagrams contributing to these cross sections are shown in figure 1.

The Born and virtual contributions are known analytically in the literature [12]. The integrated real contribution has only been obtained numerically and we present here its first fully analytic computation as an expansion in $m_b$. The different channels contributing to $\sigma^{\text{int.}}_{ij\to H}$ are:

- Born and virtual: Only the $gg$-initiated channels contributes to the Born and virtual cross sections, such that we will only consider $\sigma^{B;\text{int.}}_{gg\to H}$ and $\sigma^{V;\text{int.}}_{gg\to H}$.
- Real: There are new channels contributing to the real cross sections, namely the intial states $gg$, $qg$, $\bar{q}g$, $gq$, $g\bar{q}$, $q\bar{q}$, and $\bar{q}q$. All these contributions can be easily obtained from

  $$\sigma^{R;\text{int.}}_{gg\to gH},\quad \sigma^{R;\text{int.}}_{qg\to qH},\quad \text{and}\quad \sigma^{R;\text{int.}}_{q\bar{q}\to gH},$$

  and only these will be presented here.

Our strategy for the computation of the inclusive cross-section is based on a reduction to master integrals using integration-by-parts (IBP) identities [17–20] and on the computation of these masters integrals by the method of differential equations [19, 21–23]. The Feynman diagrams are generated using the program `FeynArts` [24] and we use the program `FeynCalc` [25] to evaluate the color and spin traces. We use our own `Mathematica` code in order to compute the square of the matrix elements. The calculation is performed in Feynman gauge and we therefore add contributions with external Faddeev-Popov ghosts. The reduction to master integrals has been performed in two independent ways – using both the program `FIRE` [26, 27] and a private `C++` code[1] – and found to agree. Differential equations with respect to the $b$-quark mass are trivial to obtain because $m_b$ enters the integrals only in internal propagators. The differential equations are then solved as an expansion with respect to $m_b^2$ using the method presented in section 4. Note that contributions from the $b$-quark loop vanish for $m_b = 0$ such that the cross-sections $\sigma^{\text{int.}}_{ij\to H}$ start at $m_b y_b \propto m_b^2$.

[1]We thank Bernhard Mistlberger who provided this code.

The inclusive real contributions can be computed in a similar way thank to the method of *reverse unitarity* [28–32]. This method is based on the usual relationship between cuts and discontinuities of Feynman propagators [33], which is given by

$$2\pi i\delta^+(p^2-m^2) = \text{Disc}\,\frac{1}{p^2-m^2+i0}. \tag{2.4}$$

In the reverse unitarity approach one takes this identity for granted and treat the phase-space integrals as loop integrals, i.e. the phase-space integrals are subject to the same IBP identities as their loop counterparts. The phase-space integrals can then be reduced to master integrals and differential equations can be obtained in the usual way. The only difference with the normal reduction is that integrals that have vanishing or negative power for any of the cut propagators must vanish. Note that powers bigger than one are allowed and correspond to derivatives of the delta function in equation (2.4), in the sense of distributions. The method has been successfully applied to a variety of cut integrals [29–32, 34] and we refer the reader to these references for further discussion.

## 3. Results

In this section we present analytic results for the first order in the small-$m_b^2$ expansion of all the relevant partonic cross sections contributing to $\sigma^{\text{int.}}_{ij\to H}$ up to order $\alpha_s^3$, as described in the previous section. Our method of expansion allows one to generate arbitrary high orders in the $m_b^2$ expansion but we will display here only the leading terms and higher-order terms can be easily generated using the method presented in section 4. Since we are not performing any detailed phenomenological study here, we will only give expressions for the unrenormalized partonic cross sections, while singling out contributions cancelled by the different renormalization counter terms. We refer the reader to references [12] for a more detailed discussion.

We start by presenting our results for the $gg$-initiated channel. The inclusive Born and virtual contributions decouple completely from the phase-space integration, which contributes as an overall pre-factor

$$\Phi_1(m_H^2) = \int \mathrm{d}\Phi(p_1, p_2; p_H) = \frac{2\pi}{s}\delta(1-z). \tag{3.1}$$

The exact result for these contributions is known analytically [12] with full dependence in $m_b^2$ and can be easily expanded[2] for small values of the bottom quark mass. We performed an independent computation and found agreement. This provides a strong check of both our method of expansion and our evaluation of the boundary conditions.

[2]To this purpose we use the package `HypExp` [35, 36].

The leading order of the small bottom mass expansion of the Born contribution reads

$$
\begin{aligned}
\sigma^{\text{B; int.}}_{gg\to H} = 2\pi\delta(1-z)\,\tilde{\alpha}_s\,\mathcal{B}\Bigg\{ & -4-6\zeta_2+L^2 \\
& +\epsilon\left(-8+6\zeta_2-6\zeta_3-2\zeta_2 L-L^2-\frac{2L^3}{3}\right) \\
& +\epsilon^2\left(-16+14\zeta_2+6\zeta_3+\frac{25\zeta_4}{2}+2\left(\zeta_2+\zeta_3\right)L+\frac{3\zeta_2 L^2}{2}+\frac{2L^3}{3}+\frac{L^4}{4}\right) \\
& +\epsilon^3\Bigg(-32+28\zeta_2+\frac{28\zeta_3}{3}-\frac{25\zeta_4}{2}+15\zeta_2\zeta_3-14\zeta_5-\left(2\zeta_3+\frac{9\zeta_4}{2}\right)L \\
& \qquad -\frac{1}{6}\left(9\zeta_2+8\zeta_3\right)L^2-\frac{2\zeta_2 L^3}{3}-\frac{L^4}{4}-\frac{L^5}{15}\Bigg) \\
& +\mathcal{O}\left(\epsilon^4\right)\Bigg\}+\mathcal{O}\left(m_b^4\right),
\end{aligned}
\tag{3.2}
$$

where we defined

$$
\mathcal{B} = \frac{c_H}{4v(1-\epsilon)^2 N_A}\frac{m_b^2}{s}, \qquad L = \log\frac{m_b^2}{s} = \log z\frac{m_b^2}{m_H^2} \in \mathbb{R},
$$

and

$$
\tilde{\alpha}_s = \left(\frac{s}{\mu}\right)^{-\epsilon} S_\epsilon\,\frac{\alpha_s}{4\pi},
$$

where $S_\epsilon = \exp\{\epsilon(\log 4\pi - \gamma_E)\}$ and $\mu$ is the 't Hooft scale. Note that because of the $\delta$-function in equation (3.1), we must have that $L = \log m_b^2/m_H^2$ here.

In contrast, the real contributions do depend explicitly on $z$ as the emission of a supplementary final-state parton allows the initial-state partons to have a center-of-mass frame energy different from the Higgs boson mass $m_H^2$. Since the real matrix element becomes singular in the soft limit $z \to 1$, where the energy of the supplementary final-state parton vanishes, it is necessary to implement some form of soft subtraction. We find that the expansion in small $m_b^2$ completely preserves the structure of the usual NLO soft subtraction[3] and we find no special difficulty there. Namely, we write

$$
\sigma^{\text{R; int.}}_{gg\to H}(z) = \left(\sigma^{\text{R; int.}}_{gg\to H}(z) - (1-z)^{-1-2\epsilon}\tilde{\sigma}^{\text{R; int.}}_{gg\to H}\right) + (1-z)^{-1-2\epsilon}\tilde{\sigma}^{\text{soft; int.}}_{gg\to H}, \tag{3.3}
$$

where $\tilde{\sigma}^{\text{R; int.}}_{gg\to H} = \lim_{z\to 1}\sigma^{\text{R; int.}}_{gg\to H}(z)/(1-z)^{-1-2\epsilon}$ is the soft limit of the real contributions and has the expected form [37]. The term in parenthesis in (3.3) is regular as $z \to 1$ and can be safely expanded in $\epsilon$ while the second term is expanded in plus-distributions using the identity

$$
(1-z)^{-1-2\epsilon} \to -\frac{\delta(1-z)}{2\epsilon} + \sum_{n=0}^{\infty}(-2\epsilon)^n\,\mathcal{D}_n(1-z),
$$

where the plus-distributions $\mathcal{D}_n$ are defined as

$$
\int \mathrm{d}z\,\mathcal{D}_n(1-z)f(z) = \int \mathrm{d}z\,\log^n(1-z)\,\frac{f(z)-f(1)}{z},
$$

[3] See section 5.2 for a more detailed discussion.

for an arbitrary function $f$.

After the soft subtraction has been performed, the total next-to-leading order contribution can be written as

$$\sigma^{\text{V; int.}}_{gg\to H} + \sigma^{\text{R; int.}}_{gg\to H} = 4\tilde{\alpha}_s \left\{ \frac{\beta_0}{2\epsilon} + \frac{3C_F}{4\epsilon}\frac{\partial}{\partial m_b^2} \right\} \sigma^{\text{B; int.}}_{gg\to H} - 8\tilde{\alpha}_s [\, \Delta^{(1)}_{gg} \otimes \sigma^{\text{B; int.}}_{gg\to H} \,] \\ + 32\pi\tilde{\alpha}_s^2 (1-\epsilon)^2 N_c \mathcal{B} \Big( \delta(1-z) C_\delta + C_+ + C_r \Big) + \mathcal{O}(m_b^4), \tag{3.4}$$

where $\beta_0$ is the first order of the QCD beta function when expanded with respect to $\alpha_s/4\pi$, that is

$$\beta_0 = \frac{11C_A - 4T_R N_F}{3},$$

the $g \to gg$ splitting kernel is given by

$$\Delta^{(1)}_{gg}(z) = \frac{\beta_0}{4\epsilon}\,\delta(1-z) + \frac{C_A}{\epsilon}\left( \mathcal{D}_0(1-z) + z(1-z) - 2 + \frac{1}{z} \right),$$

and the convolution is defined for arbitrary functions $f$ and $g$ as

$$[\, f \otimes g \,](z) = \int_0^1 \mathrm{d}x\mathrm{d}y\, f(x) g(y) \delta(xy - z).$$

To get the correct form of the PDF counter term it is important to remember that $\sigma^{\text{B; int.}}_{gg\to H}$ is proportional to $\delta(1-z)$ and that we are using $z$ and $m_H^2$ as independent variables, not $s$. The first term in (3.4) is cancelled by the $\alpha_s$ and mass renormalizations while the second term is cancelled by the introduction of the PDF counter terms. The rest of the expression is what is left after the introduction of all renormalization counter terms and can be written in terms of a $\delta$ part given by

$$\begin{aligned} C_\delta = &- \frac{7}{2} - \frac{49\zeta_2}{12} + \frac{31\zeta_3}{18} - \frac{307\zeta_4}{48} + \left( \frac{4}{9} - \frac{2\zeta_2}{9} + \frac{11\zeta_3}{18} \right) L \\ &+ \left( \frac{1}{72} + \frac{23\zeta_2}{72} \right) L^2 - \frac{L^3}{9} + \frac{5L^4}{864} \\ &+ \epsilon \Bigg( - \frac{157}{6} - \frac{19\zeta_2}{3} + \frac{307\zeta_3}{36} + \frac{763\zeta_4}{144} + \frac{121\zeta_3\zeta_2}{12} - \frac{5\zeta_5}{18} \\ &\quad + \left( \frac{4}{3} - \frac{\zeta_2}{3} + \frac{31\zeta_3}{18} + \frac{11\zeta_4}{18} \right) L + \left( \frac{85\zeta_2}{72} - \frac{29\zeta_3}{72} - \frac{11}{36} \right) L^2 + \left( -\frac{3\zeta_2}{8} - \frac{19}{216} \right) L^3 \\ &\quad - \frac{7L^5}{1440} + \frac{13L^4}{108} \Bigg) + \mathcal{O}(\epsilon^2), \end{aligned}$$

a plus-distribution part given by

$$\begin{aligned} C_+ = &\left( -4 - 6\zeta_2 + L^2 + \epsilon \left( -16 - 6\zeta_2 - 6\zeta_3 - 2\zeta_2 L + L^2 - 2L^3/3 \right) \right) \mathcal{D}_1(1-z) \\ &- \epsilon \left( -4 - 6\zeta_2 + L^2 \right) \mathcal{D}_2(1-z) + \mathcal{O}(\epsilon^2), \end{aligned}$$

and a regular parts given by

$$
\begin{aligned}
C_r =& \frac{1}{48 z\bar z}\Bigg\{4\,\log^3(\bar z)\,\bar z\Big(7\bar z^3-20\bar z^2+24\bar z-13\Big)\\
&+6\log^2(\bar z)\left(\bar z^2\left(-3\bar z^2+\bar z+2\right)-2\left(5\bar z^4-8\bar z^3+6\bar z^2-4\bar z-1\right)\log(z)\right)\\
&+4\log(\bar z)\Bigg(+3\left(-17\bar z^3+20\bar z^2-27\bar z+16\right)\bar z+3\,\log(z)\left(3\bar z^2+\bar z-4\right)\bar z^2\\
&\qquad+3\,\log^2(z)\left(5\bar z^3-9\bar z^2+12\bar z-4\right)\bar z-6\,\mathrm{Li}_2(z)\left(2\bar z^4-2\bar z^3-\bar z-1\right)\\
&\qquad+\pi^2\left(-16\bar z^4+37\bar z^3-54\bar z^2+20\bar z-1\right)\Bigg)\\
&-24\,\mathrm{Li}_3(\bar z)\Big(\bar z^4+2\bar z^3-6\bar z^2+2\bar z-1\Big)\\
&+4\,\log^3(z)\Big(7\bar z^4-16\bar z^3+24\bar z^2-12\bar z+7\Big)\\
&+2\,\log^2(z)\,\bar z\Big(-20\bar z^3+2\bar z^2+63\bar z-45\Big)\\
&-4\,\log(z)\Big(21\bar z^4-76\bar z^3+127\bar z^2+\pi^2\left(7\bar z^4-16\bar z^3+24\bar z^2-8\bar z+7\right)-48\bar z+24\Big)\\
&+4\left(25\bar z^2+26\bar z-51\right)\bar z^2+3\pi^2\,\bar z\Big(9\bar z^3+3\bar z^2-22\bar z+10\Big)\\
&+12\,\mathrm{Li}_2(z)\,(\bar z-1)\,\bar z\left(2\bar z^2\log(z)+15\right)\\
&-24\,\mathrm{Li}_3(z)\Big(3\bar z^3-6\bar z^2+2\bar z-1\Big)+24\,\zeta_3\Big(2\bar z^4-7\bar z^3+6\bar z^2-6\bar z-1\Big)\\
&+L\Bigg(12\,\log^2(\bar z)\,\bar z\Big(-3\bar z^3+10\bar z^2-12\bar z+7\Big)\\
&\qquad-24\,\log(\bar z)\,\bar z\Big(\left(2\bar z^3-3\bar z^2+6\bar z-1\right)\log(z)+(\bar z-1)\bar z\Big)\\
&\qquad-24\,\log^2(z)\Big(2\bar z^4-4\bar z^3+6\bar z^2-3\bar z+2\Big)+4\,\log(z)\,\bar z^2\Big(11\bar z^2-5\bar z-6\Big)\\
&\qquad+24\,\mathrm{Li}_2(z)\Big(\bar z^4-\bar z^3-\bar z-1\Big)+2\,\pi^2\Big(4\bar z^4-13\bar z^3+18\bar z^2-7\bar z+2\Big)\\
&\qquad+6\,\left(\bar z^2+3\bar z-4\right)\bar z^2\Bigg)\\
&+L^2\Bigg(12\,\log(\bar z)\,\bar z\Big(5\bar z^3-13\bar z^2+18\bar z-8\Big)+12\,\log(z)\Big(\bar z^4+1\Big)-9(\bar z-1)\bar z^3\Bigg)\\
&+2\,L^3\,\bar z\Big(5\bar z^2-6\bar z+5\Big)\Bigg\}\\
&-\frac{\epsilon}{720 z\bar z}\Bigg\{120\,\log^4(\bar z)\,\bar z\Big(7\bar z^3-20\bar z^2+24\bar z-13\Big)\\
&\qquad+30\,\log^3(\bar z)\Big(4\left(21\bar z^4-84\bar z^3+126\bar z^2-10\bar z+3\right)\log(z)\\
&\qquad\qquad+\bar z\left(-41\bar z^3+37\bar z^2-10\bar z+18\right)\Big)
\end{aligned}
$$

$$
\begin{aligned}
&- 30\, \log^2(\bar z)\Bigg( - 6\,\mathrm{Li}_2(\bar z)\, \bar z\Big(25\bar z^3 - 73\bar z^2 + 96\bar z - 20\Big) \\
&\qquad\qquad - 12\,\mathrm{Li}_2(z)\, \Big(8\bar z^4 - 32\bar z^3 + 48\bar z^2 - 5\bar z + 2\Big) \\
&\qquad\qquad + 3\, \log^2(z)\Big(8\bar z^4 - 33\bar z^3 + 42\bar z^2 - 4\bar z - 5\Big) \\
&\qquad\qquad + \pi^2\Big(74\bar z^4 - 202\bar z^3 + 276\bar z^2 - 90\bar z + 4\Big) \\
&\qquad\qquad + \log(z)\, \Big( - 57\bar z^4 + 57\bar z^3 - 261\bar z^2 + 267\bar z + 6\Big) \\
&+ 30\, \log(\bar z)\, \Bigg(24\,\mathrm{Li}_3(\bar z)\, \Big( - 3\bar z^4 + 6\bar z^2 + 3\bar z + 1\Big) \\
&\qquad\qquad + \log^2(z)\, \bar z\, (-61\bar z^3 + 67\bar z^2 - 60\bar z + 30) \\
&\qquad\qquad - 2\, \log(z)\, \Big(\pi^2\, (14\bar z^3 - 12\bar z^2 - 21\bar z + 19) \\
&\qquad\qquad\qquad + 2\, (42\bar z^4 - 61\bar z^3 + 91\bar z^2 - 48\bar z + 24)\,\Big) \\
&\qquad\qquad + \pi^2\Big(74\bar z^4 - 52\bar z^3 - 43\bar z^2 + 47\bar z + 2\Big) \\
&\qquad\qquad + 12\,\zeta_3\Big(44\bar z^4 - 119\bar z^3 + 150\bar z^2 - 52\bar z - 7\Big) \\
&\qquad\qquad + 2\bar z\, (230\bar z^3 - 23\bar z^2 + 81\bar z - 192)\,\Bigg) \\
&+ 360\,\mathrm{Li}_4(\bar z)\, \bar z\Big(9\bar z^3 - 9\bar z^2 - 14\Big) \\
&- 360\, S_{2,2}(\bar z)\, \Big(22\bar z^4 - 55\bar z^3 + 66\bar z^2 - 12\bar z - 5\Big) \\
&+ 180\,\mathrm{Li}_2(\bar z)\,\log^2(z)\Big(22\bar z^4 - 55\bar z^3 + 66\bar z^2 - 12\bar z - 5\Big) \\
&- 180\,\mathrm{Li}_3(\bar z)\Big(2\,(20\bar z^4 - 53\bar z^3 + 66\bar z^2 - 12\bar z - 5)\,\log(z) \\
&\qquad\qquad + 7\bar z^4 + 9\bar z^3 - 95\bar z^2 + 81\bar z + 2\Big) \\
&- 15\, \log^4(z)\, \Big(\bar z^4 + 4\bar z^3 - 6\bar z^2 - 8\bar z + 1\Big) \\
&- 10\, \log^3(z)\, \Big(80\bar z^4 - 62\bar z^3 - 93\bar z^2 + 93\bar z + 42\Big) \\
&- 10\, \log^2(z)\, \Big(3\pi^2\, (39\bar z^4 - 97\bar z^3 + 129\bar z^2 - 27\bar z + 9) \\
&\qquad\qquad - 2\, (77\bar z^4 + 163\bar z^3 - 645\bar z^2 + 297\bar z - 36)\,\Big) \\
&+ 360\, \log(z)\,\zeta_3\, \bar z\Big(25\bar z^3 - 61\bar z^2 + 78\bar z - 20\Big) \\
&+ 20\, \log(z)\, \Big(\pi^2\, (61\bar z^4 - 25\bar z^3 - 108\bar z^2 + 93\bar z + 21) \\
&\qquad\qquad + 2\, (195\bar z^4 - 646\bar z^3 + 901\bar z^2 - 324\bar z + 162)\,\Big) \\
&+ 360\,\mathrm{Li}_4(z)\, \Big(13\bar z^4 - 40\bar z^3 + 54\bar z^2 - 25\bar z + 8\Big)
\end{aligned}
$$

$$
\begin{aligned}
&- 180\,\mathrm{Li}_3(z)\Big(2\left(9\bar z^4-24\bar z^3+30\bar z^2-19\bar z+4\right)\log(z)+9\bar z^4+5\bar z^3-76\bar z^2+64\bar z+2\Big)\\
&+ 60\,\mathrm{Li}_2(z)\Big(3\left(6\bar z^4-15\bar z^3+18\bar z^2-11\bar z+2\right)\mathrm{Li}_2(z)\\
&\qquad +3\left(\bar z^3+5\bar z^2-38\bar z+32\right)\bar z\log(z)\\
&\qquad +3\left(29\bar z^4-71\bar z^3+84\bar z^2-27\bar z-3\right)\log^2(z)\\
&\qquad -6\pi^2\bar z^4+15\pi^2\bar z^3+18\bar z^3-18\pi^2\bar z^2-222\bar z^2+21\pi^2\bar z+156\bar z-4\pi^2\Big)\\
&- 360\,S_{2,2}(z)\,\bar z\Big(25\bar z^3-73\bar z^2+96\bar z-20\Big)\\
+L\Bigg(&-60\,\log^3(\bar z)\,\bar z\Big(11\bar z^3-40\bar z^2+48\bar z-29\Big)\\
&-180\,\log^2(\bar z)\Big(\left(3\bar z^4-2\bar z^3+6\bar z^2+10\bar z-1\right)\log(z)+\bar z\left(-3\bar z^3+12\bar z^2-12\bar z+5\right)\\
&\qquad +24\,\log^2(z)(\bar z-1)+36(\bar z-1)\bar z^2+\pi^2\left(7\bar z^4-22\bar z^3+30\bar z^2-8\bar z+1\right)\Big)\\
&+360\,\mathrm{Li}_3(\bar z)\Big(\bar z^4-4\bar z^3+6\bar z^2-12\bar z+1\Big)\\
&+120\,\log^3(z)\Big(3\bar z^4-6\bar z^3+9\bar z^2-7\bar z+3\Big)+360\,\log^2(z)\Big(2\bar z^4-3\bar z^3+4\bar z^2-2\bar z+2\Big)\\
&+20\,\log(z)\,\bar z\Big(-91\bar z^3+46\bar z^2+\pi^2\left(9\bar z^3-6\bar z^2+9\bar z+3\right)+27\bar z+18\Big)\\
&-540\,\zeta_3\,\bar z\Big(2\bar z^3-5\bar z^2+6\bar z-7\Big)-15\pi^2\left(11\bar z^4-19\bar z^3+12\bar z^2+4\right)\\
&+360\,\mathrm{Li}_2(z)\Big(4\bar z\log(z)-\bar z^4+3\bar z^3-4\bar z^2+\left(\bar z^4-\bar z^3-7\bar z-1\right)\log(\bar z)+3\bar z+1\Big)\\
&-2160\bar z\mathrm{Li}_3(z)-15\left(12\left(\bar z^2+11\bar z-12\right)\bar z^2\right)\Bigg)\\
+L^2\Bigg(&180\,\log^2(\bar z)\,\bar z\Big(3\bar z^3-8\bar z^2+12\bar z-5\Big)\\
&-90\,\log(\bar z)\Big(4\left(\bar z^4-3\bar z^3+6\bar z^2-\bar z-1\right)\log(z)+\bar z\left(13\bar z^3-25\bar z^2+30\bar z-14\right)\Big)\\
&+360\,\mathrm{Li}_2(z)\Big(\bar z^4-\bar z^3-1\Big)-180\,\log(z)^2\Big(3\bar z^4-6\bar z^3+9\bar z^2-5\bar z+3\Big)\\
&+30\,\log(z)\Big(5\bar z^4+\bar z^3-18\bar z^2+6\bar z-6\Big)\Bigg)\\
+L^3\Bigg(&60\,\log(\bar z)\,\bar z\Big(11\bar z^3-26\bar z^2+36\bar z-15\Big)+60\,\log(z)\Big(\bar z^4+4\bar z^3-6\bar z^2+4\bar z+1\Big)\\
&-30\,\bar z\left(3\bar z^3+\bar z^2-4\bar z+4\right)\Bigg)\\
+L^4\Bigg(&30\bar z\left(5\bar z^2-6\bar z+5\right)\Bigg)\Bigg\},
\end{aligned}
$$

where $\mathrm{Li}_i$ and $S_{i,j}$ denote the polylogarithms and the (Nielsen) generalized polylogarithms, respectively.

The virtual contribution alone has the singularity structure predicted by [37] and can be written as

$$\sigma^{\text{V; int.}}_{gg\to H} = 4\tilde{\alpha}_s \left\{ \frac{\beta_0}{2\epsilon} + \frac{3C_F}{4\epsilon}\frac{\partial}{\partial m_b^2} - \cos(\pi\epsilon)\left(\frac{C_A}{\epsilon^2} + \frac{\beta_0}{2\epsilon}\right)\right\} \sigma^{\text{B; int.}}_{gg\to H} + \mathcal{O}(\epsilon^0).$$

The real contribution must be $\epsilon$-finite before integration over the phase-space since the internal $b$-quark mass $m_b^2$ regulates the internal bottom-quark loop completely. This means that real contribution has only a $1/\epsilon$-pole, which must be proportional to the collinear splitting kernel,

$$\sigma^{\text{R; int.}}_{gg\to H} = -8\tilde{\alpha}_s [\,\Delta^{(1)}_{gg}|_{\beta_0=0} \otimes \sigma^{\text{B; int.}}_{gg\to H}\,] + \mathcal{O}(\epsilon^0).$$

Although we have the presence of $\beta_0$ in collinear and renormalization counter terms, the matrix elements do not contain it, since there is no explicit correction to the gluon propagator. In particular, this means that our result is independent of the number of light-quark flavours $N_f$ apart in the running of $\alpha_s$.

We now turn to the $qg$ and $q\bar{q}$-initiated channels. In this case, there is no corresponding virtual contribution such that the real contributions must be finite in the soft limit ($z \to 1$). This is indeed the case and no soft subtraction is needed in this case. In the case of the $qg$-initiated channel, the only divergence is of IR nature and cancelled by the PDF counterterms. The corresponding real contribution can be written as

$$\begin{aligned}
\sigma^{\text{R; int.}}_{qg\to qH} =& -4\tilde{\alpha}_s[\,\Delta^{(0)}_{gq} \otimes \sigma^{\text{B; int.}}_{gg\to H}\,] - 2\pi\tilde{\alpha}_s^2 \frac{(1-\epsilon)^2 N_A}{9z}\mathcal{B}\Bigg\{ \\
& + \log^3(\bar{z})\Big(\bar{z}^2+1\Big) - 3\log^2(\bar{z})\Big(2\log(z) + (\bar{z}-1)\bar{z}\Big) \\
& + \log(\bar{z})\Big(-6\log^2(z)\,\left(\bar{z}^2+1\right) + \pi^2\left(6\bar{z}^2+8\right) - 12\,\text{Li}_2(z) + 6\left(7\bar{z}^2-7\bar{z}+4\right)\Big) \\
& - 12\,\text{Li}_3(\bar{z}) \\
& + \log^3(z)\Big(-3\bar{z}^2-7\Big) + 9\log^2(z)\,z + 6\log(z)\Big(\pi^2\left(3\bar{z}^2+11\right) + 36\bar{z} + 12\Big) \\
& - 12\,\text{Li}_3(z) + 24\,\text{Li}_2(z)\,z + \zeta_3\Big(12\bar{z}^2+24\Big) - \pi^2 z - 18\bar{z}^2 + 6\bar{z} + 12 \\
& + L\Bigg(-3\log^2(\bar{z})\Big(\bar{z}^2+1\Big) + 6\log(\bar{z})\Big(2\log(z)\,\left(\bar{z}^2+1\right) + (\bar{z}-1)\bar{z}\Big) \\
& \qquad + 6\log^2(z)\Big(\bar{z}^2+2\Big) + 6\log(z)\,z + 12\,\text{Li}_2(z) - 2\pi^2 - 6\bar{z}^2 + 6\bar{z}\Bigg) \\
& + L^2\Bigg(-3\log(\bar{z})\Big(\bar{z}^2+1\Big) + \log(z)\Big(-3\bar{z}^2-3\Big) - 3\bar{z}^2 + 6\bar{z} - 3\Bigg) \\
& + L^3\Big(-\bar{z}^2-1\Big) \\
& + \epsilon\Bigg\{-2\log^4(\bar{z})\Big(\bar{z}^2+1\Big) + \log^3(\bar{z})\Big(112\log(z) + \bar{z}(7\bar{z}-5)\Big)
\end{aligned}$$

$$
\begin{aligned}
&+\log^2(\bar z)\Bigg(84\,\mathrm{Li}_2(\bar z)+\log^2(z)\,\left(-24+6\bar z^2+6\right)+\log(z)\,\left(60\bar z-66\right)\\
&\qquad\qquad +126\,\mathrm{Li}_2(z)+\pi^2\left(-5\bar z^2-26\right)+111\bar z-24-75\bar z^2\Bigg)\\
&+\log(\bar z)\Bigg(84\,\mathrm{Li}_3(\bar z)+\log^3(z)\left(2\bar z^2+50\right)\\
&\qquad\qquad +\log^2(z)\left(-6\bar z^2-6\bar z\right)+\log(z)\left(-8\pi^2\bar z^2-72\bar z+8\pi^2-24\right)\\
&\qquad\qquad -96\,\mathrm{Li}_3(z)+\mathrm{Li}_2(z)\,\left(120\bar z-132\right)+96\,\zeta_3\\
&\qquad\qquad +\pi^2\left(4\bar z^2-12\bar z+22\right)+234\,\bar z^2-210\,\bar z+72\Bigg)\\
&-84\,\mathrm{Li}_4(\bar z)+\mathrm{Li}_3(\bar z)\left(-132+120\bar z-96\log(z)\right)\\
&+48\,\mathrm{Li}_2(\bar z)\,\log^2(z)-96\,S_{2,2}(\bar z)\\
&+\frac{\log^4(z)}{2}\left(5-\bar z^2\right)+\log^3(z)\left(-3\bar z^2+11\bar z-18\right)\\
&+\log^2(z)\left(36\,\mathrm{Li}_2(z)+-\frac{1}{4}\,\pi^2\left(13\bar z^2+93\right)+3\left(\bar z^2-18\bar z+1\right)\right)\\
&+\log(z)\Bigg(\zeta_3\left(18\bar z^2+114\right)+48\,\mathrm{Li}_2(z)\,z++\pi^2\left(3\bar z^2-12\bar z+23\right)\\
&\qquad\qquad +168\bar z+36-12\bar z^2\Bigg)\\
&+60\,\mathrm{Li}_4(z)-168\,S_{2,2}(z)+\pi^4\,\frac{1}{60}\left(53\bar z^2-39\right)\\
&+\mathrm{Li}_3(z)\,\left(96\bar z-108\right)-114\,\zeta_3\,\left(24\bar z^2-114\bar z+126\right)\\
&+\mathrm{Li}_2(z)\left(-72\bar z+8\pi^2+24\right)+-2\,\pi^2\,\left(\bar z^2-11\bar z-1\right)-150\bar z^2+102\bar z+48\\
&+L\Bigg(\log^3(\bar z)\left(5\bar z^2+5\right)+\log^2(\bar z)\left(-6\left(2\,\log(z)\,\left(\bar z^2+3\right)+\bar z(2\bar z-1)\right)\right)\\
&\qquad +\log(\bar z)\left(-48\,\mathrm{Li}_2(z)+8\,\pi^2+\log(z)\,\left(12\bar z^2+12\bar z\right)+24\bar z^2-24\bar z\right)\\
&\qquad -48\,\mathrm{Li}_3(\bar z)+\log^3(z)\left(3\bar z^2+1\right)\\
&\qquad +\log^2(z)\,\left(6\bar z^2+12\right)+\log(z)\left(\frac{5\pi^2\bar z^2}{2}-6\bar z^2+\frac{5\pi^2}{2}+6\right)\\
&\qquad -36\mathrm{Li}_3(z)+\mathrm{Li}_2(z)(24\log(z)+12)\\
&\qquad +36\,\zeta_3+\pi^2\left(2\bar z^2-3\bar z-1\right)-24\bar z^2+24\bar z\Bigg)\\
&+L^2\Bigg(3\,\log^2(\bar z)\left(\bar z^2+1\right)-3\,\log(\bar z)\left(2\left(\bar z^2+1\right)\log(z)+\bar z(\bar z+1)\right)
\end{aligned}
$$

$$
\begin{aligned}
&- \log^2(z)\, \frac{3}{2} \left(3\bar{z}^2 + 5\right) - 3\, \log(z) \left(\bar{z}^2 - \bar{z} + 2\right) - 6\, \mathrm{Li}_2(z) \\
&+ \pi^2\, \frac{1}{4} \left(\bar{z}^2 + 5\right) - 3\, z \Bigg) \\
&+ L^3 \left(3\, \log(\bar{z}) \left(\bar{z}^2 + 1\right) + \log(z) \left(2\bar{z}^2 + 2\right) + 2\bar{z}^2 - 6\bar{z} + 2\right) \\
&+ L^4 \left(\bar{z}^2 + 1\right) \Bigg\} \\
&+ \mathcal{O}\left(\epsilon^2\right) \Bigg\} + \mathcal{O}(m_b^4),
\end{aligned}
$$

where $\bar{z} \equiv 1 - z$ and the $qg \to q$ splitting kernel is given by

$$
\Delta_{gq}^{(1)}(z) = \frac{2}{3\epsilon} \frac{1 + (1-z)^2}{z} \,.
$$

The $q\bar{q}$-initiated channel does not develop any IR singularity, since all the contributing diagrams must have the final-state gluon attached to the $b$-quark loop, and is therefore completely $\epsilon$-finite. The corresponding contribution reads

$$
\begin{aligned}
\sigma_{q\bar{q}\to gH}^{\text{R; int.}} = \tilde{\alpha}_s^2\, \frac{(2(1-\epsilon)^2 N_A)^2}{36} \mathcal{B} \, \Bigg\{ \\
&- \frac{2}{3} \left(\bar{z}^2 \log^2(z) - 4\bar{z} \log(z) - 4\bar{z}^2\right) + L\, \frac{4}{3}\, \log(z)\, \bar{z}^2 \\
&+ \epsilon \, \Bigg\{ \log(\bar{z}) \left( \left( \frac{4}{3} \log^2(z)\, \bar{z}^2 - \frac{16}{3}\, \log(z)\, \bar{z} - \frac{16\bar{z}^2}{3} \right) \right) \\
&\quad + \frac{2}{9} \bar{z}^2 \log^3(z) + \log^2(z) \left( - \frac{4\bar{z}^2}{9} - \frac{4\bar{z}}{3} \right) \\
&\quad + \log(z) \left( - \frac{16}{3} \zeta_2 \bar{z}^2 - \frac{8\bar{z}^2}{3} + \frac{88\bar{z}}{9} \right) + \frac{88\bar{z}^2}{9} \\
&\quad + L \left( - \frac{8}{3}\, \log(\bar{z})\, \bar{z}^2 \log(z) + \frac{8}{9}\, \log(z)\, \bar{z}^2 \right) \\
&\quad + L^2 \left( - \frac{2}{3}\, \log(z)\, \bar{z}^2 \right) \Bigg\} \\
&+ \mathcal{O}(\epsilon^2) \Bigg\} + \mathcal{O}(m_b^4).
\end{aligned} \tag{3.5}
$$

Although we have obtained results at all orders in $\epsilon$ (see the ancillary file attached to this article), we presented our results up to order $\epsilon$ only as these terms enter the renormalization of the finite part of the finite bottom-quark mass effects at NNLO.

## 4. Systematic expansion of Feynman integrals

In this section we introduce a method for the expansion of Feynman integrals with respect to a small parameter. It is a simple extension and systematisation of the method of expansion developed recently in [1,38,39] and we show here that it can be formulated as an explicit algorithm and that a closed-form formula in the form of a Dyson expansion can be obtained. The procedure presented here is completely general and can be carried out with any family of integrals and any external parameter. It is based on the general method of differential equations [19, 21, 40] as well as more recent ideas presented in [41] and we refer the reader to references [42, 43] for a modern introduction to the subject.

The usual strategy is to seek solutions of the differential equations in the form of an expansion in $\epsilon$, see for example [44, 45]. Many different types of functions appear in such expansions: most notably polylogarithms, harmonic polylogarithms [46] and Goncharov polylogarithms [47]. These functions are generally embedded with a powerful algebraic structure [48, 49] that enables spectacular simplifications [50]. However, for integrals with many different external parameters, more complicated functions appear, such as elliptic functions [51], and no useful algebraic structure generating relations among them has been found so far. Handling such functions properly in a physical computation remains a difficult challenge. It might be easier to obtain solutions of the differential equations as expansions in terms of an external parameter. In such expansions, only powers and logarithms of the expansion parameter appear, hence removing the hassle of dealing with complicated functions[4]. The hope is that single terms in the expansion are simpler to compute than the complete result, and that by summing a sufficient number of these terms it might be possible to obtain a satisfactory approximation.

Let us consider an arbitrary Feynman integral $I$ that depends on the small bottom-mass $m_b^2$. Extracting a pre-factor of $(-s)^{2d-\nu}$, where $\nu = \sum_i \nu_i$ denotes the total inverse power of propagators, the $m_b$-dependence of this integral can be expressed using the dimensionless variable

$$r = -\frac{m_b^2}{s}. \tag{4.1}$$

In what follows, we will set $s = -1$ and consider only the variable $r$ from now on. In general, Feynman integrals have non-trivial analytic properties when considered as functions of their external parameters and possess branch cuts whose starting points are determined by the Landau equations. If one tries to expand the Feynman integral around a point which is not a solution of the Landau equations, the integral is analytic there and a naive expansion of the integrand will give the correct Taylor expansion. However, if the expansion point $r = 0$ is a solution of the Landau equations then the integral is not analytic there and can not be naively expanded. In this case the expansion can be performed using the prescription of expansion by regions [52, 53] and the general analytic behaviour of the integral in the

[4]Note that if there are many scales, the boundary conditions might be as well complicated functions of the remaining parameters.

vicinity of the expansion point $r = 0$ is found to be of the form

$$I(r, \epsilon) \sim \sum_{n,m \in \mathbb{N}} r^{-n\epsilon} \log(r)^m \, I^{(n,m)}(r, \epsilon), \qquad \text{as} \quad r \to 0, \tag{4.2}$$

where the functions $I^{(n,m)}$ have a *Laurent* expansion around $r = 0$. Only some of the $I^{(n,m)}$ are non-vanishing such that the sum over $n$ and $m$ contains only a finite number of terms. Such an expansion makes the analytic behaviour of the Feynman integral in the vicinity of the expansion point explicit since the functions $I^{(n,m)}$ do not posses any branch point at $r = 0$. We call contributions with different $n$ as having different *scalings* while the term proportional to $I^{(0,0)}$ has a trivial analytic structure and will be denoted the *hard region*.

In order to solve a system of differential equations in the form of an expansion, it is necessary to resolve the analytic structure as shown explicitly in (4.2). In particular, one has to understand how the explicit logarithms, known in the literature as integrating factors, appear in such an expansion [19, 22, 54]. In the rest of this section, we will see that once the analytic structure of the solution of the differential equations in the vicinity of the expansion point has been exposed, one can easily generate an expansion in the form of (4.2) explicitly.

Note that although we have introduced the parameter $r$ as (4.1) for definiteness, the algorithm developed in what follows can be applied to *any* small parameter and is not restricted to expansions with respect to internal masses. Such a parameter can be the square of any linear combination of the external momenta, any internal mass, or any combination thereof. For example, while computing the virtual corrections in section 5.1 we use the (related) variable $x = (\sqrt{1 + 4r} - 1)/(\sqrt{1 + 4r} + 1)$ as our expansion parameter.

### 4.1 Analytic behaviour near a singular point of the differential equations

Let us first comment briefly on the link between the structure of a set of differential equations and the analytic behaviour of its solution in the vicinity of a regular singular point. A similar discussion can be found in reference [54]. The considerations presented here are not new nor very advanced and we refer the reader to textbook literature for more detailed discussion. Consider a family of master integrals,

$$\mathbf{M}(r, \epsilon) = (I_1(r, \epsilon), \ldots, I_m(r, \epsilon))^{\mathsf{T}},$$

that depends on an external parameter $r$ and further assume that it satisfies a system of differential equations with a regular singular point at $r = 0$, that is

$$\partial_r \mathbf{M} = \frac{\mathrm{R}(\epsilon)}{r} \mathbf{M} + \ldots \,, \tag{4.3}$$

where the dots indicate contributions less singular at $r = 0$ and the residue matrix $\mathrm{R}(\epsilon)$ is *not* nilpotent. As we will discuss later, if the residue matrix is nilpotent the pole can be removed altogether by a suitable change of master integrals and the solution of equation (4.3) is analytic at $r = 0$. Forgetting about non-singular contributions, the solution of equation (4.3) can be readily written in terms of a matrix exponential as

$$\mathbf{M}(r) = \exp\left\{ \int^r \mathrm{d}r' \, \frac{\mathrm{R}(\epsilon)}{r'} \right\} \mathbf{B}(\epsilon) = \exp\left\{ \log r \, \mathrm{R}(\epsilon) \right\} \mathbf{B}(\epsilon) \,, \tag{4.4}$$

where $\mathbf{B}(\epsilon)$ denote some vector of boundary conditions. Matrix exponentials are best computed using the Jordan decomposition, that we recall here for completeness and to fix notation. For every square matrix R with eigenvalues $\lambda_1, \dots, \lambda_n$ there exists a similarity transformation S[R] such that

$$\mathrm{J}[\mathrm{R}] = \mathrm{S}[\mathrm{R}]^{-1}\, \mathrm{R}\ \mathrm{S}[\mathrm{R}] \tag{4.5}$$

is of block-diagonal form and can be written as

$$\mathrm{J}[\mathrm{R}] = \begin{pmatrix} \mathrm{J}_{a_1}^{d_1} & & \\ & \ddots & \\ & & \mathrm{J}_{a_n}^{d_n} \end{pmatrix}, \qquad \text{where} \quad \mathrm{J}_a^d = \begin{pmatrix} a & 1 & & \\ & a & \ddots & \\ & & \ddots & 1 \\ & & & a \end{pmatrix}, \tag{4.6}$$

with $a_i \in \{\lambda_1, \dots, \lambda_n\}$ and $d_i \in \mathbb{N}$. The matrix J[R] is the Jordan normal form of R and the $d_i$-dimensional square matrices $\mathrm{J}_{a_i}^{d_i}$ are its Jordan blocks. For a given eigenvalue $\lambda_i$, the number of Jordan blocks with $a_j = \lambda_i$ is called the geometric multiplicity of this eigenvalue, while the sum of the sizes of the Jordan blocks is called the algebraic multiplicity. A matrix R is diagonalizable if and only if the geometric and algebraic multiplicities are equal for each eigenvalue, that is, if all the Jordan blocks are $1 \times 1$ matrices. Using the very definition of the matrix exponential, we obtain that $\exp\{c\,\mathrm{R}\} = \mathrm{S}[\mathrm{R}]\, \exp\{c\,\mathrm{J}[\mathrm{R}]\}\, \mathrm{S}[\mathrm{R}]^{-1}$, where $c \in \mathbb{C}$ is an arbitrary constant, such that only the exponential of J[R] needs to be computed. The exponential acts on each Jordan block independently and the exponential of a single Jordan block can be easily obtained[5] as

$$\exp\left\{c\,\mathrm{J}_a^d\right\} = \mathrm{e}^{c\,a} \begin{pmatrix} 1 & c & \frac{c^2}{2} & \dots & \frac{c^{n-1}}{(n-1)!} \\ 0 & 1 & c & \dots & \frac{c^{n-2}}{(n-2)!} \\ \vdots & \ddots & \ddots & & \vdots \\ 0 & \dots & 0 & 1 & c \\ 0 & & \dots & 0 & 1 \end{pmatrix}. \tag{4.7}$$

We are now ready to understand how solution (4.4), which we now write as

$$\mathbf{M}(r, \epsilon) = \mathrm{S}[\mathrm{R}(\epsilon)] \exp\{\log r\ \mathrm{J}[\mathrm{R}(\epsilon)]\}\, \mathrm{S}[\mathrm{R}(\epsilon)]^{-1}\, \mathbf{B}(\epsilon),$$

develops non-trivial analytic behaviour at $r = 0$. The only non-trivial analytic structure comes from the exponentiation of each Jordan block and appears in two distinct forms: (i) the exponential pre-factor in (4.7), which is of the form $r^{\lambda_i}$ for $c = \log r$, and (ii) the logarithms arising from the exponentiation of non-trivial Jordan blocks. This corresponds exactly to our decomposition (4.2). Hence explicit logarithms appear by the exponentiation of non-trivial Jordan blocks of the residue matrix and the highest power of the logarithms is given by the dimension of the corresponding Jordan block.

[5] Set $\mathrm{J}_a^d = a\mathbb{1}_d + \mathrm{N}$, where N is the (nilpotent) matrix with 1 on the super-diagonal and 0 everywhere else, and compute $\exp\{c\,\mathrm{N}\}$ explicitly.

In turn, the fact that the general structure of the asymptotic expansion of a Feynman integral must be of the form given by equation (4.2) puts constraints on the system of differential equations satisfied by the master integrals. In particular, it means that

- Feynman integrals do not have essential singularities, such that the system of differential equations has a *regular singular point* at $r = 0$.
- If the system is in a form where it can be expanded as in equation (4.3), then the residue matrix $\mathrm{R}(\epsilon)$ must have eigenvalues of the form $\lambda_i = v_i - n_i\epsilon$ with $v_i \in \mathbb{Z}$ and $n_i \in \mathbb{N}$.

At first sight, the fact that the system has a regular singular point does not imply that it can be directly expanded as in equation (4.3) but there are known methods to achieve such a task, as we discuss below.

### 4.2 Description of the algorithm

We now present our algorithm in detail. Let us consider a family of master integrals $\mathbf{M}(r,\epsilon) = (M_1(r,\epsilon), M_2(r,\epsilon), \ldots)$ satisfying a linear system of differential equations given by

$$\partial_r \mathbf{M}(r,\epsilon) = \mathrm{A}(r,\epsilon)\,\mathbf{M}(r,\epsilon), \tag{4.8}$$

where A is a $m \times m$ matrix whose entries are rational functions of $r$ and $\epsilon$. Under a linear redefinition of the master integrals $\mathbf{M}(r,\epsilon) \to \mathbf{M}'(r,\epsilon) = \mathrm{T}(r,\epsilon)\mathbf{M}(r,\epsilon)$ the system of differential equation becomes

$$\partial_r \mathbf{M}'(r,\epsilon) = \mathrm{A}'(r,\epsilon)\mathbf{M}'(r,\epsilon) \tag{4.9}$$

where the transformed matrix is given by

$$\mathrm{A}'(r,\epsilon) = \mathcal{D}_r^{\mathrm{T}}[\mathrm{A}(r,\epsilon)] \equiv \mathrm{T}(r,\epsilon)^{-1}\left(\mathrm{A}(r,\epsilon) - \partial_r\right)\mathrm{T}(r,\epsilon)\ . \tag{4.10}$$

The first step of our algorithm is to find a transformation such that the system can be expanded as in equation (4.3) and solved by (4.4) in the vicinity of $r = 0$. In general, even if the system of differential equations has a regular singular point at $r = 0$, the Laurent expansion of the matrix $\mathrm{A}(r,\epsilon)$ around $r = 0$ can have poles of order higher than one. In this case, the residue matrix of the highest pole must be nilpotent, hinting at a ‘fake’ singularity. This issue was investigated long ago by Moser [55] who showed that it is *always* possible to construct a transformation – that we will call $\mathrm{T}_{\mathrm{rank}}$ in what follows – such that the transformed matrix $\mathcal{D}_x^{\mathrm{T}_{\mathrm{rank}}}[\mathrm{A}]$ has a singularity of order one as in equation (4.3). When A is rational, such a transformation has further been shown to be rational as well by Barkatou and Pflüegel [56,57], who also provided a practical algorithm[6] for its computation. It should be noted that such a transformation is not unique. This algorithm has been recently used in reference [41] in order to reduce maximally the degree of all the singular points with respect to a given variable *simultaneously*, whenever it is possible. If this is achieved, the solution of the differential equation can be easily obtained in an $\epsilon$-expansion [23, 58]. However, the very existence of such a form is a tremendous mathematical problem closely related to Hilbert's 21st problem [41]. Here we completely avoid such considerations because reducing the apparent order of a *single* singular point is always possible and can be automatized, possibly at the price of worsening the behaviour of the matrix A at other singular points.

[6]This algorithm, called *rational Moser reduction*, is for example implemented in the standard `Maple` package `DEtools` as `moser_reduce` and `super_reduce`.

The second step of our algorithm is to normalize the eigenvalues of the residue matrix of the transformed system to be proportional to $\epsilon$ as first suggested in [41]. After performing the transformation $\mathrm{T}_{\mathrm{rank}}$, the system must have the following expansion

$$\mathcal{D}_r^{\mathrm{T}_{\mathrm{rank}}}[\mathrm{A}(r,\epsilon)] = \frac{\mathrm{R}_0(\epsilon)}{r} + \dots\,,$$

where the dots indicate less singular contributions. The eigenvalues of the residue matrix $\mathrm{R}_0(\epsilon)$ can be normalized by the following standard trick. Let $\mathrm{S}[\mathrm{R}_0]$ be the similarity transformation of $\mathrm{R}_0$ into its Jordan normal form and further denote by $\vec{u}_i$ the $i$-th column vector of this transformation and by $\vec{v}_i$ the $i$-th row vector of its inverse, that is

$$\mathrm{S}[\mathrm{R}_0] = (\vec{u}_1, \dots, \vec{u}_m), \qquad \mathrm{S}[\mathrm{R}_0]^{-1} = (\vec{v}_1, \dots, \vec{v}_m)^{\mathsf{T}}. \tag{4.11}$$

These vectors are the generalized eigenvectors of the matrix $\mathrm{R}_0(\epsilon)$. It is possible to shift the $i$-th eigenvalue of the residue matrix from $\lambda_i$ to $\lambda_i + 1$ using the transformation

$$\mathrm{T}_i(r,\epsilon) = \mathbb{1} + \frac{1-r}{r}\,\vec{u}_i\,\vec{v}_i^{\mathsf{T}}, \tag{4.12}$$

with inverse

$$\mathrm{T}_i^{-1}(r,\epsilon) = \mathbb{1} - (1-r)\,\vec{u}_i\,\vec{v}_i^{\mathsf{T}}. \tag{4.13}$$

In reference [41], such a transformation is called a balance between 0 and $\infty$. Since all the eigenvalues of $\mathrm{R}_0$ must be of the form $\lambda_i = v_i - n_i\epsilon$ with $v_i \in \mathbb{Z}$ and $n_i \in \mathbb{N}$, we can normalize $\mathrm{R}_0$ by applying the total transformation

$$\mathrm{T}_{\mathrm{norm.}}(r,\epsilon) = \prod_i \mathrm{T}_i^{v_i}(r,\epsilon). \tag{4.14}$$

The residue matrix of the transformed system $\mathcal{D}_x^{\mathrm{T}_{\mathrm{tot.}}}[\mathrm{A}(x,\epsilon)]$ , where $\mathrm{T}_{\mathrm{tot.}} = \mathrm{T}_{\mathrm{rank}}\mathrm{T}_{\mathrm{norm.}}$, then has eigenvalues proportional to $\epsilon$. Note that this is independent from the order by which we construct this transformation, since the matrices $\vec{u}_i\,\vec{v}_i^{\mathsf{T}}$ commute with each other for different $i$.

The third and final step of our algorithm is to construct the expansion explicitly. After the two first steps have been carried out, we must have by construction that

$$\mathcal{D}_r^{\mathrm{T}_{\mathrm{tot.}}}[\mathrm{A}(r,\epsilon)] = \frac{\mathrm{R}(\epsilon)}{r} + \sum_{n=0}^{\infty} \mathrm{A}_n(\epsilon) r^n,$$

where the eigenvalues of the residue matrix R are all proportional to $\epsilon$. Let us further define the vector $\tilde{\mathbf{M}}$ given by

$$\mathbf{M}(r,\epsilon) = r^{\mathrm{R}(\epsilon)}\,\tilde{\mathbf{M}}(r,\epsilon), \tag{4.15}$$

where $r^{\mathrm{R}(\epsilon)} = \exp\{\log(r)\mathrm{R}(\epsilon)\}$ is the fundamental matrix of the system for which only the pole is retained. Following our previous discussion, we know that the matrix exponential contains the whole non-trivial analytic structure and we expect $\tilde{\mathbf{M}}$ to have a Taylor expansion at $r = 0$. Indeed by using the transformation rule (4.9), we see that $\tilde{\mathbf{M}}$ satisfies a differential equation without any singularity at $r = 0$, namely

$$\partial_r \tilde{\mathbf{M}}(r,\epsilon) = r^{-\mathrm{R}(\epsilon)} \left( \sum_{n=0}^{\infty} \mathrm{A}_n(\epsilon) r^n \right) r^{\mathrm{R}(\epsilon)}\,\tilde{\mathbf{M}}(r,\epsilon) = \left( \sum_{n=0}^{\infty} \tilde{\mathrm{A}}_n(r,\epsilon) r^n \right) \tilde{\mathbf{M}}(r,\epsilon), \tag{4.16}$$

where we defined $\tilde{\mathrm{A}}_i(r,\epsilon) = r^{-\mathrm{R}(\epsilon)} \mathrm{A}_i(\epsilon)\, r^{\mathrm{R}(\epsilon)}$. Because we normalized the eigenvalues of the matrix $\mathrm{R}(r,\epsilon)$, every entry of the matrices $\tilde{\mathrm{A}}_i$ must be proportional to $r^{n\epsilon}$ with $n \in \mathbb{Z}$, such that every term in (4.16) contributes to a single order in $r$ for $\epsilon = 0$. This fact is of crucial importance because it allows us to generate terms of the expansion order by order in $r$. The sought after expansion can be obtained by means of a Dyson series as follows. Turning (4.16) into an integral equation, we get

$$\tilde{\mathbf{M}}(r,\epsilon) = \mathbf{B}(\epsilon) + \int_0^r \mathrm{d}r_1 \left( \sum_{n=0}^{\infty} \tilde{\mathrm{A}}_n(r_1,\epsilon)\, r_1^n \right) \tilde{\mathbf{M}}(r_1,\epsilon), \tag{4.17}$$

where $\mathbf{B}(\epsilon)$ is an arbitrary vector of boundary conditions. Substituting this very equation into itself iteratively, we get the expansion

$$\tilde{\mathbf{M}}(r,\epsilon) = \sum_{n=0}^{\infty} \mathrm{D}_n(r,\epsilon)\,\mathbf{B}(\epsilon), \tag{4.18}$$

where the matrix coefficients of the Dyson series are given by

$$\begin{aligned}
\mathrm{D}_0(r,\epsilon) &= \mathbb{1}, \\
\mathrm{D}_1(r,\epsilon) &= \int_0^r \mathrm{d}r_1\; \tilde{\mathrm{A}}_0(r_1,\epsilon), \\
\mathrm{D}_2(r,\epsilon) &= \int_0^r \mathrm{d}r_1\; \tilde{\mathrm{A}}_0(r_1,\epsilon) \int_0^{r_1} \mathrm{d}r_2\; \tilde{\mathrm{A}}_0(r_2,\epsilon) + \int_0^r \mathrm{d}r_1\; r_1 \tilde{\mathrm{A}}_1(r_1,\epsilon), \\
\mathrm{D}_3(r,\epsilon) &= \int_0^r \mathrm{d}r_1\; \tilde{\mathrm{A}}_0(r_1,\epsilon) \int_0^{r_1} \mathrm{d}r_2\; \tilde{\mathrm{A}}_0(r_2,\epsilon) \int_0^{r_2} \mathrm{d}r_3\; \tilde{\mathrm{A}}_0(r_3,\epsilon) \\
&\quad + \int_0^r \mathrm{d}r_1\; \tilde{\mathrm{A}}_0(r_1,\epsilon) \int_0^{r_1} \mathrm{d}r_2\; r_2 \tilde{\mathrm{A}}_1(r_2,\epsilon) \\
&\quad + \int_0^r \mathrm{d}r_1\; r_1 \tilde{\mathrm{A}}_1(r_1,\epsilon) \int_0^{r_1} \mathrm{d}r_2\; \tilde{\mathrm{A}}_0(r_2,\epsilon) \\
&\quad + \int_0^r \mathrm{d}r_1\; r_1^2 \tilde{\mathrm{A}}_2(r_1,\epsilon), \\
&\dots
\end{aligned} \tag{4.19}$$

where the matrices $\tilde{\mathrm{A}}_i$ are defined as in (4.16). Note that every entry of $\mathrm{D}_m$ is proportional to $r^{m+n\epsilon}$ with $n \in \mathbb{Z}$, such that the expansion (4.18) indeed generates a Taylor expansion for each scaling. Note however that the boundary conditions $\mathbf{B}(\epsilon)$ are common to all orders and have to be obtained from the first non-vanishing order of each master integral.

Summarizing, the solution of the differential equation (4.8) can be written as an expansion in the small parameter $r$ by the following steps:

- Reduce the singular rank of the matrix A at the point $r = 0$ to its minimal value using the rational Moser algorithm [56, 57]. This produces a change of basis $\mathrm{T}_{\mathrm{rank}}$.
- Normalize the eigenvalues of the residue matrix of $\mathcal{D}_r^{\mathrm{T}_{\mathrm{rank}}}[\mathrm{A}]$ at $r = 0$ using equation (4.14). This produces a further change of basis $\mathrm{T}_{\mathrm{norm.}}$.
- The solution of the system of differential equations can then be written as

$$\mathbf{M}(r, \epsilon) = \mathrm{F}(r, \epsilon)\, \mathbf{B}(\epsilon), \tag{4.20}$$

with

$$\mathrm{F}(r, \epsilon) = \mathrm{T}_{\mathrm{tot.}}(r, \epsilon)\, r^{\mathrm{R}(\epsilon)} \sum_{n=0}^{\infty} \mathrm{D}_n(r, \epsilon), \tag{4.21}$$

where $\mathrm{T}_{\mathrm{tot.}}(r, \epsilon) = \mathrm{T}_{\mathrm{rank}}(r, \epsilon)\mathrm{T}_{\mathrm{norm.}}(r, \epsilon)$ is the total transformation, $\mathrm{R}(\epsilon)$ is the residue matrix of $\mathcal{D}_r^{\mathrm{T}_{\mathrm{tot.}}}[\mathrm{A}(r, \epsilon)]$ at $r = 0$, and the coefficients $\mathrm{D}_n(r, \epsilon)$ of the Dyson series are given by equation (4.19).

Note that even if the Dyson series can be obtained order by order, the transformation matrices $\mathrm{T}_{\mathrm{rank}}$, $\mathrm{T}_{\mathrm{norm}}$ and $\mathrm{S}[\mathrm{A}_0]$ are in general not homogeneous in $r$ and will mix the different powers of the expansion. In practice one has to be careful and track which terms of the Dyson series have been truncated.

The computation of the Jordan normal form in $r^{\mathrm{R}(\epsilon)}$ can be computationally intensive and become untractable for big systems of differential equations. In this case, a simple extension of our algorithm presented in appendix A can be used to take advantage of the fact that most systems of differential equations ecountered in practice are not completely coupled and can be solved iteratively, avoiding the hassle of dealing with large matrices $\mathrm{R}(\epsilon)$.

### 4.3 Example: Expansion of a family of factorizable two-loop integrals

We illustrate the method of expansion presented in the previous section by considering the following family $I[\nu_1, \nu_2, \nu_3, \nu_4]$ of factorizable two-loop integrals:

$$\int \frac{\mathrm{d}^d k}{i\pi^{d/2}} \frac{\mathrm{d}^d l}{i\pi^{d/2}} \frac{1}{\left[(k-l)^2 - m_b^2\right]^{\nu_1} \left[l^2 - m_b^2\right]^{\nu_2} \left[(l+p_1)^2 - m_b^2\right]^{\nu_3} \left[(l+p_1+p_2)^2 - m_b^2\right]^{\nu_4}},$$

with $p_1^2 = p_2^2 = 0$ and $(p_1 + p_2)^2 = s$. We choose the following vector of master integrals:

$$\mathbf{M} = (I[1,1,0,0], I[1,1,0,1], I[1,1,1,1])^{\mathsf{T}}.$$

This family corresponds to the closed sub-system $T_1^{\rm V}[\nu_2, \nu_3, \nu_4, \nu_1, 0, 0, 0]$ of the two-loop integrals appearing in the computation of virtual contributions that we will consider in section 5.1. We want to expand these integrals with respect to $m_b^2$. As explained earlier, we set $s = -1$ and consider only the variable $r = -m_b^2/s$ from now on. Note that these integrals can be factorized as products of one-loop integrals such that their computation using our method of expansion is somewhat artificial since a direct computation of the one-loop integrals would be easier. However, the very fact that these integrals are easy to compute allows us to illustrate our method pedagically in details.

Differential equations with respect to $r$ for the vector of master integrals $\mathbf{M}$ can be easily obtained and read

$$\partial_r \mathbf{M} = \mathrm{A}(r,\epsilon)\mathbf{M}, \qquad \text{with} \quad \mathrm{A}(r,\epsilon) = \begin{pmatrix} \frac{2(1-\epsilon)}{r} & 0 & 0 \\ -\frac{2(1-\epsilon)}{r(4r+1)} & \frac{1+6r-\epsilon(1+8r)}{r(4r+1)} & 0 \\ \frac{1-\epsilon}{r^2(4r+1)} & -\frac{1-2\epsilon}{r(4r+1)} & \frac{1-2\epsilon}{r} \end{pmatrix}. \tag{4.22}$$

We see that the matrix A has a second order pole at $r = 0$ and the first step of our algorithm is to find the transformation $\mathrm{T_{rank}}(r)$. This can be done either by using the rational Moser algorithm or, in this simple case, by inspection. The second transformation matrix $\mathrm{T_{norm}}(r)$ can then be obtained using equation (4.14), and we have

$$\mathrm{T_{rank}}(r) = \begin{pmatrix} r & 0 & 0 \\ 0 & 1 & 0 \\ 0 & 0 & 1 \end{pmatrix}, \qquad \text{and} \quad \mathrm{T_{norm.}}(r) = \begin{pmatrix} r & 0 & 0 \\ 0 & r & 0 \\ 0 & 0 & r \end{pmatrix}.$$

Here the transformation matrices have a very simple diagonal form but in general the two first steps of the algorithm can be computationally intensive. Under the transformation $\mathrm{T_{tot.}}(r) = \mathrm{T_{rank}}(r)\mathrm{T_{norm.}}(r)$, the matrix A becomes

$$\mathcal{D}_r^{\mathrm{T_{tot.}}}[\mathrm{A}(r,\epsilon)] = \frac{\mathrm{R}(\epsilon)}{r} + \mathrm{A}_0(\epsilon) + \mathcal{O}(r),$$

with

$$\mathrm{R}(\epsilon) = \begin{pmatrix} -2\epsilon & 0 & 0 \\ 0 & -\epsilon & 0 \\ 1-\epsilon & -1+2\epsilon & -2\epsilon \end{pmatrix}, \qquad \mathrm{A}_0(\epsilon) = 2\begin{pmatrix} 0 & 0 & 0 \\ -(1-\epsilon) & 1-2\epsilon & 0 \\ -2(1-\epsilon) & 2-4\epsilon & 0 \end{pmatrix}.$$

The matrix R has the following Jordan decomposition

$$\mathrm{S[R]} = \begin{pmatrix} 0 & \frac{1}{1-\epsilon} & 0 \\ 0 & 0 & -\frac{\epsilon}{1-2\epsilon} \\ 1 & 0 & 1 \end{pmatrix}, \qquad \mathrm{J[R]} = \mathrm{S[R]}^{-1}\mathrm{R}\,\mathrm{S[R]} = \begin{pmatrix} -2\epsilon & 1 & 0 \\ 0 & -2\epsilon & 0 \\ 0 & 0 & -\epsilon \end{pmatrix}, \tag{4.23}$$

and we see that its eigenvalues are properly normalized. We are now ready to perform the expansion. The solution of the differential equation is then given by (4.20) and reads

$$\mathbf{M}(r,\epsilon) = \mathrm{T_{tot.}}(r)\, r^{\mathrm{R}(\epsilon)} \left(\mathbb{1} + \mathrm{D}_1(r,\epsilon) + \mathcal{O}(r^2)\right)\mathbf{B}(\epsilon) \tag{4.24}$$

where $r^{\mathrm{R}(\epsilon)} = \exp\{\log r\,\mathrm{R}(\epsilon)\}$ can be easily obtained from (4.23), using the explicit formula (4.7), and is given by

$$r^{\mathrm{R}(\epsilon)} = \mathrm{S}[\mathrm{R}]\,r^{\mathrm{J}[\mathrm{R}]}\,\mathrm{S}[\mathrm{R}]^{-1} = \begin{pmatrix} r^{-2\epsilon} & 0 & 0 \\ 0 & r^{-\epsilon} & 0 \\ r^{-2\epsilon}(1-\epsilon)\log r & \left(r^{-\epsilon}-r^{-2\epsilon}\right)\left(2-\frac{1}{\epsilon}\right) & r^{-2\epsilon} \end{pmatrix}.$$

The first order of the Dyson series can be easily computed as

$$\mathrm{D}_1(\epsilon, r) = \int_0^r \mathrm{d}r_1\, r_1^{-\mathrm{R}(\epsilon)}\mathrm{A}_0(r_1,\epsilon) r_1^{\mathrm{R}(\epsilon)} = r\mathrm{D}_1^{(0)}(\epsilon) + r^{1-\epsilon}\mathrm{D}_1^{(-1)}(\epsilon),$$

where

$$\mathrm{D}_1^{(0)}(\epsilon) = \begin{pmatrix} 0 & 0 & 0 \\ 0 & 2-4\epsilon & 0 \\ \frac{2(\epsilon-1)}{\epsilon} & -\frac{2(1-2\epsilon)^2}{\epsilon} & 0 \end{pmatrix}, \quad \text{and} \quad \mathrm{D}_1^{(-1)}(\epsilon) = \begin{pmatrix} 0 & 0 & 0 \\ -2 & 0 & 0 \\ \frac{2-4\epsilon}{\epsilon} & 0 & 0 \end{pmatrix}.$$

Higher orders can be obtained similarly using (4.19). Setting $\mathbf{B}(\epsilon) \equiv (B_1(\epsilon), B_2(\epsilon), B_3(\epsilon))$ we obtain the solution (4.24) explicitly as

$$\begin{aligned} M_1(r,\epsilon) =& r^{2-2\epsilon}\left(B_1 + \mathcal{O}\left(r^2\right)\right), \\ M_2(r,\epsilon) =& r^{1-\epsilon}\left(1+(2-4\epsilon)r\right)B_2 - 2r^{2-2\epsilon}B_1 + \mathcal{O}\left(r^2\right) \\ M_3(r,\epsilon) =& -r^{1-\epsilon}\left(\left(\frac{(1-2\epsilon)}{\epsilon} - \frac{(2-8\epsilon^2)}{1+\epsilon}r\right)B_2 + \mathcal{O}\left(r^2\right)\right) \\ & - r^{1-2\epsilon}\left(\frac{2\epsilon-1}{\epsilon}B_2 - B_3 - \left((1-\epsilon)\log r - 2r\right)B_1 + \mathcal{O}\left(r^2\right)\right). \end{aligned} \tag{4.25}$$

We finally have to determine the boundary conditions $B_i$. The three two-loop integrals considered here are factorizable, i.e. they can be written as the product of two one-loop integrals, for which all order in $\epsilon$ results are known. In particular they have the following asymptotic behavior

$$\begin{aligned} M_1(r,\epsilon) &= r^{2(1-\epsilon)}\left(\frac{\Gamma(1+\epsilon)}{\epsilon(1-\epsilon)} + \mathcal{O}(r)\right)^2, \\ M_2(r,\epsilon) &= \left(\frac{\Gamma(1-\epsilon)^2\Gamma(\epsilon)}{\Gamma(2-2\epsilon)} + \mathcal{O}(r)\right) r^{1-\epsilon}\left(\frac{\Gamma(1+\epsilon)}{\epsilon(1-\epsilon)} + \mathcal{O}(r)\right), \\ M_3(r,\epsilon) &= r^{1-\epsilon}\left(\frac{\Gamma(1+\epsilon)}{\epsilon(1-\epsilon)} + \mathcal{O}(r)\right) \\ &\quad \times\left(-r^{-\epsilon}\,\Gamma(\epsilon)(\gamma_E - \log(r) + \psi(\epsilon) + \mathcal{O}(r)) - \frac{\Gamma(1+\epsilon)\Gamma(1-\epsilon)^2}{\epsilon^2\Gamma(1-2\epsilon)} + \mathcal{O}(r)\right), \end{aligned}$$

where $\psi$ denotes the digamma function and $\gamma_E$ is the Euler-Mascheroni constant. We see that the structure of these asymptotic expansions matches the solution (4.25) exactly at

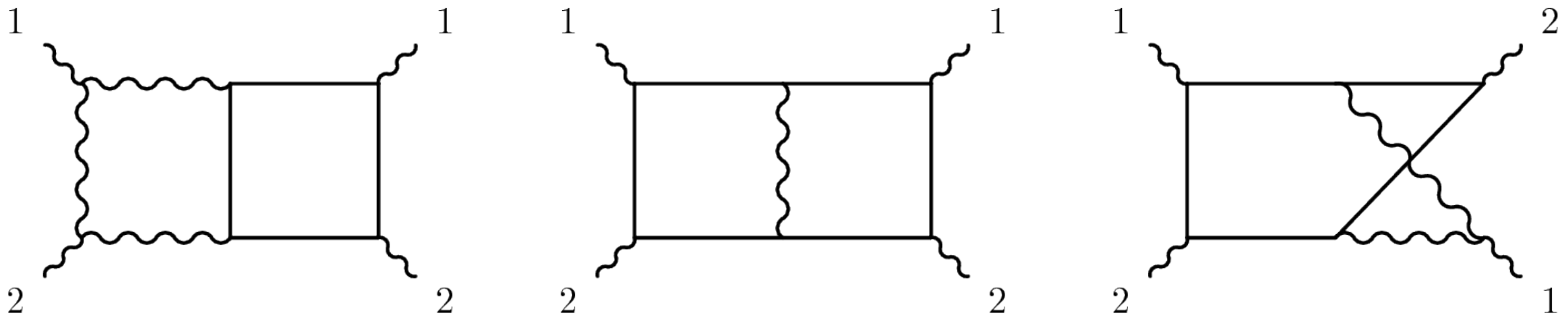

**Figure 2:** Topologies $T_i^{\mathrm{V}}[\nu_1,\ldots,\nu_7]$, $i=1,2,3$, appearing in the computation of the virtual contributions. Wavy lines indicate massless internal propagators and external legs, while plain lines indicate massive internal propagators and external legs.

lowest order in $r$, such that we can easily determine the boundary conditions to be

$$
\begin{aligned}
B_1(\epsilon) &= \left(\frac{\Gamma(1+\epsilon)}{\epsilon(1-\epsilon)}\right)^2, \\
B_2(\epsilon) &= \frac{\Gamma(1-\epsilon)^2\Gamma(1+\epsilon)\Gamma(\epsilon)}{\epsilon(1-\epsilon)\Gamma(2-2\epsilon)}, \\
B_3(\epsilon) &= -\frac{\Gamma(\epsilon+1)\left(\epsilon^2\Gamma(1-2\epsilon)\Gamma(\epsilon)(\psi(\epsilon)+\gamma_E)+\Gamma(1-\epsilon)^2\Gamma(\epsilon+1)\right)}{(1-\epsilon)\epsilon^3\Gamma(1-2\epsilon)}.
\end{aligned}
$$

The higher order terms in (4.25) are then completely determined as well.

## 5. Details on the computation of the master integrals

### 5.1 Master integrals appearing in the virtual contributions

In this section we detail the computation of the two-loop scalar integrals needed in the virtual contributions $\sigma^{\mathrm{V;int.}}_{gg\to H}$. There are 17 master integrals falling into the three distinct topologies shown in figure 2 and defined by

$$
\begin{aligned}
T_1^{\mathrm{V}}[\nu_1,\ldots,\nu_7] &= \int \tilde{\mathrm{d}}^d k \tilde{\mathrm{d}}^d l D_0^{\nu_1}(k) D_0^{\nu_2}(k_1) D_0^{\nu_3}(k_{12}) D_{m_b}^{\nu_4}(k-l) D_{m_b}^{\nu_5}(l_{12}) D_{m_b}^{\nu_6}(l_1) D_{m_b}^{\nu_7}(l), \\
T_2^{\mathrm{V}}[\nu_1,\ldots,\nu_7] &= \int \tilde{\mathrm{d}}^d k \tilde{\mathrm{d}}^d l D_{m_b}^{\nu_1}(k) D_{m_b}^{\nu_2}(k_2) D_{m_b}^{\nu_3}(k_{12}) D_0^{\nu_4}(k-l) D_{m_b}^{\nu_5}(l_{12}) D_{m_b}^{\nu_6}(l_2) D_{m_b}^{\nu_7}(l), \\
T_3^{\mathrm{V}}[\nu_1,\ldots,\nu_7] &= \int \tilde{\mathrm{d}}^d k \tilde{\mathrm{d}}^d l D_{m_b}^{\nu_1}(k) D_{m_b}^{\nu_2}(k-l_1) D_{m_b}^{\nu_3}(k_{12}) D_{m_b^2}^{\nu_4}(l_{12}) D_{m_b}^{\nu_5}(l_1) D_{m_b}^{\nu_6}(l) D_0^{\nu_7}(k-l),
\end{aligned}
$$

where $\tilde{\mathrm{d}}^d p \equiv -i\pi^{-d/2}\mathrm{e}^{-\epsilon\gamma_E}\mathrm{d}^d p$ and we defined the scalar propagator

$$
D_m(p) = \frac{1}{p^2-m^2}.
$$

Although every topology is closed under IBP reduction, such that the corresponding differential equations could be treated separately, we found easier to work with all of them at the same time, avoiding the duplication of simple master integrals shared by different topologies. We follow reference [12] and choose for our master integrals the following:

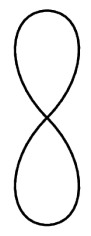 $= T_1^{\mathrm{V}}[0,0,0,1,0,1,0] = M_1^{\mathrm{V}},$

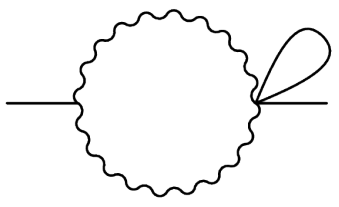

$$= T_1^{\mathrm{V}}[1,0,1,1,0,0,0] = M_2^{\mathrm{V}},$$

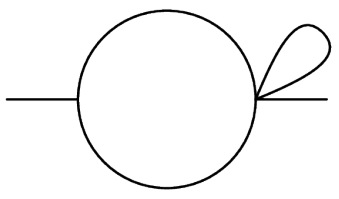

$$= T_1^{\mathrm{V}}[0,0,0,1,1,0,1] = M_3^{\mathrm{V}},$$

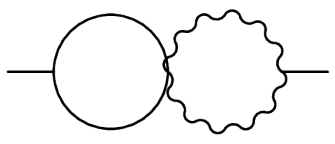

$$= T_1^{\mathrm{V}}[1,0,1,0,1,0,1] = M_4^{\mathrm{V}},$$

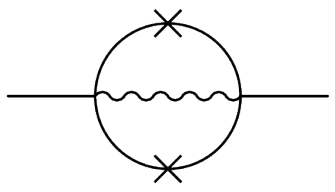

$$= T_1^{\mathrm{V}}[1,0,0,2,2,0,0] = M_5^{\mathrm{V}},$$

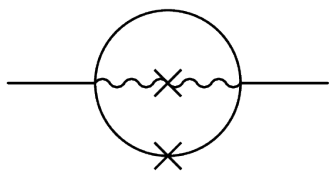

$$= T_1^{\mathrm{V}}[2,0,0,2,1,0,0] = M_6^{\mathrm{V}},$$

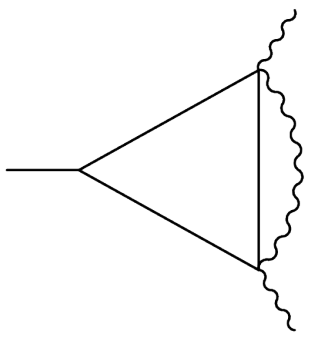

$$= T_1^{\mathrm{V}}[0,1,0,1,1,0,1] = M_7^{\mathrm{V}},$$

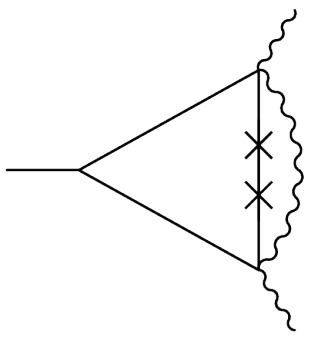

$$= T_1^{\mathrm{V}}[0,1,0,3,1,0,1] = M_8^{\mathrm{V}},$$

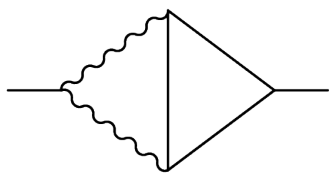

$$= T_1^{\mathrm{V}}[1,0,1,1,1,0,1] = M_9^{\mathrm{V}},$$

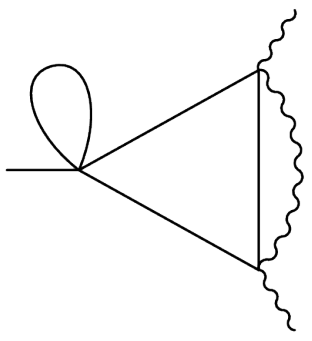

$$= T_1^{\mathrm{V}}[0,0,0,1,1,1,1] = M_{10}^{\mathrm{V}},$$

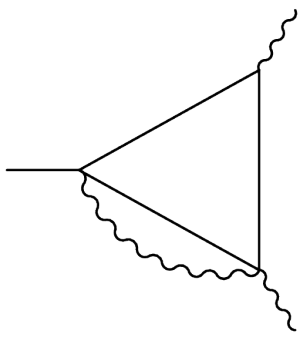

$$= T_1^{\mathrm{V}}[0,0,1,1,0,1,1] = M_{11}^{\mathrm{V}},$$

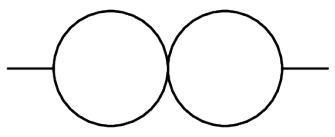

$$= T_2^{\mathrm{V}}[1,0,1,0,1,0,1] = M_{13}^{\mathrm{V}},$$

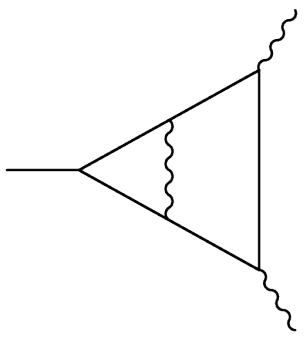

$$= T_2^{\mathrm{V}}[1,0,1,1,1,1,1] = M_{14}^{\mathrm{V}},$$

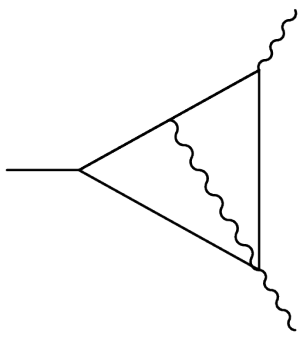

$$= T_2^{\mathrm{V}}[0,1,1,1,1,0,1] = M_{15}^{\mathrm{V}},$$

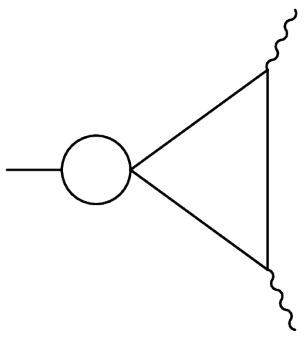

$$= T_2^{\mathrm{V}}[1,0,1,0,1,1,1] = M_{16}^{\mathrm{V}},$$

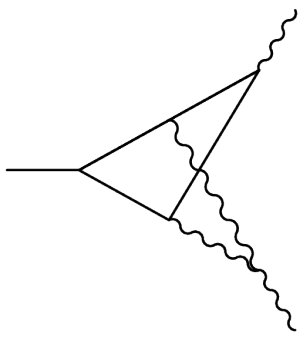

$$= T_3^{\mathrm{V}}[1,0,1,1,1,1,1] = M_{17}^{\mathrm{V}},$$

where we omitted master $M_{12}^{\mathrm{V}}$ which has a non-trivial numerator and can be written as

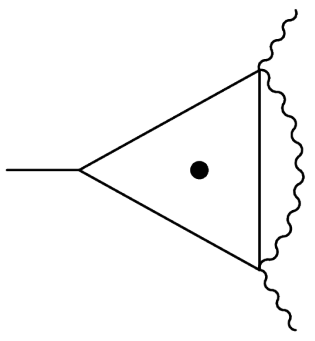

$$= \frac{1}{2}\left(T_2^{\mathrm{V}}[0,1,0,1,1,-1,1] - T_2^{\mathrm{V}}[0,0,0,1,1,0,1]\right) = M_{12}^{\mathrm{V}}.$$

Up to a pre-factor of $(-s)^{2d-\nu}$, where $\nu$ denotes the total inverse power of propagators, these integrals can be written using the dimensionless variable

$$x = \frac{\sqrt{1-4m_b^2/s}-1}{\sqrt{1-4m_b^2/s}+1}, \tag{5.1}$$

such that $m_b^2 \to 0$ corresponds to $x = -m_b^2/s + \mathcal{O}(m_b^4) \to 0$. Since this pre-factor can be easily reconstructed from dimensional analysis, we will set $s = -1$ and use only the variable $x$ to present our results. Introducing such a variable is necessary when considering the system at arbitrary values of the mass $m_b^2$ but is not mandatory in our case. We use it here to simplify comparison with the known full results. Analytic continuation is unambiguously defined by the causal prescription and we have

$$-\frac{m_b^2}{s} \to -\frac{m_b^2 - i0}{s+i0} = -\frac{m_b^2}{s} + i0, \quad \text{as} \qquad m_b \to 0^+ \quad \text{with} \quad s > 0,$$

such that $x \to x + i0$ as well.

We solve the corresponding differential equations for the vector of master integrals $\mathbf{M}^{\mathrm{V}} = (M_1^{\mathrm{V}}, \ldots, M_{17}^{\mathrm{V}})$ as an expansion in the parameter $x$ using the methods presented in the previous section. The reduction to master integrals and the differential equations have been obtained in two different ways and found to agree. The differential equation are then normalized and expanded using our algorithm and the solution can be written as

$$\mathbf{M}^{\mathrm{V}}(x,\epsilon) = \mathrm{F}^{\mathrm{V}}(x,\epsilon)\,\mathbf{B}^{\mathrm{V}}(\epsilon), \tag{5.2}$$

where the fundamental matrix of the system is given by

$$\mathrm{F}^{\mathrm{V}}(x,\epsilon) = \mathrm{T}^{\mathrm{V}}_{\mathrm{tot.}}(x,\epsilon)\,x^{\mathrm{R}^{\mathrm{V}}(\epsilon)}\sum_{n=0}^{\infty}\mathrm{D}^{\mathrm{V}}_{n}(x,\epsilon),$$

where $\mathrm{T}^{\mathrm{V}}_{\mathrm{tot.}} = \mathrm{T}^{\mathrm{V}}_{\mathrm{rank}}\mathrm{T}^{\mathrm{V}}_{\mathrm{norm.}}$ is the transformation used to achieve the normalized form presented in section 4, $\mathrm{R}^{V}$ is the residue matrix at $x = 0$ of the properly normalized differential equation matrix and $\mathbf{B}^{\mathrm{V}} = (B^{\mathrm{V}}_{1}, \ldots, B^{\mathrm{V}}_{17})$ is the vector of boundary conditions. The Dyson series is defined as in (4.18). The differential equations, transformation matrix $\mathrm{T}^{V}_{\mathrm{tot.}}$ and the corresponding boundary conditions can be found in the ancillary file attached to this article. The expansion (5.2) can then be unambiguously reconstructed using the method presented in section 4.

Since we are expanding around $x = 0$, the boundary conditions must be obtained from the asymptotic expansion of the integrals at $x = 0$, which are using the strategy of region [52,53]. Note that there are 7 factorizable integrals that can be written as a product of the following one-loop integrals with asymptotic behaviour

$$M^{\mathrm{B}}_{1} =$$

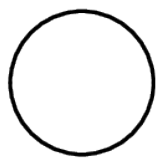

$$= x^{1-\epsilon}\left(\frac{\Gamma(1+\epsilon)}{\epsilon(1-\epsilon)} + \mathcal{O}(x)\right),$$

$$M^{\mathrm{B}}_{2} =$$

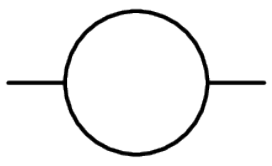

$$= \frac{\Gamma(1-\epsilon)^2\Gamma(\epsilon)}{\Gamma(2-2\epsilon)} + \mathcal{O}(x),$$

$$M^{\mathrm{B}}_{3} =$$

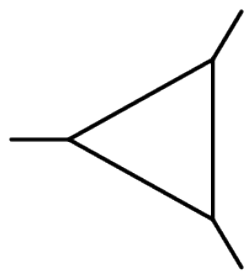

$$= -x^{-\epsilon}\Gamma(\epsilon)\left(\gamma_E - \log x + \psi(\epsilon)\right) - \frac{\Gamma(1+\epsilon)\Gamma(1-\epsilon)^2}{\epsilon^2\Gamma(1-2\epsilon)} + \mathcal{O}(x), \tag{5.3}$$

such that the corresponding boundary conditions can be easily obtained. We were able to obtain the all the boundary conditions at all orders in $\epsilon$ and we see that they can be expressed using gamma functions and hypergeometric functions with unit argument, which can be easily expanded in $\epsilon$ using the package `HypExp` [35,36].

When expanded in $\epsilon$, our solution reproduces the results of reference [12] in the small

$x$ limit and are given by

$$
\begin{aligned}
M_1^V(x,\epsilon) = {} & \frac{1}{\epsilon^2}\left\{x^2+4x^3+10x^4+\mathcal{O}\left(x^5\right)\right\} \\
& + \frac{1}{\epsilon}\left\{x^2(2-2\log x)+x^3(4-8\log x)+x^4(2-20\log x)+\mathcal{O}\left(x^5\right)\right\} \\
& + x^2\left(\zeta_2+2\log^2 x-4\log x+3\right)+x^3\left(4\zeta_2+8\log^2 x-8\log x+4\right) \\
& + x^4\left(10\zeta_2+20\log^2 x-4\log x+2\right)+\mathcal{O}\left(x^5\right) \\
& + \epsilon\left\{x^2\left(2\zeta_2-\frac{2\zeta_3}{3}-2\zeta_2\log x-\frac{4}{3}\log^3 x+4\log^2 x-6\log x+4\right)\right. \\
& \qquad + x^3\left(4\zeta_2-\frac{8\zeta_3}{3}-8\zeta_2\log x-\frac{16}{3}\log^3 x+8\log^2 x-8\log x+4\right) \\
& \left. + x^4\left(2\zeta_2-\frac{20\zeta_3}{3}-20\zeta_2\log x-\frac{40}{3}\log^3 x+4\log^2 x-4\log x+2\right)+\mathcal{O}\left(x^5\right)\right\} \\
& + \mathcal{O}\left(\epsilon^2\right)
\end{aligned}
$$

$$
\begin{aligned}
M_2^V(x,\epsilon) = {} & \frac{1}{\epsilon^2}\left\{x+2x^2+3x^3+\mathcal{O}\left(x^4\right)\right\} \\
& + \frac{1}{\epsilon}\left\{x(3-\log x)+x^2(4-2\log x)+x^3(4-3\log x)+\mathcal{O}\left(x^4\right)\right\} \\
& + x\left(\frac{\log^2 x}{2}-3\log x+7\right)+x^2\left(\log^2 x-4\log x+8\right)+x^3\left(\frac{3\log^2 x}{2}-4\log x+8\right) \\
& + \mathcal{O}\left(x^4\right)+\epsilon\left\{x\left(-\frac{8\zeta_3}{3}-\frac{1}{6}\log^3 x+\frac{3\log^2 x}{2}-7\log x+15\right)\right. \\
& \qquad + x^2\left(-\frac{16\zeta_3}{3}-\frac{1}{3}\log^3 x+2\log^2 x-8\log x+16\right) \\
& \left. + x^3\left(-8\zeta_3-\frac{1}{2}\log^3 x+2\log^2 x-8\log x+16\right)+\mathcal{O}\left(x^4\right)\right\}+\mathcal{O}\left(\epsilon^2\right)
\end{aligned}
$$

$$
\begin{aligned}
M_3^V(x,\epsilon) = {} & \frac{1}{\epsilon^2}\left\{x+2x^2+3x^3+\mathcal{O}\left(x^4\right)\right\}+\frac{1}{\epsilon}\left\{x(3-\log x)+2x^2+x^3(3\log x-1)+\mathcal{O}\left(x^4\right)\right\} \\
& + x\left(\frac{\log^2 x}{2}-3\log x+7\right)+x^2\left(-2\zeta_2-2\log^2 x+2\log x+4\right) \\
& + x^3\left(-6\zeta_2-\frac{15}{2}\log^2 x+4\log x+\frac{13}{2}\right)+\mathcal{O}\left(x^4\right) \\
& + \epsilon\left\{x\left(-\frac{8\zeta_3}{3}-\frac{1}{6}\log^3 x+\frac{3\log^2 x}{2}-7\log x+15\right)\right. \\
& \qquad + x^2\left(-4\zeta_2-\frac{28\zeta_3}{3}+4\zeta_2\log x+2\log^3 x-5\log^2 x+2\log x+10\right) \\
& \left. + x^3\left(-3\zeta_2-20\zeta_3+12\zeta_2\log x+\frac{13\log^3 x}{2}-5\log^2 x-7\log x+\frac{73}{4}\right)+\mathcal{O}\left(x^4\right)\right\} \\
& + \mathcal{O}\left(\epsilon^2\right)
\end{aligned}
$$

$$
\begin{aligned}
M_4^V(x,\epsilon) = {} & \frac{1}{\epsilon^2}\left\{1+\mathcal{O}\left(x^3\right)\right\} + \frac{1}{\epsilon}\left\{4 + x(2\log x - 2) + x^2(2\log x - 1) + \mathcal{O}\left(x^3\right)\right\} + (12-\zeta_2) \\
& + x\left(-2\zeta_2 - \log^2 x + 6\log x - 6\right) + x^2\left(-2\zeta_2 - \log^2 x + \log x + \frac{3}{2}\right) + \mathcal{O}\left(x^3\right) \\
& + \epsilon\left\{\left(-4\zeta_2 - \frac{14\zeta_3}{3} + 32\right) + x\left(-4\zeta_2 - 4\zeta_3 + \frac{\log^3 x}{3} - 3\log^2 x + 14\log x - 14\right)\right. \\
& \left. + x^2\left(-4\zeta_3 + \frac{\log^3 x}{3} - \frac{\log^2 x}{2} - \frac{3\log x}{2} + \frac{27}{4}\right) + \mathcal{O}\left(x^3\right)\right\} + \mathcal{O}\left(\epsilon^2\right)
\end{aligned}
$$

$$
\begin{aligned}
M_5^V(x,\epsilon) = {} & -\log^2 x + \mathcal{O}\left(x^3\right) + \epsilon\left\{\left(6\zeta_3 + 2\zeta_2\log x + \log^3 x\right) + x\left(6\log^2 x + 4\log x - 12\right)\right. \\
& \left. + x^2\left(3\log^2 x - 5\log x + \frac{9}{2}\right) + \mathcal{O}\left(x^3\right)\right\} + \mathcal{O}\left(\epsilon^2\right)
\end{aligned}
$$

$$
\begin{aligned}
M_6^V(x,\epsilon) = {} & \frac{1}{\epsilon}\left\{-\log x + 2x\log x - 2x^2\log x + \mathcal{O}\left(x^3\right)\right\} + \left(\zeta_2 + \frac{\log^2 x}{2}\right) \\
& + x\left(-2\zeta_2 + 8\log x - 4\right) + x^2\left(2\zeta_2 - 18\log x + 10\right) + \mathcal{O}\left(x^3\right) + \epsilon\left\{\left(8\zeta_3 - \frac{\log^3 x}{6}\right)\right. \\
& + x\left(-8\zeta_2 - 22\zeta_3 - 2\zeta_2\log x - \frac{2}{3}\log^3 x - 3\log^2 x + 6\log x - 10\right) \\
& \left. + x^2\left(18\zeta_2 + 22\zeta_3 + 2\zeta_2\log x + \frac{2\log^3 x}{3} - \frac{3\log^2 x}{2} - \frac{99\log x}{2} + \frac{247}{4}\right) + \mathcal{O}\left(x^3\right)\right\} \\
& + \mathcal{O}\left(\epsilon^2\right)
\end{aligned}
$$

$$
\begin{aligned}
M_7^V(x,\epsilon) = {} & \frac{1}{\epsilon^2}\left\{\frac{1}{2} + \mathcal{O}\left(x^3\right)\right\} + \frac{1}{\epsilon}\left\{\frac{5}{2} + x(2\log x - 2) + x^2(2\log x - 1) + \mathcal{O}\left(x^3\right)\right\} \\
& + \left(\frac{\zeta_2}{2} + \frac{19}{2}\right) + x\left(4\zeta_3 + 2\zeta_2\log x + \frac{\log^3 x}{6} - \frac{5\log^2 x}{2} + 9\log x - 9\right) \\
& + x^2\left(8\zeta_3 + 4\zeta_2\log x + \frac{\log^3 x}{3} - \frac{5\log^2 x}{2} + \frac{7\log x}{2} - \frac{15}{4}\right) + \mathcal{O}\left(x^3\right) \\
& + \epsilon\left\{\left(\frac{5\zeta_2}{2} - \frac{13\zeta_3}{3} + \frac{65}{2}\right) + x\left(-\zeta_2 + \zeta_3 - 11\zeta_4 - \frac{7}{2}\zeta_2\log^2 x + 5\zeta_2\log x\right.\right. \\
& \left. - 11\zeta_3\log x - \frac{5}{24}\log^4 x + \frac{13\log^3 x}{6} - \frac{21\log^2 x}{2} + 29\log x - 29\right) \\
& + x^2\left(\frac{7\zeta_2}{2} - 19\zeta_3 - 22\zeta_4 - 7\zeta_2\log^2 x - 5\zeta_2\log x - 22\zeta_3\log x - \frac{5}{12}\log^4 x\right. \\
& \left.\left. + \frac{4\log^3 x}{3} - 3\log^2 x + 2\log x + 7\right) + \mathcal{O}\left(x^3\right)\right\} + \mathcal{O}\left(\epsilon^2\right)
\end{aligned}
$$

$$
\begin{aligned}
M_8^V(x,\epsilon) = {} & \frac{\log^2 x}{4x} - \frac{\log^2 x}{2} + \frac{1}{4}x\log^2 x + \mathcal{O}\left(x^2\right) + \epsilon\left\{\frac{-\frac{9\zeta_3}{2} - \frac{3}{2}\zeta_2\log x - \frac{5}{12}\log^3 x}{x}\right. \\
& + \left(9\zeta_3 + 3\zeta_2\log x + \frac{5\log^3 x}{6} - \frac{\log^2 x}{2} - 3\log x + 5\right) \\
& \left. + x\left(-\frac{9\zeta_3}{2} - \frac{3}{2}\zeta_2\log x - \frac{5}{12}\log^3 x + \frac{3\log^2 x}{4} + \frac{23\log x}{4} - \frac{79}{8}\right) + \mathcal{O}\left(x^2\right)\right\} + \mathcal{O}\left(\epsilon^2\right)
\end{aligned}
$$

$$
\begin{aligned}
M_9^V(x,\epsilon) = & -6\zeta_3 + x\left(2\log^2 x - 6\log x + 6\right) + x^2\left(\log^2 x - \frac{3\log x}{2} + \frac{3}{4}\right) + \mathcal{O}\left(x^3\right) \\
& + \epsilon\Bigg\{(-12\zeta_3 - 9\zeta_4) + x\left(6\zeta_2 - 4\zeta_2\log x - 2\log^3 x + 9\log^2 x - 22\log x + 28\right) \\
& + x^2\left(\frac{3\zeta_2}{2} - 2\zeta_2\log x - \log^3 x - \frac{23\log^2 x}{4} + \frac{61\log x}{4} - \frac{25}{4}\right) + \mathcal{O}\left(x^3\right)\Bigg\} + \mathcal{O}\left(\epsilon^2\right)
\end{aligned}
$$

$$
\begin{aligned}
M_{10}^V(x,\epsilon) = & \frac{1}{\epsilon}\left\{-\frac{1}{2}x\log^2 x - x^2\log^2 x - \frac{3}{2}x^3\log^2 x + \mathcal{O}\left(x^4\right)\right\} \\
& + x\left(3\zeta_3 + \zeta_2\log x + \frac{5\log^3 x}{6} - \frac{\log^2 x}{2}\right) \\
& + x^2\left(6\zeta_3 + 2\zeta_2\log x + \frac{5\log^3 x}{3} + \log^2 x + 2\log x - 4\right) \\
& + x^3\left(9\zeta_3 + 3\zeta_2\log x + \frac{5\log^3 x}{2} + \frac{7\log^2 x}{2} + \frac{7\log x}{2} - \frac{15}{2}\right) + \mathcal{O}\left(x^4\right) \\
& + \epsilon\Bigg\{x\Bigg(3\zeta_3 + \frac{5\zeta_4}{4} - 2\zeta_2\log^2 x + \zeta_2\log x - 4\zeta_3\log x - \frac{17}{24}\log^4 x + \frac{5\log^3 x}{6} \\
& \qquad - \frac{\log^2 x}{2}\Bigg) + x^2\Bigg(-2\zeta_2 - 6\zeta_3 + \frac{5\zeta_4}{2} - 4\zeta_2\log^2 x - 2\zeta_2\log x - 8\zeta_3\log x \\
& \qquad - \frac{17}{12}\log^4 x - \frac{5\log^3 x}{3} - 2\log^2 x + 8\log x - 4\Bigg) \\
& \quad + x^3\Bigg(-\frac{7\zeta_2}{2} - 21\zeta_3 + \frac{15\zeta_4}{4} - 6\zeta_2\log^2 x - 7\zeta_2\log x - 12\zeta_3\log x - \frac{17}{8}\log^4 x \\
& \qquad - \frac{35\log^3 x}{6} - \frac{23\log^2 x}{4} + \frac{23\log x}{4} + \frac{21}{2}\Bigg) + \mathcal{O}\left(x^4\right)\Bigg\} + \mathcal{O}\left(\epsilon^2\right)
\end{aligned}
$$

$$
\begin{aligned}
M_{11}^V(x,\epsilon) = & \frac{1}{\epsilon^2}\left\{\frac{1}{2} + \mathcal{O}\left(x^3\right)\right\} + \frac{1}{\epsilon}\left\{\left(\frac{1}{2} - \log x\right) - 2x - x^2 + \mathcal{O}\left(x^3\right)\right\} \\
& + \left(\frac{\zeta_2}{2} + \frac{\log^2 x}{2} - 3\log x - \frac{5}{2}\right) + x\left(-4\zeta_3 + \frac{\log^3 x}{3} - 2\right) \\
& + x^2\left(-8\zeta_3 + \frac{2\log^3 x}{3} + 2\log^2 x - 6\log x + 7\right) + \mathcal{O}\left(x^3\right) \\
& + \epsilon\Bigg\{\left(\frac{5\zeta_2}{2} + \frac{8\zeta_3}{3} - \frac{1}{6}\log^3 x + \frac{3\log^2 x}{2} - 7\log x - \frac{35}{2}\right) + x\Bigg(2\zeta_2 - 4\zeta_3 + 4\zeta_4 \\
& \qquad - \zeta_2\log^2 x + 2\zeta_3\log x - \frac{5}{12}\log^4 x + \frac{\log^3 x}{3} + \log^2 x - 2\log x - 4\Bigg) \\
& \quad + x^2\Bigg(7\zeta_2 - 4\zeta_3 + 8\zeta_4 - 2\zeta_2\log^2 x - 4\zeta_2\log x + 4\zeta_3\log x - \frac{5}{6}\log^4 x - \frac{8\log^3 x}{3} \\
& \qquad + \frac{11\log^2 x}{2} - \frac{9\log x}{2} + \frac{13}{4}\Bigg) + \mathcal{O}\left(x^3\right)\Bigg\} + \mathcal{O}\left(\epsilon^2\right)
\end{aligned}
$$

$$
\begin{aligned}
M_{12}^{V}(x,\epsilon) = {}& \frac{1}{\epsilon^2}\left\{\frac{1}{8}+\mathcal{O}\left(x^3\right)\right\}+\frac{1}{\epsilon}\left\{\frac{9}{16}-\frac{x}{2}+x^2\left(-\frac{3\log x}{2}-\frac{1}{4}\right)+\mathcal{O}\left(x^3\right)\right\}+\left(\frac{\zeta_2}{8}+\frac{63}{32}\right) \\
& -\frac{5x}{2}+x^2\left(-2\zeta_3-\zeta_2\log x-\frac{1}{12}\log^3 x+\frac{15\log^2 x}{8}-\frac{43\log x}{8}-\frac{21}{16}\right)+\mathcal{O}\left(x^3\right) \\
& +\epsilon\left\{\left(\frac{9\zeta_2}{16}-\frac{13\zeta_3}{12}+\frac{405}{64}\right)+\left(-\frac{\zeta_2}{2}-\frac{19}{2}\right)x+x^2\left(-\frac{9\zeta_2}{8}+\frac{5\zeta_3}{4}+\frac{11\zeta_4}{2}+\frac{7}{4}\zeta_2\log^2 x\right.\right. \\
& \left.-\frac{11}{4}\zeta_2\log x+\frac{11}{2}\zeta_3\log x+\frac{5\log^4 x}{48}-\frac{37\log^3 x}{24}+\frac{101\log^2 x}{16}-\frac{245\log x}{16}-\frac{219}{32}\right) \\
& \left.+\mathcal{O}\left(x^3\right)\right\}+\mathcal{O}\left(\epsilon^2\right)
\end{aligned}
$$

$$
\begin{aligned}
M_{13}^{V}(x,\epsilon) = {}& \frac{1}{\epsilon^2}\left\{1+\mathcal{O}\left(x^3\right)\right\}+\frac{1}{\epsilon}\left\{4+x(4\log x-4)+x^2(4\log x-2)+\mathcal{O}\left(x^3\right)\right\}+(12-\zeta_2) \\
& +x\left(-4\zeta_2-2\log^2 x+12\log x-12\right)+x^2\left(-4\zeta_2+2\log^2 x-6\log x+7\right)+\mathcal{O}\left(x^3\right) \\
& +\epsilon\left\{\left(-4\zeta_2-\frac{14\zeta_3}{3}+32\right)+x\left(-8\zeta_2-8\zeta_3+\frac{2\log^3 x}{3}-6\log^2 x+28\log x-28\right)\right. \\
& \left.+x^2\left(8\zeta_2-8\zeta_3-8\zeta_2\log x-\frac{10}{3}\log^3 x+11\log^2 x-19\log x+\frac{43}{2}\right)+\mathcal{O}\left(x^3\right)\right\} \\
& +\mathcal{O}\left(\epsilon^2\right)
\end{aligned}
$$

$$
\begin{aligned}
M_{14}^{V}(x,\epsilon) = {}& \left(9\zeta_4+2\zeta_2\log^2 x+8\zeta_3\log x+\frac{\log^4 x}{24}\right) \\
& +x\left(-18\zeta_4-4\zeta_2\log^2 x-16\zeta_3\log x-\frac{1}{12}\log^4 x+3\log^2 x-18\right) \\
& +x^2\left(18\zeta_4+4\zeta_2\log^2 x+16\zeta_3\log x+\frac{\log^4 x}{12}-\frac{13\log^2 x}{4}-8\log x+\frac{351}{8}\right)+\mathcal{O}\left(x^3\right) \\
& +\epsilon\left\{\left(-12\zeta_2\zeta_3+7\zeta_5-\frac{5}{2}\zeta_2\log^3 x-\frac{13}{2}\zeta_3\log^2 x-10\zeta_4\log x-\frac{1}{20}\log^5 x\right)\right. \\
& +x\left(24\zeta_2\zeta_3-38\zeta_3-54\zeta_4-14\zeta_5+5\zeta_2\log^3 x-12\zeta_2\log^2 x+13\zeta_3\log^2 x-2\zeta_2\log x\right. \\
& \left.-48\zeta_3\log x+20\zeta_4\log x+\frac{\log^5 x}{10}-\frac{\log^4 x}{4}-\frac{10\log^3 x}{3}+5\log^2 x-4\log x+38\right) \\
& +x^2\left(12\zeta_2-24\zeta_2\zeta_3+\frac{117\zeta_3}{2}+99\zeta_4+14\zeta_5-5\zeta_2\log^3 x+22\zeta_2\log^2 x\right. \\
& -13\zeta_3\log^2 x-\frac{9}{2}\zeta_2\log x+88\zeta_3\log x-20\zeta_4\log x-\frac{1}{10}\log^5 x+\frac{11\log^4 x}{24} \\
& \left.\left.+\frac{11\log^3 x}{3}-\frac{161\log^2 x}{8}-\frac{23\log x}{4}+\frac{873}{16}\right)+\mathcal{O}\left(x^3\right)\right\}+\mathcal{O}\left(\epsilon^2\right)
\end{aligned}
$$

$$
\begin{aligned}
M_{15}^{V}(x,\epsilon) = {} & \left(-\frac{5\zeta_4}{2} - \frac{1}{2}\zeta_2\log^2 x\right) + x\left(2\zeta_2 - 2\zeta_2\log x - \frac{1}{2}\log^3 x + \frac{\log^2 x}{2}\right) \\
& + x^2\left(\frac{\zeta_2}{2} - \zeta_2\log x - \frac{1}{4}\log^3 x - \frac{7\log^2 x}{8} + 2\log x - 1\right) + \mathcal{O}\left(x^3\right) \\
& + \epsilon\left\{\left(-3\zeta_2\zeta_3 + 9\zeta_5 + \frac{1}{2}\zeta_2\log^3 x - \frac{1}{2}\zeta_3\log^2 x + 4\zeta_4\log x\right) + x\left(-6\zeta_2 + \zeta_3 + 17\zeta_4\right.\right. \\
& \quad \left. + \frac{13}{2}\zeta_2\log^2 x - \zeta_2\log x + 9\zeta_3\log x + \frac{5\log^4 x}{8} - \frac{5\log^3 x}{6} + \frac{\log^2 x}{2} - 2\log x + 7\right) \\
& \quad + x^2\left(-\frac{57\zeta_2}{4} + \frac{41\zeta_3}{4} + \frac{17\zeta_4}{2} + \frac{13}{4}\zeta_2\log^2 x + \frac{47}{4}\zeta_2\log x + \frac{9}{2}\zeta_3\log x + \frac{5\log^4 x}{16}\right. \\
& \qquad \left.\left. + \frac{77\log^3 x}{24} - \frac{33\log^2 x}{16} - \frac{21\log x}{8} - \frac{245}{32}\right) + \mathcal{O}\left(x^3\right)\right\} + \mathcal{O}\left(\epsilon^2\right)
\end{aligned}
$$

$$
\begin{aligned}
M_{16}^{V}(x,\epsilon) = {} & \frac{1}{\epsilon}\left\{-\frac{\log^2 x}{2} + \mathcal{O}\left(x^3\right)\right\} + \left(3\zeta_3 + \zeta_2\log x + \frac{\log^3 x}{3} - \log^2 x\right) \\
& + x\left(-\log^3 x + 2\log^2 x + 2\log x - 4\right) + x^2\left(-\log^3 x + \log^2 x - \frac{\log x}{2} + \frac{1}{2}\right) + \mathcal{O}\left(x^3\right) \\
& + \epsilon\left\{\left(6\zeta_3 + \frac{5\zeta_4}{4} - \frac{1}{2}\zeta_2\log^2 x + 2\zeta_2\log x - \zeta_3\log x - \frac{1}{8}\log^4 x + \frac{2\log^3 x}{3} - 2\log^2 x\right)\right. \\
& \quad + x\left(-2\zeta_2 - 12\zeta_3 + 3\zeta_2\log^2 x - 4\zeta_2\log x + 6\zeta_3\log x + \frac{7\log^4 x}{6} - \frac{7\log^3 x}{3}\right. \\
& \quad \left. + 2\log^2 x + 6\log x - 8\right) + x^2\left(\frac{\zeta_2}{2} - 6\zeta_3 + 3\zeta_2\log^2 x - 2\zeta_2\log x + 6\zeta_3\log x\right. \\
& \qquad \left.\left. + \frac{7\log^4 x}{6} + \frac{17\log^3 x}{6} + \frac{\log^2 x}{2} - \frac{73\log x}{4} + 19\right) + \mathcal{O}\left(x^3\right)\right\} + \mathcal{O}\left(\epsilon^2\right)
\end{aligned}
$$

$$
\begin{aligned}
M_{17}^{V}(x,\epsilon) = {} & \frac{1}{\epsilon}\left\{\left(-12\zeta_3 - 4\zeta_2\log x + \frac{2\log^3 x}{3}\right) + x(16 - 8\log x) + x^2(2\log x - 2) + \mathcal{O}\left(x^3\right)\right\} \\
& + \left(-32\zeta_4 - 2\zeta_2\log^2 x - 20\zeta_3\log x - \frac{5}{6}\log^4 x\right) \\
& + x\left(24\zeta_2 - 8\log^2 x + 32\right) + x^2\left(-2\zeta_2 + 26\log x - 44\right) + \mathcal{O}\left(x^3\right) \\
& + \epsilon\left\{\left(60\zeta_2\zeta_3 - 44\zeta_5 + \frac{14}{3}\zeta_2\log^3 x + 16\zeta_3\log^2 x + 9\zeta_4\log x + \frac{17\log^5 x}{30}\right)\right. \\
& \quad + x\left(-16\zeta_2 + 168\zeta_3 + 24\zeta_2\log x + 18\log^3 x - 44\log^2 x + 80\log x - 144\right) \\
& \left. + x^2\left(-40\zeta_2 - 22\zeta_3 - 2\zeta_2\log x - \frac{1}{6}\log^3 x + \frac{33\log^2 x}{2} + 23\log x - \frac{233}{2}\right) + \mathcal{O}\left(x^3\right)\right\} \\
& + \mathcal{O}\left(\epsilon^2\right)
\end{aligned}
$$

### 5.2 Master integrals appearing in the real contributions

In this section we describe the computation of the integrals contributing to $\sigma_{gg\to H}^{\mathrm{R;int.}}$, which are unknown in analytic form in the literature. They can be classified into the two topolo-

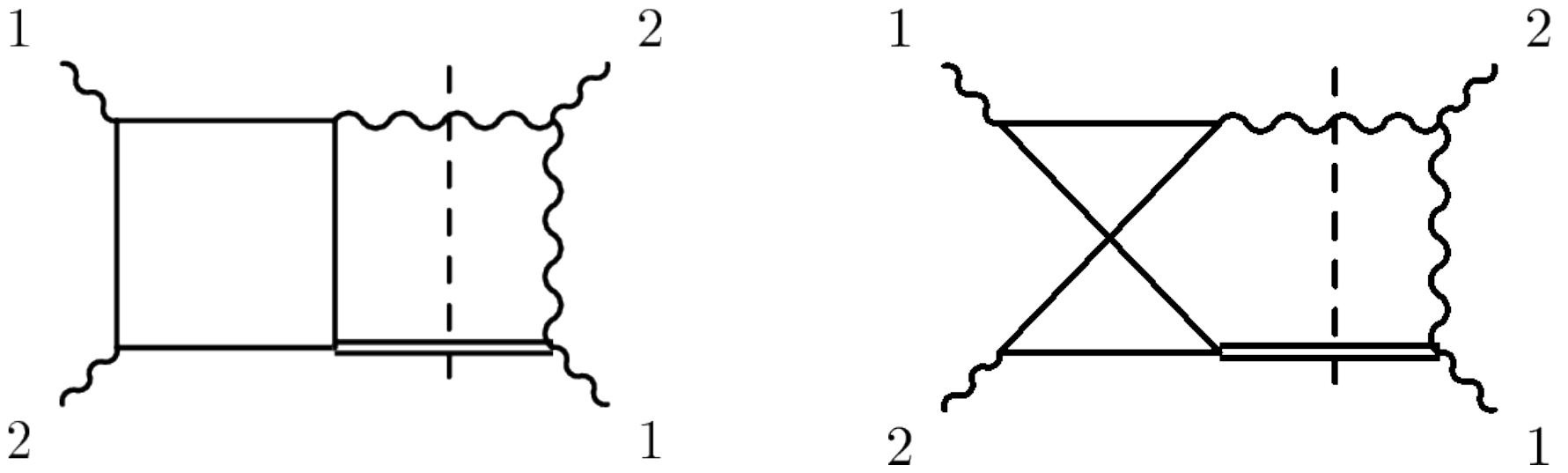

**Figure 3:** The two independent topologies appearing in the computation of the real contributions.

gies shown in figure 3, which are defined as

$$
\begin{aligned}
T_1^{\mathrm{R}}[\nu_1,\ldots,\nu_5] &= \mathrm{e}^{-\epsilon\gamma_E}\int \tilde{\mathrm{d}}^d k\, \mathrm{d}\Phi_{12\to Hg} D_{m_b}^{\nu_1}(k) D_{m_b}^{\nu_2}(k_1) D_{m_b}^{\nu_3}(k_{12}) D_{m_b}^{\nu_4}(k-l) D_0^{\nu_5}(l_2),\\
T_2^{\mathrm{R}}[\nu_1,\ldots,\nu_5] &= \mathrm{e}^{-\epsilon\gamma_E}\int \tilde{\mathrm{d}}^d k\, \mathrm{d}\Phi_{12\to Hg} D_{m_b}^{\nu_1}(k) D_{m_b}^{\nu_2}(k_1) D_{m_b}^{\nu_3}(k-l_2) D_{m_b}^{\nu_4}(k-l) D_0^{\nu_5}(l_2).
\end{aligned}
$$

where $\mathrm{d}\Phi_{12\to Hg} = \mathrm{d}\Phi(p_1,p_2;p_3,p_H)$ denotes the $2\to 2$ phase space. We parametrize it as

$$
\int \mathrm{d}\Phi(p_1,p_2;p_3,p_H) = \mathcal{N}\int_0^1 \mathrm{d}x_1\,(x_1(1-x_1))^{-\epsilon}, \quad \text{with} \quad \mathcal{N} = \frac{s^{-\epsilon}\Omega_{d-2}}{4(2\pi)^{d-2}}(1-z)^{1-2\epsilon}, \tag{5.4}
$$

with invariants given by

$$
s_{13} = (p_1-p_3)^2 = -s(1-z)x_1, \qquad s_{23} = (p_2-p_3)^2 = -s(1-z)(1-x_1).
$$

Performing the reduction, we find 16 master integrals that we choose as follows:

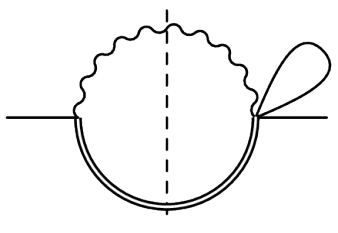

$$= T_1^{\mathrm{R}}[0,0,1,0,0] = M_1^{\mathrm{R}},$$

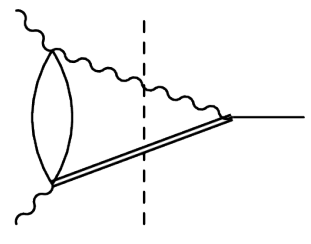

$$= T_1^{\mathrm{R}}[0,1,0,1,0] = M_2^{\mathrm{R}},$$

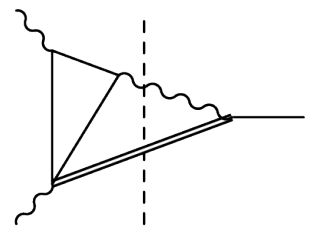

$$= T_1^{\mathrm{R}}[0,1,1,1,0] = M_3^{\mathrm{R}},$$

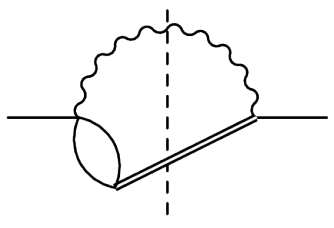

$$= T_1^{\mathrm{R}}[1,0,0,1,0] = M_4^{\mathrm{R}},$$

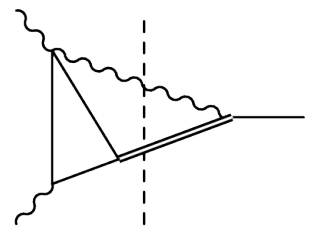

$$= T_1^{\mathrm{R}}[1,1,0,1,0] = M_5^{\mathrm{R}},$$

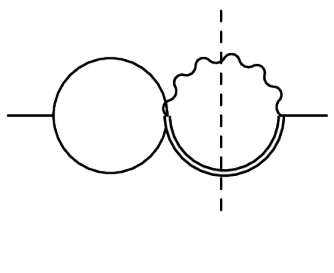

$$= T_1^{\mathrm{R}}[1,0,1,0,0] = M_6^{\mathrm{R}},$$

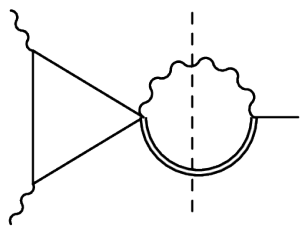 $= T_1^{\rm R}[1,1,1,0,0] = M_7^{\rm R}$,

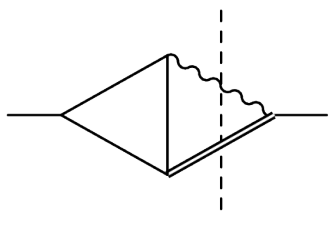 $= T_1^{\rm R}[1,0,1,1,0] = M_8^{\rm R}$,

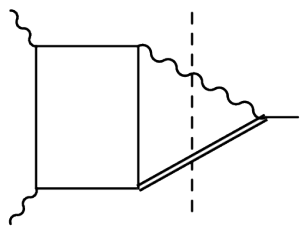 $= T_1^{\rm R}[1,1,1,1,0] = M_9^{\rm R}$,

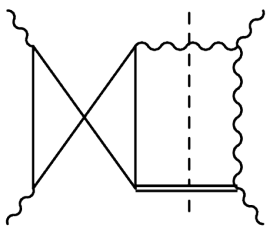 $= T_1^{\rm R}[1,1,1,1,1] = M_{10}^{\rm R}$,

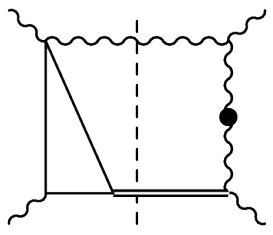 $= T_1^{\rm R}[1,1,0,1,-1] = M_{11}^{\rm R}$,

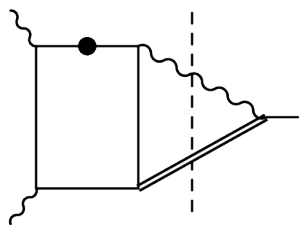 $= T_1^{\rm R}[1,1,-1,1,0] = M_{12}^{\rm R}$,

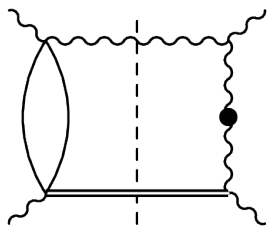 $= T_1^{\rm R}[0,1,0,1,-1] = M_{13}^{\rm R}$,

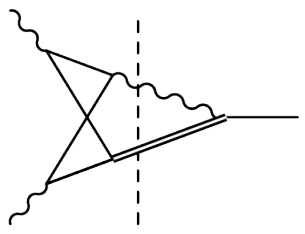 $= T_2^{\rm R}[1,1,1,1,0] = M_{14}^{\rm R}$,

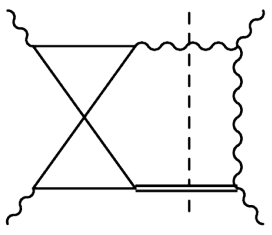 $= T_2^{\rm R}[1,1,1,1,1] = M_{15}^{\rm R}$,

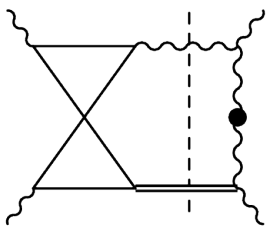 $= T_2^{\rm R}[1,1,1,1,-1] = M_{16}^{\rm R}$.

We solve the corresponding differential equations for the vector of master integrals $\mathbf{M}^{\rm V} = (M_1^{\rm R}, \ldots, M_{16}^{\rm R})$ as an expansion in the parameter $r = m_b^2/s$ using the methods presented in section 4. The reductions to master integrals and the differential equations have been obtained in two different ways and found to agree. The differential equation are then normalized and expanded using our algorithm and the solution can be written as

$$\mathbf{M}^{\rm R}(r,\epsilon) = {\rm F}^{\rm R}(r,\epsilon)\,\mathbf{B}^{\rm R}(z,\epsilon), \tag{5.5}$$

where the fundamental matrix of the system is given by

$${\rm F}^{\rm R}(r,\epsilon) = {\rm T}^{\rm R}_{\rm tot.}(r,\epsilon)\, r^{{\rm R}(\epsilon)} \sum_{n=0}^{\infty} {\rm D}_n^{\rm R}(r,\epsilon),$$

where $\mathrm{T}^{\mathrm{R}}_{\mathrm{tot.}} = \mathrm{T}^{\mathrm{R}}_{\mathrm{rank}}\mathrm{T}^{\mathrm{R}}_{\mathrm{norm.}}$ is the transformation used to achieve the normalized form presented in section 4, $\mathrm{R}^R$ is the residue matrix at $r = 0$ of the properly normalized differential equation matrix and $\mathbf{B}^{\mathrm{R}} = (B^{\mathrm{R}}_1, \ldots, B^{\mathrm{R}}_{16})$ is the vector of boundary conditions. Note that the boundary conditions now depends on the variable $z$ as well, making their evaluation much more difficult than in the case of the virtual master integrals. The differential equations, transformation matrix $\mathrm{T}^R_{\mathrm{tot.}}$ and the corresponding boundary conditions can be found in the ancillary file attached to this article. The expansion (5.5) can then be unambiguously reconstructed using the method presented in section 4.

The main difficulty in evaluating the boundary conditions $B^{\mathrm{R}}_i(z, \epsilon)$ is the fact that the phase-space integral (5.4) induces a non-trivial analytic behaviour of the integrals as $m_b \to 0$. In the corresponding two-loop integrals, where no propagator is cut, one expects the integral to have scalings of the type $(m_b^2)^{-n\epsilon}$, with $n = 0, 1, 2$. However, one-loop integrals can only develop scalings with $n = 0, 1$ such that the supplementary scaling with $n = 2$ must only appear after the phase-space integral has been performed. There is however no such prescription as the expansion by regions in the case where some of the propagators are cut. How can one then obtain the relevant boundary conditions for the $(m_b^2)^{-2\epsilon}$ scalings that appear in the solution of the differential equations? We tackle this problem by introducing a Mellin-Barnes representation [59, 60] and, owing to the simple structure of the one-loop integrals appearing in this computation, we are able to resolve the general asymptotic behavior of all the real master integrals.

Let us consider a generic one-loop integrals with $n$-external legs and whose propagators are all regulated by the same mass $m_b$,

$$I[\nu_1, \ldots, \nu_n] = \int \frac{\mathrm{d}^d k}{i\pi^{d/2}} D^{\nu_1}_{m_b}(k) D^{\nu_2}_{m_b}(k + p_1) \cdot \ldots \cdot D^{\nu_n}_{m_b}(k + p_1 + \ldots + p_{n-1}).$$

The general parametric representation for such an integral is given by

$$I[\nu_1, \ldots, \nu_n] = (-1)^{\nu} \frac{\Gamma(\nu - d/2)}{\prod_i \Gamma(\nu_i)} \int_0^1 \left( \prod_{i=1}^{n} \mathrm{d}x\, x^{\nu_i - 1} \right) \delta(1 - \sum_i x_i) \mathcal{F}^{-\nu + d/2}, \tag{5.6}$$

where $\mathcal{F}$ is the first Symanzik polynomial and $\nu = \sum_i \nu_i$ is the sum of the inverse powers of the propagators. Since all internal propagators have mass $m_b^2$, the first Symanzik polynomial must be of the form $\mathcal{F} = \mathcal{V} + m_b^2$, with $\mathcal{V} = \mathcal{F}|_{m_b=0}$. We can then introduce a Mellin-Barnes integral as

$$\begin{aligned} \mathcal{F}^{-\nu+d/2} &= (\mathcal{V} + m_b^2)^{-\nu+d/2} \\ &= \frac{1}{2\pi i} \int_{-i\infty}^{i\infty} \mathrm{d}\zeta\, \frac{\Gamma(-\zeta)\Gamma(\zeta + \nu - d/2)}{\Gamma(\nu - d/2)} (m_b^2)^{-\zeta-\nu+d/2}\, \mathcal{V}^{\zeta}, \end{aligned} \tag{5.7}$$

where the integral is performed along the usual Barnes contour separating the poles at $\zeta = n$ from those at $\zeta = d/2 - \nu - n$, $n \in \mathbb{N}$. Plugging this Mellin-Barnes representation in the parametric representation (5.6), one can reinterpret the integrand as being the very same one-loop integral but with $m_b = 0$ and a shifted dimension $d' = 2(\zeta + \nu)$. Denoting this integral by $I_0^{(2(\zeta+\nu))}$ and absorbing normalization factors, we obtain the general relation

$$I[\nu_1, \ldots, \nu_n] = \frac{1}{2\pi i} \int_{-i\infty}^{i\infty} \mathrm{d}\zeta\, \Gamma(\zeta + \nu - d/2)\, (m_b^2)^{-\zeta-\nu+d/2}\, I_0^{(2(\zeta+\nu))}[\nu_1, \ldots, \nu_n]. \tag{5.8}$$

Note that one can not obtain a representation such as (5.8) at higher loops because in this case the second Symanzik polynomial $\mathcal{U}$ is not equal to one and the exponents of $\mathcal{U}$ and $\mathcal{F}$ do not allow one to recast the integrand after introducing the Mellin-Barnes representation (5.7) in a form where the original integral with $m_b = 0$ appears.

Equation (5.8) completely elucidates the structure of this class of integrals in the limit where $m_b^2 \to 0$. Applying Cauchy's theorem, the integral must be given by a sum over the residues of poles lying at the left of the contour, that is with $\Re(\zeta) < 0$. Each of these poles contributes to a specific power of $m_b^2$ such that the complete expansion in $m_b^2$ can be obtained by inspection of the poles of the integrand of (5.8) in the complex $z$-plane. There are two types of poles to be considered:

- The poles coming from the overall gamma function $\Gamma(\zeta + \nu - d/2)$ that are located at
$$\zeta + \nu - d/2 = -n, \quad \Rightarrow \quad \zeta = 2 - n - \epsilon - \nu,$$
with $n \in \mathbb{N}$. The corresponding residues will have an overall factor of $(m_b^2)^{-(\zeta+\nu-d/2)} = (m_b^2)^n$ such that they all contribute to the hard region.

- The poles coming from the integral $I_0^{(2(\zeta+\nu))}$ itself. It is well known that Feynman integrals can develop poles only at even integer values of their dimension, such that these poles must be located at
$$d' = 2(\zeta + \nu) = 2n, \quad \Rightarrow \quad \zeta = 2 - n - \nu,$$
with $n \in \mathbb{Z}$. The corresponding factor will then be $(m_b^2)^{-(\zeta+\nu-d/2)} = (m_b^2)^{n-\epsilon}$, such that these poles contribute to the $(m_b^2)^{-\epsilon}$ scaling.

Note that only poles lying at the left of the contour need to be considered, that is we must require $n > 2 - \nu$.

This means that the small $m_b^2$ expansion of the massive integral $I$ can be unambiguously obtained from the $\epsilon$-pole structure of the massless integral with shifted dimension $I_0^{(2(\zeta+\nu))}$. As an illustration, we consider the first poles with $n = 0$. The residue of the overall $\Gamma$-function at $\zeta = 2 - \epsilon - \nu$ is easily seen to be $I_0^{(4-2\epsilon)}$, i.e. the correct hard region. Further assume that the massless integral has an $\epsilon$-expansion of the form

$$I_0^{(4-2\epsilon)} = \frac{\alpha}{\epsilon^2} + \frac{\beta}{\epsilon} + \mathcal{O}(\epsilon^0),$$

where $\beta$ and $\alpha$ are the coefficients of the first and second order poles at $\epsilon = 0$. It is then easy to compute the residue at $\zeta = 2 - \nu$ and obtain the first order in the small $m_b^2$ expansion as

$$I[\nu_1, \ldots, \nu_n] = I_0^{(4-2\epsilon)} - (m_b^2)^{-\epsilon}\,\Gamma(\epsilon)\left(\beta + \alpha\left(\log(m^2) - \psi(\epsilon)\right)\right) + \mathcal{O}(m_b^2), \tag{5.9}$$

where $\psi(\epsilon)$ is the digamma function. Validity of this equation can be easily checked with the one-loop triangle integrals shown in equation (5.3). Note that in order for the pole at $z = 2 - \nu$ to contribute we need $\nu > 2$ such that the tadpole and bubble integrals (with standard powers of the propagators) do not satisfy equation (5.9).

Generalization to higher orders is straightforward, such that we obtain the small $m_b$ expansion

$$I[\nu_1, \dots, \nu_n] = \sum_{i=0}^{\infty} (m_b^2)^i \left\{ I_0^{(4-2(\epsilon-i))} - (m_b)^{-\epsilon}\, \Gamma(\epsilon - i) \left(\beta_i + \alpha_i(\log m_b^2 - \psi(\epsilon - i))\right) \right\},$$

where $\beta_i$ and $\alpha_i$ are the first and second order poles of $I_0^{(d')}$ at $(4-2i-d')/2 = 0$, respectively. Note that $I_0^{(d')}$ can not develop poles of order higher than two, because its integrand has only two independent singular limits (soft and collinear).

The Mellin-Barnes representation (5.8) also allows one to obtain the scalings $(m_b^2)^{-2\epsilon}$ created by the phase-space integral. Taking the phase-space into account, one gets integrals of the form

$$\frac{1}{2\pi i}\int_{-i\infty}^{i\infty} \mathrm{d}\zeta\; \Gamma(\zeta + \nu - d/2)\, (m_b^2)^{-\zeta-\nu+d/2} \int \mathrm{d}\Phi_{12\to Hg}\, I_0^{(2(\zeta+\nu))}. \tag{5.10}$$

Note that the dimension ($d$) of the phase-space integral given in (5.4) does not coincide with the dimension of the shifted one-loop integral ($d' = 2(\zeta + \nu)$). The trick is to perform the phase-space integrals *before* the Mellin-Barnes integral, creating new poles in the $\zeta$ complex plane. These poles must be summed and give rise to the aforementioned scaling.

As an illustration, let us consider a simple one-loop triangle $I_3(\epsilon, m_b^2)$ similar to the one shown in equation (5.3) but with $s \to s_{13}$. Introducing our Mellin-Barnes representation, we get for the corresponding phase-space integral

$$\int \mathrm{d}\Phi_{12\to Hg}\, I_3(\epsilon, m_b^2) = \frac{1}{2\pi i}\int_{-i\infty}^{i\infty} \mathrm{d}\zeta\; \Gamma(1+\epsilon+\zeta)\, (m_b^2)^{-1-\epsilon-\zeta} \int \mathrm{d}\Phi_{12\to Hg}\; I_3(\epsilon', 0), \tag{5.11}$$

where we have $\nu = 3$ for the triangle and we set $d' = 4 - 2\epsilon' = 2(\zeta + \nu)$, i.e $\epsilon' = -1 - \zeta$. We can directly compute

$$\begin{aligned}
&\int \mathrm{d}\Phi_{12\to Hg}\; I_3(\epsilon', 0) \\
&\quad = \mathcal{N} \int_0^1 \mathrm{d}x_1\, (x_1(1-x_1))^{-\epsilon} ((1-z)x_1)^{-1-\epsilon'} \left( -\frac{1}{\epsilon'^2} \frac{\Gamma(1+\epsilon')\Gamma^2(1-\epsilon')}{\Gamma(1-2\epsilon')} \right) \\
&\quad = \mathcal{N}(1-z)^{-1-\epsilon'} \frac{\Gamma(1-\epsilon)\Gamma(-\epsilon-\epsilon')}{\Gamma(1-2\epsilon-\epsilon')} \left( -\frac{1}{\epsilon'^2} \frac{\Gamma(1+\epsilon')\Gamma^2(1-\epsilon')}{\Gamma(1-2\epsilon')} \right),
\end{aligned} \tag{5.12}$$

where $\mathcal{N}$ is the phase-space normalization factor given in (5.4) and the expression in parenthesis is the corresponding massless integral with shifted dimension $d'$. Note the appearance of the gamma function $\Gamma(-\epsilon - \epsilon')$ in equation (5.12) which creates a pole at

$$(4 - d')/2 = \epsilon' = -\epsilon = (d-4)/2,$$

corresponding to a value of $\zeta = d'/2 - \nu = 2 - \epsilon' - \nu = 2 + \epsilon - \nu$ and an overall scaling of $(m_b^2)^{-(\zeta+\nu-d/2)} = (m_b^2)^{-2\epsilon}$. Hence we see that the phase-space integral has created the new scaling. Plugging (5.12) in our Mellin-Barnes representation (5.11) we can collect the first order poles and obtain

$$
\begin{aligned}
\int \mathrm{d}\Phi_{12\to Hg}\, I_3(\epsilon, m_b^2) = \mathcal{N}\frac{\Gamma(1-\epsilon)}{(1-z)^{1+\epsilon}}\Bigg\{ & \frac{\Gamma(1-\epsilon)^2\Gamma(\epsilon+1)}{2\epsilon^3\Gamma(1-3\epsilon)} \\
+ \left(\frac{m_b^2}{1-z}\right)^{-\epsilon} & \frac{\Gamma(1-\epsilon)\Gamma(\epsilon)\left(\gamma_E + \psi^{(0)}(-\epsilon) + \psi^{(0)}(\epsilon) - \psi^{(0)}(1-2\epsilon) - \log\frac{m_b^2}{1-z}\right)}{\epsilon\Gamma(1-2\epsilon)} \\
- \left(\frac{m_b^2}{1-z}\right)^{-2\epsilon} & \frac{\Gamma(\epsilon+1)^2}{2\epsilon^3} \\
\Bigg\} + \mathcal{O}(m_b^2). &
\end{aligned}
\tag{5.13}
$$

The two first terms corresponding to the scalings $(m_b^2)^{-0\epsilon}$ and $(m_b^2)^{-\epsilon}$ can be obtained by first expanding $I_3(\epsilon, m_b^2)$ for small $m_b^2$ and then integrating over the phase-space term by term. The new scaling $(m_b^2)^{-2\epsilon}$ corresponds to non-commuting contributions and the phase-space integral must be performed first in this case. In some sense, the poles coming from the virtual and real integrals do not overlap in the Melling-Barnes representation. Further note that result (5.13) is finite in $\epsilon$. This is expected because all the IR divergences are regulated by the internal mass $m_b^2$.

Such a trick has allowed us to recover all the scalings missing from a naive application of the strategy of regions. In the general case, the one-loop integral $I_0^{(2(\zeta+\nu))}$ under consideration might be a complicated function of the phase-space variables $s_{1g}$ and $s_{2g}$ and carrying the phase-space integration explicitly might be very difficult. However, since we only need to compute the coefficient of the $\epsilon$-poles created by the phase-space integral, it is sufficient to look at the asymptotic limit of the massless integrals for small $x_1$ (and $\bar{x}_1$), where we must have that

$$
I_0^{(d'=4-2\epsilon')} \sim x_1^{-m}(A(\epsilon') + \mathcal{O}(x_1)) + x_1^{-n-\epsilon'}(B(\epsilon') + \mathcal{O}(x_1)),
$$

with $m, n \in \mathbb{N}$. This expression can then be integrated over the phase-space and it is easy to see that only the second term contributes to the $-2\epsilon$ scaling. In conclusion, all contributions to the $-2\epsilon$ scaling can be obtained by first taking the asymptotic expansions of the massless integral $I_0^{(d')}$ in the limits where $x_1$ and $\bar{x}_1$ are small, integrating over the phase space and finally taking the corresponding residue in the $\zeta$ plane.

As previously said, equation (5.8) allows us to obtain all the boundary conditions from the corresponding massless one-loop integrals. The most difficult massless one-loop integral to consider is the massless one-loop box with one off-shell leg ($p_4^2 = m_H^2$) given by

$$
\begin{aligned}
T_1^R[1,1,1,1,0]\Big|_{m_b^2=0} = \frac{2\Gamma(1-\epsilon)^2\Gamma(1+\epsilon)}{\epsilon^2\Gamma(1-2\epsilon)}\frac{1}{st}\Bigg\{ & (-t)^{-\epsilon}\,{}_2F_1\left(1,-\epsilon;1-\epsilon;-\frac{u}{s}\right) \\
& + (-s)^{-\epsilon}\,{}_2F_1\left(1,-\epsilon;1-\epsilon;-\frac{u}{t}\right) \\
& - (-m_H^2)^{-\epsilon}\,{}_2F_1\left(1,-\epsilon;1-\epsilon;-\frac{m_H^2 u}{ts}\right)\Bigg\},
\end{aligned}
\tag{5.14}
$$

with $t = s_{23}$, $u = s_{13}$. See for example [61] for a derivation. Note that the box appearing in the second topology shown in figure 3 can be expressed from (5.14) by simple crossing.

Even if the hard region are generally simple, the integral over the phase space can lead to complicated hypergoemetric functions. We were able to obtain all the boundary conditions at all orders in $z$ and $\epsilon$ using the hypergeometric functions ${}_2F_1$ and ${}_3F_2$, which can be easily expanded by using the `Mathematica` package `HypExp` [35, 36] , together with the Meijer $G$ function with arguments:

$$M_1 = G_{3,3}^{3,2}\left(\begin{matrix} 0,0,1-2\epsilon \\ 0,0,-\epsilon \end{matrix} \,\middle|\, \frac{z}{z-1}\right), \quad \text{and} \quad M_2 = G_{3,3}^{3,2}\left(\begin{matrix} 0,0,2-2\epsilon \\ 0,0,1-\epsilon \end{matrix} \,\middle|\, \frac{z}{z-1}\right).$$

The Meijer $G$ function is best defined via its Mellin-Barnes representation as

$$G_{p,q}^{m,n}\left(\begin{matrix} a_1,\ldots,a_p \\ b_1,\ldots,b_q \end{matrix} \,\middle|\, z\right) = \frac{1}{2\pi i}\int_{-i\infty}^{i\infty} \mathrm{d}s\, \frac{\prod_{j=1}^{m}\Gamma(b_j - s)\prod_{j=1}^{n}\Gamma(1-a_j+s)}{\prod_{j=m+1}^{q}\Gamma(1-b_j+s)\prod_{j=n+1}^{p}\Gamma(a_j - s)}\, z^s,$$

where the integral is performed along the usual Barnes contour. This representation allows one to obtain the $\epsilon$-expansion of such integrals easily. As an example, let us consider the following Mellin-Barnes integral

$$G_{3,3}^{3,2}\left(\begin{matrix} 1 & 1 & 1+\epsilon \\ 1 & 1 & 2\epsilon \end{matrix} \,\middle|\, z\right) = \frac{1}{2\pi i}\int_\gamma \mathrm{d}s\, \frac{\Gamma(1-s)^2\Gamma(2\epsilon - s)\Gamma(s)^2}{\Gamma(1+\epsilon-s)}\, z^s, \tag{5.15}$$

where $\gamma$ denotes the usual Barnes contour. Representation (5.15) is not suitable for expanding in $\epsilon$, because the right-pole at $s = 2\epsilon$ and the left-pole at $s = 0$ would pinch the contour $\gamma$ as $\epsilon \to 0$. It is thus necessary to define a modified contour $\gamma'$ that goes to the right of the pole at $s = 2\epsilon$, such that (5.15) becomes

$$\frac{1}{2\pi i}\int_{\gamma'} \mathrm{d}s \frac{\Gamma(1-s)^2\Gamma(2\epsilon-s)\Gamma(s)^2}{\Gamma(1+\epsilon-s)}\,(-z)^s + \frac{\Gamma(1-2\epsilon)^2\Gamma(2\epsilon)^2}{\Gamma(1-\epsilon)}\,(-z)^{2\epsilon}.$$

The integrand of the Mellin-Barnes integral can now be safely expanded in $\epsilon$. Integrating order by order produces harmonic sums that can be rewritten in terms of harmonic polylogarithms using standard techniques [62, 63][7]. We obtain

$$G_{3,3}^{3,2}\left(\begin{matrix} 1 & 1 & 1+\epsilon \\ 1 & 1 & 2\epsilon \end{matrix} \,\middle|\, z\right) = \mathrm{e}^{-\epsilon\gamma_E}\left(\frac{1}{4\epsilon^2} + \frac{\log(z)}{2\epsilon} + \frac{31}{8}\zeta_2 + \log(z)\log\left(\frac{z}{1-z}\right)\right.$$
$$\left. - \operatorname{Li}_2(z) + \mathcal{O}(\epsilon)\right). \tag{5.16}$$

A complete $\epsilon$-expansions of the Meijer $G$-functions appearing in this computation can be found in the ancillary file attached to this article.

Before presenting our results for the master integrals let us comment briefly on the issue of the soft limit $z \to 1$. In principle, we want to obtain the small $m_b^2$ expansion of the total *hadronic* cross section (2.2) such that the integral over $z$ must still be taken into account.

[7]We thank Falko Dulat who provided us with personal code to perform this task.

Extending the ideas presented in this section, we can expect this last integral to create a new non-trivial analytic behaviour as $m_b \to 0$ and induce a new scaling $(m_b^2)^{-3\epsilon}$. Indeed we find that this is actually the case for single terms in the matrix elements. In particular, we looked at the master integral $M_3^{\mathrm{R}}$ (multiplied by the its coefficient from the $gg \to gH$ matrix element) and we see using the explicit Mellin-Barnes representation introduced earlier that the $z$ integral indeed creates a non-vanishing $(m_b^2)^{-3\epsilon}$ scaling. However, we do not need to worry about such subtleties when considering the full matrix element because the correct soft limit (with massive $b$-quarks) is recovered at all orders in $\epsilon$ by simply expanding under the $z$ integral, see section 3 for more details. This means that even if individual masters multiplied by their coefficients do develop terms proportional to $(m_b^2)^{-3\epsilon}$ when integrated over $z$, the physical matrix element does not have such contributions because a soft subtraction can be implemented as usual. Such a cancellation should also work at $\mathrm{N^2LO}$ and in particular the correct soft limit $z \to 1$ should also be obtained there by simply expanding under the $z$ integral.

Since we recover the correct soft limit, we can simply expand the real master integrals in $\epsilon$ without preserving terms of the form $(1-z)^{n\epsilon}$, $n \in \mathbb{Z}$. We obtain

$$
\begin{aligned}
M_1^R(r,z,\epsilon) = &\frac{1}{\epsilon}\left\{r + \mathcal{O}\left(r^2\right)\right\} + r(1-\log(r)) + \mathcal{O}\left(r^2\right) \\
&+ \epsilon\left\{\frac{r}{2}\left(\zeta_2 + \log^2(r) - 2\log(r) + 2\right) + \mathcal{O}\left(r^2\right)\right\} \\
&+ \epsilon^2\left\{\frac{r}{6}\left(3\zeta_2 - 2\zeta_3 + (-3\zeta_2 - 6)\log(r) - \log^3(r) + 3\log^2(r) + 6\right) + \mathcal{O}\left(r^2\right)\right\} \\
&+ \mathcal{O}\left(\epsilon^3\right)
\end{aligned}
$$

$$
\begin{aligned}
M_2^R(r,z,\epsilon) = &\frac{1}{\epsilon}\{1\} + (3 - \log(\bar{z})) + \frac{r}{\bar{z}}\left(\log(r)(2\log(\bar{z}) + 2) - \log^2(r) - \log^2(\bar{z}) - 2\log(\bar{z}) - 2\right) \\
&+ \mathcal{O}\left(r^2\right) + \epsilon\left\{\frac{1}{2}\left(-3\zeta_2 + \log^2(\bar{z}) - 6\log(\bar{z}) + 18\right)\right. \\
&\quad + \frac{r}{\bar{z}}\left(-4\zeta_2 + 4\zeta_3 + \log(r)\left(4\zeta_2 + \log^2(\bar{z}) + 2\right) + \log^2(r)(-2\log(\bar{z}) - 1)\right. \\
&\quad \left.\left. + \log^3(r) - 4\zeta_2\log(\bar{z}) + \log^2(\bar{z}) - 2\right) + \mathcal{O}\left(r^2\right)\right\} \\
&+ \epsilon^2\left\{\frac{1}{6}\left(-27\zeta_2 - 38\zeta_3 + 9\zeta_2\log(\bar{z}) - \log^3(\bar{z}) + 9\log^2(\bar{z}) - 54\log(\bar{z}) + 162\right)\right. \\
&+ \frac{r}{12\bar{z}}\left(-12\zeta_2 - 144\zeta_3 + 24\zeta_4 + \log(r)\left(12\zeta_2 + 48\zeta_3 + 60\zeta_2\log(\bar{z}) + 4\log^3(\bar{z}) + 24\right)\right. \\
&\quad + \log^2(r)\left(-54\zeta_2 - 12\log^2(\bar{z}) - 12\right) + \log^3(r)(16\log(\bar{z}) + 4) - 7\log^4(r) \\
&\left.\left. - 6\zeta_2\log^2(\bar{z}) + 36\zeta_2\log(\bar{z}) - 96\zeta_3\log(\bar{z}) - \log^4(\bar{z}) - 4\log^3(\bar{z}) - 24\right) + \mathcal{O}\left(r^2\right)\right\} \\
&+ \mathcal{O}\left(\epsilon^3\right)
\end{aligned}
$$

$$
\begin{aligned}
M_3^R(r,z,\epsilon) = {}& \frac{1}{6\bar{z}}\left(-12\zeta_3 - 3\log^2(r)\log(\bar{z}) + 3\log(r)\log^2(\bar{z}) + \log^3(r) - \log^3(\bar{z})\right)\\
&+ \frac{r}{\bar{z}^2}(-2\log(r) - 2(-\log(\bar{z}) - 1)) + \mathcal{O}\left(r^2\right)\\
&+ \epsilon\left\{\frac{1}{6\bar{z}}\left(24\zeta_3 - 12\zeta_4 + \log^2(r)\left(-6\zeta_2 - 3\log^2(\bar{z}) + 6\log(\bar{z})\right)\right.\right.\\
&\quad + \log(r)\left(12\zeta_2\log(\bar{z}) + \log^3(\bar{z}) - 6\log^2(\bar{z})\right) + \log^3(r)(3\log(\bar{z}) - 2) - \log^4(r)\\
&\quad \left. - 6\zeta_2\log^2(\bar{z}) + 12\zeta_3\log(\bar{z}) + 2\log^3(\bar{z})\right)\\
&\left. + \frac{r}{\bar{z}^2}\left(2\log(r)(\log(\bar{z})+3) + 2\left(2\zeta_2 - \log^2(\bar{z}) - 4\log(\bar{z}) - 2\right)\right) + \mathcal{O}\left(r^2\right)\right\} + \mathcal{O}\left(\epsilon^2\right)
\end{aligned}
$$

$$
\begin{aligned}
M_4^R(r,z,\epsilon) = {}& \frac{1}{\epsilon}\{1\} + (-\log(z) + i\pi + 2) + \mathcal{O}\left(r^1\right)\\
&+ \epsilon\left\{\frac{1}{2}\left(-7\zeta_2 + \log^2(z) - 2i\pi\log(z) - 4\log(z) + 4i\pi + 8\right) + \mathcal{O}\left(r^1\right)\right\}\\
&+ \epsilon^2\left\{\frac{1}{12}\left(-84\zeta_2 - 28\zeta_3 + 42\zeta_2\log(z) - 2\log^3(z) + 6i\pi\log^2(z) + 12\log^2(z)\right.\right.\\
&\quad \left.\left. - 24i\pi\log(z) - 48\log(z) - 3i\pi^3 + 48i\pi + 96\right) + \mathcal{O}\left(r^1\right)\right\} + \mathcal{O}\left(\epsilon^3\right)
\end{aligned}
$$

$$
\begin{aligned}
M_6^R(r,z,\epsilon) = {}& \frac{1}{\epsilon}\{1\} + (2 + i\pi) + r(-2\log(r) - 2i\pi + 2) + \mathcal{O}\left(r^2\right)\\
&+ \epsilon\left\{\left(-\frac{7\zeta_2}{2} + 2i\pi + 4\right) + r\left(8\zeta_2 + \log^2(r) - 2\log(r) + 2\right) + \mathcal{O}\left(r^2\right)\right\} + \mathcal{O}\left(\epsilon^2\right)
\end{aligned}
$$

$$
\begin{aligned}
M_7^R(r,z,\epsilon) = {}& \left(-3\zeta_2 + \frac{\log^2(r)}{2} + i\pi\log(r)\right) + r(2\log(r) + 2i\pi) + \mathcal{O}\left(r^2\right)\\
&+ \epsilon\left\{\left(-3\zeta_3 - \zeta_2\log(r) - \frac{1}{3}\log^3(r) - \frac{1}{2}i\pi\log^2(r) - \frac{i\pi^3}{3}\right)\right.\\
&\quad \left. + r\left(-8\zeta_2 - \log^2(r) + 2\log(r) + 2i\pi - 4\right) + \mathcal{O}\left(r^2\right)\right\} + \mathcal{O}\left(\epsilon^2\right)
\end{aligned}
$$

$$
\begin{aligned}
M_8^R(r,z,\epsilon) = {}& \frac{1}{2\bar{z}}\left(2\log(r)\log(z) - \log^2(z) + 2i\pi\log(z)\right) + \frac{r}{-z\bar{z}}(2\bar{z}\log(r) + 2(-\log(z) + i\pi\bar{z}))\\
&+ \mathcal{O}\left(r^2\right) + \epsilon\left\{\frac{1}{6\bar{z}}\left(-3\log^2(r)\log(z) - 24\zeta_2\log(z) + \log^3(z) - 3i\pi\log^2(z)\right)\right.\\
&+ \frac{r}{-z\bar{z}}\left(-\bar{z}\log^2(r) + 2\bar{z}\log(r) + \log^2(z) - 2i\pi\log(z) - 2\log(z) - 8\zeta_2\bar{z} + 2i\pi\bar{z}\right.\\
&\left.\left. - 4\bar{z}\right) + \mathcal{O}\left(r^2\right)\right\} + \epsilon^2\left\{\frac{1}{24\bar{z}}\left(12\zeta_2\log(r)\log(z) + 4\log^3(r)\log(z) + 42\zeta_2\log^2(z)\right.\right.\\
&\left. - 48\zeta_3\log(z) - \log^4(z) + 4i\pi\log^3(z) - 6i\pi^3\log(z)\right) + \frac{r}{-6z\bar{z}}\left(\log(r)\left(6\zeta_2\bar{z} + 12\bar{z}\right)\right.\\
&\quad + 2\bar{z}\log^3(r) - 6\bar{z}\log^2(r) + 42\zeta_2\log(z) - 2\log^3(z) + 6i\pi\log^2(z) + 6\log^2(z)\\
&\left.\left. - 12i\pi\log(z) + 12\log(z) - 48\zeta_2\bar{z} - 24\zeta_3\bar{z} - 3i\pi^3\bar{z} - 12i\pi\bar{z}\right) + \mathcal{O}\left(r^2\right)\right\} + \mathcal{O}\left(\epsilon^3\right)
\end{aligned}
$$

$$
\begin{aligned}
M_{13}^R(r,z,\epsilon) = & \frac{1}{\epsilon}\left\{-\frac{\bar{z}}{2}\right\} + \left(\frac{1}{2}\bar{z}\log(\bar{z}) - \frac{7\bar{z}}{4}\right) + r\left(-2\log(r)\log(\bar{z}) + \log^2(r) + \log^2(\bar{z}) + 2\right) \\
& + \mathcal{O}\left(r^2\right) + \epsilon\left\{\frac{1}{8}\left(6\zeta_2\bar{z} - 45\bar{z} - 2\bar{z}\log^2(\bar{z}) + 14\bar{z}\log(\bar{z})\right)\right. \\
& + r\left(-4\zeta_3 + \log(r)\left(-4\zeta_2 - \log^2(\bar{z})\right) + 2\log^2(r)\log(\bar{z}) - \log^3(r) + 4\zeta_2\log(\bar{z})\right. \\
& \left.\left. - 2\log(\bar{z}) + 6\right) + \mathcal{O}\left(r^2\right)\right\} \\
& + \epsilon^2\left\{\frac{1}{48}\left(126\zeta_2\bar{z} + 152\zeta_3\bar{z} - 36\zeta_2\bar{z}\log(\bar{z}) - 837\bar{z} + 4\bar{z}\log^3(\bar{z}) - 42\bar{z}\log^2(\bar{z})\right.\right. \\
& \left. + 270\bar{z}\log(\bar{z})\right) + \frac{r}{12}\left(-36\zeta_2 - 24\zeta_4 + \log(r)\left(-48\zeta_3 - 60\zeta_2\log(\bar{z}) - 4\log^3(\bar{z})\right)\right. \\
& + \log^2(r)\left(54\zeta_2 + 12\log^2(\bar{z})\right) - 16\log^3(r)\log(\bar{z}) + 7\log^4(r) + 6\zeta_2\log^2(\bar{z}) \\
& \left.\left. + 96\zeta_3\log(\bar{z}) + \log^4(\bar{z}) + 12\log^2(\bar{z}) - 72\log(\bar{z}) + 216\right) + \mathcal{O}\left(r^2\right)\right\} + \mathcal{O}\left(\epsilon^3\right).
\end{aligned}
$$

The remaining master integrals $M_5^R$, $M_9^R$, $M_{10}^R$, $M_{11}^R$, $M_{12}^R$, $M_{14}^R$, $M_{15}^R$ and $M_{16}^R$ are too big to be displayed here and can be found in the ancillary file attached to this article.

## 6. Matrix elements

### 6.1 Matrix element appearing in the virtual contributions

In this section we present our results for the unrenormalized squared matrix element for the process $gg \to H$ in terms of master integrals needed for the computation of the virtual contributions. The amplitude consists of two pieces: the first piece is the leading order of this process in the full theory with bottom quarks running in the loop interfered with one loop correction to the $ggH$ vertex in the effective theory, while the second piece is the two-loop correction to the to the same process in the full theory interfered with the Higgs-gluon-gluon vertex in the effective theory. See section 2 for more details.

We write the amplitude for the process $gg \to H$ generically as

$$
\mathcal{A}_{gg\to H} = \frac{1}{v}\delta_{ab}\,\mathcal{P}_{ab}\,\frac{2A}{(1-\epsilon)\,s^2},
$$

where $\mathcal{P}$ is the following helicity projector:

$$
\mathcal{P}_{ab} = -\epsilon_a(p_1)\cdot\epsilon_b(p_2)\,p_1\cdot p_2 + \epsilon_a(p_1)\cdot p_2\,\epsilon_b(p_2)\cdot p_1,
$$

where $\epsilon_\mu(p_1)$ and $\epsilon_\nu(p_2)$ are the polarisation vectors for incoming gluons with momenta $p_1$ and $p_2$ respectively, and $a$ and $b$ are the colour indices of the gluons. Then the interference of two amplitudes can be written as

$$
\begin{aligned}
\sum_{\text{pols}} 2\Re\,\mathcal{A}_{gg\to H}\left(\mathcal{B}_{gg\to H}\right)^* = & \frac{1}{v}4\,C_A C_F\sum_{\text{pols}}\mathcal{P}^2\,\frac{4A\,B^*}{(1-\epsilon)^2\,s^4} \\
= & \frac{1}{v}\left(N_C^2-1\right)\frac{4A\,B^*}{(1-\epsilon)\,s^2},
\end{aligned}
$$

where we used

$$\sum_{\text{pols}} \mathcal{P}^2 = \frac{1}{2}\,(1-\epsilon)\,s^2.$$

The first piece, the leading order for the production of a Higgs boson via gluon fusion in the full theory interfered with the NLO matrix element in the effective theory is then given by

$$\sum_{\text{pols}} 2\Re\,\mathcal{A}^{(0)}_{gg\to H}\left(\mathcal{B}^{(1)}_{gg\to H}\right)^* = \frac{1}{v}\,c_H g_s^4\,\frac{4\left(N_C^2-1\right)}{(1-\epsilon)\,s^2} A_{\text{LO}}\,B^*_{\text{NLO}},$$

where

$$\begin{aligned}
A_{\text{LO}} &= \frac{2x\left(x^2(\epsilon-1)-2x(\epsilon+1)+\epsilon-1\right)}{(x-1)^4} M_3^{\text{B}}(\epsilon) + \frac{4x\epsilon}{(x-1)^2} M_2^{\text{B}}(\epsilon), \\
B_{\text{NLO}} &= C_A\left(-\frac{(4-2\epsilon)^3-16(4-2\epsilon)^2+68(4-2\epsilon)-88}{8\epsilon}\right) M_0^{\text{B}}(\epsilon),
\end{aligned}$$

and $M_3^{\text{B}}(\epsilon)$ and $M_2^{\text{B}}(\epsilon)$ are the massive one-loop master integrals given in (5.3) while $M_0^{\text{B}}(\epsilon)$ is the one loop massless bubble master. These results are in agreement with the literature [64]. The second piece, the double virtual amplitude for the production of Higgs boson via gluon fusion in the full theory interfered with the LO matrix element in the effective theory is given by

$$\sum_{\text{pols}} 2\Re\,\mathcal{A}^{(1)}_{gg\to H}\left(\mathcal{B}^{(0)}_{gg\to H}\right)^* = \frac{1}{v}\,c_H g_s^4\,\frac{4\left(N_C^2-1\right)}{(1-\epsilon)\,s^2} A_{\text{V}}.$$

We present $A_{\text{V}}$ below in terms of master integrals. For convenience we have set $s=1$.

$$
\begin{aligned}
A_{\mathrm{V}} = \Bigg\{ & -\frac{\epsilon-1}{2x^2(x+1)^2(1-4\epsilon)^2(1-2\epsilon)^2\epsilon(3\epsilon-1)}\Bigg(C_A\Big[(x \\
& +1)^2\left(128(x-1)^2\left(10x^2-17x+10\right)\epsilon^8-32(x-1)^2\left(88x^2-173x+88\right)\epsilon^7\right. \\
& +16\left(107x^4-510x^3+1094x^2-510x+107\right)\epsilon^6+8\left(25x^4-255x^3-932x^2-255x+25\right)\epsilon^5 \\
& +\left(-538x^4+6992x^3+1428x^2+6992x-538\right)\epsilon^4 \\
& +\left(185x^4-4720x^3-2034x^2-4720x+185\right)\epsilon^3-8\left(3x^4-204x^3-206x^2-204x+3\right)\epsilon^2 \\
& \left.+\left(x^4-304x^3-482x^2-304x+1\right)\epsilon+24x(x+1)^2\right)\Big] \\
& +C_F\Big[-4\epsilon\left(4\epsilon^2-5\epsilon+1\right)\Big(x^6-12x^5-25x^4+8x^3-5\left(x^2-1\right)^2\left(5x^2-52x+5\right)\epsilon^3 \\
& +16\left(x^2-1\right)^2\left(10x^2-11x+10\right)\epsilon^5-4\left(x^2-1\right)^2\left(33x^2-17x+33\right)\epsilon^4-25x^2 \\
& -4\left(3x^6-25x^5-37x^4+38x^3-37x^2-25x+3\right)\epsilon \\
& +\left(44x^6-282x^5-236x^4+564x^3-236x^2-282x+44\right)\epsilon^2-12x+1\Big)\Big]\Bigg)\Bigg\}M_1^{\mathrm{V}} \\
+\Bigg\{ & -\frac{2}{(x-1)^2(x+1)^2(1-4\epsilon)^2(\epsilon-1)(2\epsilon-1)}\Bigg(C_A\Big[(x+1)^2\left(-16\left(13x^2-50x+13\right)\epsilon^5\right. \\
& +10\left(17x^2-114x+17\right)\epsilon^4+\left(66x^2+796x+66\right)\epsilon^3-\left(127x^2+362x+127\right)\epsilon^2 \\
& \left.+\left(49x^2+102x+49\right)\epsilon+32(x-1)^2\epsilon^7+24(x-1)^2\epsilon^6-6(x+1)^2\right)\Big] \\
& +C_F\Big[-2\left(4\epsilon^2-5\epsilon+1\right)\Big(24\left(x^2-1\right)^2\epsilon^5-30\left(x^2-1\right)^2\epsilon^4+37\left(x^2-1\right)^2\epsilon^3 \\
& +(x-1)^2\left(x^2-6x+1\right)-4\left(7x^4+12x^3-22x^2+12x+7\right)\epsilon^2 \\
& +2\left(x^4+24x^3-34x^2+24x+1\right)\epsilon\Big)\Big]\Bigg)\Bigg\}M_3^{\mathrm{V}}
\end{aligned}
$$

$$+\Bigg\{C_A\Big[-\frac{8x\left(\epsilon^3+2\epsilon^2-3\epsilon+1\right)}{(x-1)^2(\epsilon-1)}\Big]\Bigg\}M_4^{\mathrm{V}}$$

$$-\Bigg\{\Bigg(C_A\Big[-64(x-1)^2\left(5x^2-14x+5\right)\epsilon^8-2x\left(23x^2+42x+23\right)\epsilon$$

$$+16\left(39x^4-192x^3+298x^2-192x+39\right)\epsilon^7$$

$$-16\left(17x^4-106x^3+188x^2-106x+17\right)\epsilon^6$$

$$-2\left(59x^4-112x^3+786x^2-112x+59\right)\epsilon^5$$

$$+\left(105x^4-224x^3+2894x^2-224x+105\right)\epsilon^4$$

$$-2\left(10x^4+131x^3+678x^2+131x+10\right)\epsilon^3$$

$$+\left(x^4+192x^3+398x^2+192x+1\right)\epsilon^2+4x(x+1)^2\Big]$$

$$+C_F\Big[4\epsilon\left(4\epsilon^2-5\epsilon+1\right)\Big(8(x-1)^2\left(5x^2-14x+5\right)\epsilon^5$$

$$+\left(-23x^4+136x^3-194x^2+136x-23\right)\epsilon^4$$

$$-4\left(3x^4-3x^3+14x^2-3x+3\right)\epsilon^3+\left(8x^4-8x^3+68x^2-8x+8\right)\epsilon^2$$

$$-\left(x^4+8x^3+22x^2+8x+1\right)\epsilon+2x(x+1)^2\Big)\Big]\Bigg)\Bigg\}M_5^{\mathrm{V}}$$

$$+\Bigg\{-\frac{2(x+1)^2\epsilon}{(x-1)^4(1-2\epsilon)^2(\epsilon-1)(3\epsilon-1)(4\epsilon-1)}\Bigg(C_A\Big[$$

$$-16\left(7x^2-15x+7\right)\epsilon^4+4\left(8x^2-33x+8\right)\epsilon^3+6\left(5x^2-18x+5\right)\epsilon^2$$

$$-5\left(3x^2-26x+3\right)\epsilon+x^2+64(x-1)^2\epsilon^5-26x+1\Big]$$

$$+C_F\Big[-4\left(4\epsilon^2-5\epsilon+1\right)\Big(\left(-3x^2+10x-3\right)\epsilon^2$$

$$+\left(-3x^2+x-3\right)\epsilon+8(x-1)^2\epsilon^3+(x-1)^2\Big)\Big]\Bigg)\Bigg\}M_6^{\mathrm{V}}$$

$$-\Bigg\{\frac{2}{(x-1)^4(1-4\epsilon)^2(\epsilon-1)\epsilon}\Bigg(C_A\Big[32(x-1)^2\left(x^2-4x+1\right)\epsilon^7$$

$$-8(x-1)^2\left(5x^2-28x+5\right)\epsilon^6-6(x+1)^2\left(x^2-4x+1\right)$$

$$+16\left(x^4-8x^3-34x^2-8x+1\right)\epsilon^5-2\left(61x^4-160x^3-602x^2-160x+61\right)\epsilon^4$$

$$+2\left(119x^4-310x^3-546x^2-310x+119\right)\epsilon^3$$

$$+\left(-171x^4+422x^3+730x^2+422x-171\right)\epsilon^2$$

$$+\left(53x^4-118x^3-270x^2-118x+53\right)\epsilon\Big]$$

$$+C_F\Big[-2\epsilon\left(4\epsilon^2-5\epsilon+1\right)\big(8(x-1)^2\left(x^2-4x+1\right)\epsilon^4$$

$$+2(x-1)^2\left(3x^2+4x+3\right)\epsilon^3-(x-1)^2\left(7x^2-12x+7\right)\epsilon^2$$

$$+2(x+1)^2\left(x^2-4x+1\right)+\left(-3x^4+4x^3+30x^2+4x-3\right)\epsilon\big)\Big]\Bigg)\Bigg\}M_7^{\mathrm{V}}$$

$$+\Bigg\{\frac{4x^2}{(x-1)^6(1-4\epsilon)^2\epsilon^2(2\epsilon-1)}\Bigg(C_A\Big[(2\epsilon-1)\big(4\left(3x^2+2x+3\right)\epsilon^3$$

$$-\left(47x^2+58x+47\right)\epsilon^2+\left(26x^2+44x+26\right)\epsilon+24(x-1)^2\epsilon^5$$

$$-6(x-1)^2\epsilon^4-4(x+1)^2\big)\Big]+C_F\Big[-2\epsilon(4\epsilon-1)\big(-\left(x^2+6x+1\right)\epsilon$$

$$+12(x-1)^2\epsilon^4-5(x-1)^2\epsilon^3-4(x-1)^2\epsilon^2+(x+1)^2\big)\Big]\Bigg)\Bigg\}M_8^{\mathrm{V}}$$

$$+\Bigg\{C_A\Big[-\frac{4x\left(\epsilon^3+2\epsilon^2-3\epsilon+1\right)\left((x-1)^2\epsilon-(x+1)^2\right)}{(x-1)^4(\epsilon-1)\epsilon}\Big]\Bigg\}M_9^{\mathrm{V}}$$

$$+\Bigg\{-\frac{2(\epsilon-1)}{(x-1)^2\left(8\epsilon^2-6\epsilon+1\right)}\Bigg(C_A\Big[(2\epsilon-1)\left(-x^2\right.$$

$$\left.+16(x-1)^2\epsilon^3-4(x-1)^2\epsilon^2-4(x-1)^2\epsilon-6x-1\right)\Big]$$

$$+C_F\Big[-2(4\epsilon-1)\left(-8\left(x^2-x+1\right)\epsilon^2+2\left(x^2+8x+1\right)\epsilon\right.$$

$$\left.-x^2+10(x-1)^2\epsilon^3-14x-1\right)\Big]\Bigg)\Bigg\}M_{10}^{\mathrm{V}}$$

$$-\Bigg\{\frac{\epsilon-1}{(x-1)^2\left(8\epsilon^2-6\epsilon+1\right)}\Bigg(C_A\Big[-2\left(7x^2-38x+7\right)\epsilon+x^2$$

$$-64(x-1)^2\epsilon^4+48(x-1)^2\epsilon^3+16(x-1)^2\epsilon^2-18x+1\Big]$$

$$+C_F\Big[4(4\epsilon-1)\left(x^2+8(x-1)^2\epsilon^3-3(x-1)^2\epsilon^2-3(x-1)^2\epsilon-6x+1\right)\Big]\Bigg)\Bigg\}M_{11}^{\mathrm{V}}$$

$$-\Bigg\{-\frac{4(3\epsilon-2)}{(x-1)^2(1-4\epsilon)^2(\epsilon-1)\epsilon}C_A\Big[2\left(x^2+94x+1\right)\epsilon^4$$

$$-4\left(15x^2+46x+15\right)\epsilon^3+\left(89x^2+134x+89\right)\epsilon^2$$

$$-\left(41x^2+70x+41\right)\epsilon+16(x-1)^2\epsilon^6-12(x-1)^2\epsilon^5+6(x+1)^2\Big]$$

$$+C_F\Big[$$

$$-2\epsilon\left(4\epsilon^2-5\epsilon+1\right)\left(4(x-1)^2\epsilon^3+5(x-1)^2\epsilon^2-(x-1)^2\epsilon-2(x+1)^2\right)\Big]\Bigg\}M_{12}^{\mathrm{V}}$$

$$+\left\{C_F\left[\frac{8x\left(x^2+(x+1)^2\epsilon^2-2(x+1)^2\epsilon+1\right)}{\left(x^2-1\right)^2}\right]\right\}M^{\rm V}_{13}$$

$$-\left\{\frac{4x}{(x-1)^6\left(8\epsilon^2-6\epsilon+1\right)}\left(C_A\Big[(x+1)^2\epsilon(2\epsilon\right.\right.$$

$$\left.-1)\left(4(x-1)^2\epsilon^3-(x-1)^2\epsilon^2-(x+1)^2\right)\Big]$$

$$-C_F\Big[(4\epsilon-1)\Big(4\left(x^2-1\right)^2\epsilon^4-3\left(x^2-1\right)^2\epsilon^3$$

$$\left.\left.+(x+1)^2\left(x^2+1\right)-2\left(x^4+x^3+4x^2+x+1\right)\epsilon\Big)\Big]\right)\right\}M^{\rm V}_{14}$$

$$+\left\{\Big(C_A-2C_F\Big)\left[\frac{4x\epsilon\left(2\epsilon^2-\epsilon-1\right)}{(x-1)^2}\right]\right\}M^{\rm V}_{15}$$

$$-\left\{\frac{4x}{(x-1)^4(4\epsilon-1)}\left(C_A\Big[(2\epsilon-1)\left(4(x-1)^2\epsilon^3-(x-1)^2\epsilon^2-(x+1)^2\right)\Big]\right.\right.$$

$$\left.\left.-C_F\Big[\left(4\epsilon^2-5\epsilon+1\right)\left(4(x-1)^2\epsilon^2+2(x-1)^2\epsilon-(x+1)^2\right)\Big]\right)\right\}M^{\rm V}_{16}$$

$$+\left\{C_A\left[\frac{x\left(-\left(7x^2+10x+7\right)\epsilon+4(x-1)^2\epsilon^2+2(x+1)^2\right)}{(x-1)^4(4\epsilon-1)}\right]\right\}M^{\rm V}_{17}.$$

### 6.2 Matrix elements appearing in the real contributions

The QCD corrections to the Higgs production via real emission involves partons in the final state, which means we have to take into account the processes involving the light quarks in the initial and final states.

We first consider the process $g(p_1)+g(p_2)\to H(p_H)+g(p_3)$ in the full theory with bottom quarks running in the loop and interfere it with the same process in the effective theory. As stated before we calculate the squared matrix element for real emission processes with reverse unitarity. We write the squared matrix element as

$$\sum_{\rm pols}2\Re\,\mathcal{A}^{(0)}_{gg\to gH}\left(\mathcal{B}^{(0)}_{gg\to gH}\right)^*=-\frac{c_H\,g_s^4N_c\left(N_c^2-1\right)}{v}A_{\rm R},\qquad(6.1)$$

where

$$
\begin{aligned}
A_{\rm R} = \Bigg\{ & -\frac{1}{(z-1)z(z+1)\epsilon(2\epsilon-1)(4\epsilon-1)\left(1-4r\right)^2\left(4r-z+1\right)} \\
& \Bigg[4\Big(z(z+1)\left(1-4r\right)^2\left(\left(5z^2+2z+1\right)r+8(z-1)r^2+z\left(-2z^2+3z-3\right)\right) \\
& +z\epsilon\left(-384\left(z^2-1\right)r^4-16\left(39z^3+9z^2-7z+71\right)r^3+8\left(29z^4+31z^3-3z^2+65z+54\right)r^2\right. \\
& \left.+\left(-124z^4+33z^3+7z^2-261z-23\right)r+15z^4-8z^3-4z^2+34z-5\right) \\
& +16\epsilon^4\left(128(z-1)^2\left(z^2+3z+2\right)r^3+2\left(12z^5-27z^3+10z^2+21z-8\right)r\right. \\
& \left.-4\left(11z^5+16z^4-19z^3-26z^2+10z+16\right)r^2-3z^5+z^4+8z^3-6z^2-6z+4\right) \\
& -2\epsilon^3\left(128\left(5z^4+9z^3-27z^2+z+20\right)r^3-8\left(43z^5+68z^4-109z^3-46z^2+116z+80\right)r^2\right. \\
& \left.+4\left(52z^5-2z^4-115z^3+106z^2+103z-40\right)r-25z^5+4z^4+70z^3-86z^2-47z+40\right) \\
& +\epsilon^2\left(256z\left(z^2-1\right)r^4+32\left(41z^4+11z^3-77z^2+105z+32\right)r^3\right. \\
& -8\left(71z^5+84z^4-123z^3+124z^2+204z+32\right)r^2 \\
& +2\left(166z^5-71z^4-131z^3+429z^2+87z-32\right)r-39z^5+23z^4+40z^3-131z^2+7z \\
& \left.+16\right)\Big)\Bigg]\Bigg\} M_1^{\rm R}
\end{aligned}
$$

$$
\begin{aligned}
&+\Bigg\{-\frac{1}{\epsilon(2\epsilon-1)(4\epsilon-1)\left(4r-z+1\right)(z-1)(z+1)\left(z-4rz\right)^2}\Bigg[512(1-2\epsilon)^2z^2\left(z^2-1\right)r^5\\
&-64\left(51z^5-43z^4+21z^3+51z^2-32z+256\epsilon^4(z-1)^3\left(2z^2+3z+1\right)\right.\\
&-8\epsilon\left(39z^5-34z^4+23z^3+40z^2-20z+24\right)-16\epsilon^3\left(45z^5-61z^4-13z^3+73z^2+16z-12\right)\\
&\left.+4\epsilon^2\left(161z^5-165z^4+79z^3+189z^2-16z+56\right)+32\right)r^4\\
&+32\left(34z^6-11z^5+10z^4+55z^3+16z+16\epsilon^4\left(24z^6-9z^5-33z^4+17z^3+33z^2-8\right)\right.\\
&-4\epsilon^3\left(158z^6-49z^5-57z^4+129z^3+187z^2+96z-24\right)\\
&-\epsilon\left(230z^6-53z^5+97z^4+345z^3+45z^2+144z+96\right)\\
&\left.+2\epsilon^2\left(269z^6-62z^5+92z^4+330z^3+163z^2+208z+56\right)+16\right)r^3-4\left(16z^7+96z^6-141z^5\right.\\
&+153z^4+117z^3-97z^2+128z+32\epsilon^4\left(2z^7+49z^6-63z^5-15z^4+101z^3-26z^2-8z+8\right)\\
&-8\epsilon^3\left(28z^7+276z^6-251z^5+31z^4+447z^3+29z^2+8z+24\right)\\
&-4\epsilon\left(29z^7+159z^6-173z^5+232z^4+158z^3-85z^2+200z-48\right)\\
&\left.+2\epsilon^2\left(130z^7+787z^6-636z^5+750z^4+866z^3+23z^2+592z-112\right)-32\right)r^2\\
&+2\left(8\left(8z^7+51z^6-76z^5-18z^4+124z^3-25z^2-32z+16\right)\epsilon^4\right.\\
&-2\left(106z^7+197z^6-152z^5-142z^4+422z^3+305z^2-288z+48\right)\epsilon^3\\
&+\left(245z^7+57z^6+122z^5+26z^4+33z^3+717z^2-192z-112\right)\epsilon^2\\
&-\left(113z^7-35z^6+44z^5+136z^4-81z^3+199z^2+48z-96\right)\epsilon\\
&\left.+2\left(8z^7-3z^6-3z^5+16z^4-3z^3+5z^2+8z-8\right)\right)r\\
&-z\left(z^2-1\right)\left(4\left(5z^4-5z^3+19z^2-27z+16\right)\epsilon^4\right.\\
&+\left(-61z^4+101z^3-267z^2+315z-176\right)\epsilon^3+2\left(33z^4-70z^3+153z^2-158z+84\right)\epsilon^2\\
&\left.+\left(-29z^4+69z^3-133z^2+125z-64\right)\epsilon+2\left(2z^4-5z^3+9z^2-8z+4\right)\right)\Bigg]\Bigg\}M_2^{\mathrm{R}}
\end{aligned}
$$

$$+\Bigg\{ -\frac{1}{(8\epsilon^2-6\epsilon+1)\,(1-4r)^2}\Bigg[2(z-1)\epsilon r\left(4\epsilon^2\left(8r-z-1\right)\right.$$

$$\left.\left.\left.+\epsilon\left(64(z-7)r^2-8(z-20)r+5z-19\right)-48(z-3)r^2+2(9z-41)r+128r^3-3z+11\right)\right]\right\} M_3^{\mathrm{R}}$$

$$+\Bigg\{\frac{1}{(z-1)^2z(z+1)(\epsilon-1)\epsilon(2\epsilon-3)(4\epsilon-1)\,(1-4r)^2}\Bigg[2r\,\Big(16z\epsilon^5\left(16\left(9z^4+22z^3-20z^2-6z+27\right)r^2\right.$$

$$\left.-8\left(9z^4+20z^3-16z^2-8z+27\right)r+7z^4+24z^3-18z^2-8z+27\right)$$

$$-4\epsilon^4\left(16\left(93z^5+266z^4-216z^3-42z^2+219z+64\right)r^2\right.$$

$$\left.-8\left(97z^5+224z^4-144z^3-76z^2+219z+64\right)r+67z^5+284z^4-174z^3-76z^2+219z+64\right)$$

$$+4\epsilon^3\left(16\left(91z^5+241z^4-177z^3+43z^2+18z+176\right)r^2\right.$$

$$-8\left(102z^5+172z^4-69z^3-10z^2+21z+176\right)r+128(z-1)^2z(z+1)r^3+63z^5+251z^4$$

$$\left.-109z^3-11z^2+22z+176\right)-2\epsilon^2\left(32\left(49z^5+74z^4-52z^3+80z^2-115z+136\right)r^2\right.$$

$$-8\left(115z^5+75z^4+9z^3+85z^2-212z+272\right)r$$

$$\left.+768(z-1)^2z(z+1)r^3+75z^5+161z^4-37z^3+79z^2-206z+272\right)$$

$$+\epsilon\left(16\left(3z^5+86z^4-20z^3+10z^2-79z+96\right)r^2-8\left(9z^5+89z^4-11z^3-41z^2-46z+96\right)r\right.$$

$$\left.+1408(z-1)^2z(z+1)r^3-3z^5+124z^4-34z^3-52z^2-35z+96\right)$$

$$\left.\left.-3(z-1)^2z(z+1)\,(1-4r)^2\,(8r-3z+7)\Big)\right]\right\} M_4^{\mathrm{R}}$$

$$
+\Bigg\{\frac{1}{(\epsilon-1)\epsilon(2\epsilon-1)(4\epsilon-1)\left(1-4r\right)^2(z-1)^2z^2(z+1)}
$$

$$
\Bigg[2r\Big(128(z+1)\Big(32(z-1)^4(z+1)\epsilon^5-2\left(13z^5-79z^4+75z^3+31z^2-100z+28\right)\epsilon^4
$$

$$
+\left(7z^5-159z^4+213z^3-125z^2-76z-4\right)\epsilon^3-\left(z^5-102z^4+169z^3-200z^2+72z-52\right)\epsilon^2
$$

$$
+\left(-33z^4+62z^3-89z^2+52z-28\right)\epsilon+4\left(z^2-z+1\right)^2\Big)r^3
$$

$$
-16\Big(8\left(9z^7-11z^6-58z^5+38z^4+57z^3-59z^2-24z+16\right)\epsilon^5
$$

$$
+\left(-34z^7+286z^6+756z^5-508z^4-514z^3+926z^2+656z-224\right)\epsilon^4
$$

$$
+\left(13z^7-437z^6-566z^5+374z^4-143z^3-769z^2-632z-16\right)\epsilon^3
$$

$$
+\left(-3z^7+254z^6+302z^5-172z^4+341z^3+406z^2+152z+208\right)\epsilon^2
$$

$$
-\left(71z^6+84z^5-58z^4+164z^3+99z^2-24z+112\right)\epsilon+8\left(z^2-z+1\right)^2\left(z^2+3z+2\right)\Big)r^2
$$

$$
+2\Big(16\left(19z^7-50z^6-60z^5+94z^4+17z^3-92z^2+8\right)\epsilon^5
$$

$$
-4\left(45z^7-418z^6-248z^5+454z^4-141z^3-484z^2-224z+56\right)\epsilon^4
$$

$$
+2\left(43z^7-1015z^6-4z^5+284z^4-871z^3-269z^2-784z-8\right)\epsilon^3
$$

$$
+\left(-18z^7+1077z^6-126z^5+184z^4+1056z^3+35z^2+848z+208\right)\epsilon^2
$$

$$
-\left(287z^6-70z^5+120z^4+342z^3-55z^2+192z+112\right)\epsilon+16\left(z^2-z+1\right)^2\left(2z^2+3z+1\right)\Big)r
$$

$$
-z\Big(8\left(9z^6-26z^5-26z^4+52z^3-7z^2-42z+8\right)\epsilon^5
$$

$$
+\left(-34z^6+384z^5+188z^4-472z^3+342z^2+344z+80\right)\epsilon^4
$$

$$
+\left(13z^6-468z^5+100z^4+86z^3-513z^2+70z-312\right)\epsilon^3
$$

$$
+\left(-3z^6+259z^5-122z^4+118z^3+229z^2-121z+232\right)\epsilon^2-\left(71z^5-50z^4+60z^3+58z^2-51z+72\right)\epsilon
$$

$$
+8(z+1)\left(z^2-z+1\right)^2\Big)\Big)\Bigg]\Bigg\}M_5^{\rm R}
$$

$$
+\Bigg\{\frac{1}{(z-1)^2(\epsilon-1)(2\epsilon-3)\left(1-4r\right)^2}
$$

$$
\Bigg[2r\left(4\epsilon^2\left(16\left(z^2-6z+1\right)r^2-4\left(z^3-8z^2+5z-6\right)r-3z^3+3z^2-z-3\right)\right.
$$

$$
+\epsilon\left(-192\left(z^2-4z+1\right)r^2+8\left(5z^3-19z^2-z-9\right)r+11z^3-13z^2+17z+9\right)
$$

$$
\left.-4(z-1)^2\epsilon^3\left(8r-z-1\right)+(z-1)^2\left(8r+1\right)\left(16r-3z-1\right)\right)\Bigg]\Bigg\}M_6^{\rm R}
$$

$$+\left\{\frac{2r\left(\epsilon\left(-8r+z+1\right)+2r\left(16r-3z-1\right)\right)}{\left(1-4r\right)^2}\right\}M_7^{\mathrm{R}}$$

$$+\left\{-\frac{1}{(z-1)(\epsilon-1)\epsilon(2\epsilon-3)\left(1-4r\right)^2}\right.$$

$$\left[2r\left(-6\left(1-4r\right)^2\left(8\left(z^2-z+1\right)r+z\left(-2z^2+3z-3\right)\right)\right.\right.$$

$$+8\epsilon^4\left(4\left(5z^3-z^2-z+5\right)r^2-2\left(5z^3-2z^2+z+4\right)r+z^3+1\right)$$

$$+\epsilon^3\left(-32\left(20z^3-z^2+12z+25\right)r^2+4\left(83z^3-45z^2+49z+65\right)r\right.$$

$$\left.+384(z+1)^2r^3-35z^3+13z^2-13z-29\right)$$

$$-2\epsilon^2\left(64\left(11z^2+8z+11\right)r^3-8\left(61z^3+5z^2+55z+91\right)r^2\right.$$

$$\left.+\left(259z^3-189z^2+233z+185\right)r-29z^3+23z^2-23z-17\right)$$

$$+\epsilon\left(256\left(7z^2-3z+7\right)r^3-16\left(43z^3+11z^2+15z+75\right)r^2\right.$$

$$\left.\left.\left.+\left(362z^3-326z^2+366z+222\right)r-43z^3+51z^2-51z-13\right)\right)\right]\right\}M_8^{\mathrm{R}}$$

$$+\left\{-\frac{1}{\left(1-4r\right)^2}\left[4r\left(\epsilon\left(4\left(3z^2-12z+5\right)r^2+\left(-6z^2+22z-8\right)r+z^2-3z+1\right)\right.\right.\right.$$

$$\left.\left.\left.-2\left(9z^2+2z-11\right)r^2+\left(7z^2-10z+3\right)r+96(z-1)r^3-(z-1)^2\right)\right]\right\}M_9^{\mathrm{R}}$$

$$+\left\{2r\left(8\left(z^2-z+1\right)r+z\left(2\left(z^2-2z+2\right)\epsilon-2z^2+3z-3\right)\right)\right\}M_{10}^{\mathrm{R}}$$

$$+\left\{-\frac{1}{(z-1)^2z^2(z+1)(\epsilon-1)\epsilon(4\epsilon-1)\left(1-4r\right)^2}\left[2r\left(256(z+1)r^3\left(2\left(z^2-z+1\right)^2\right.\right.\right.\right.$$

$$+\left(-25z^4+50z^3-29z^2-24z+12\right)\epsilon^3+2\left(13z^4-26z^3+34z^2-14z+7\right)\epsilon^2$$

$$\left.+\left(-13z^4+26z^3-39z^2+24z-12\right)\epsilon+16(z-1)^3(z+1)\epsilon^4\right)$$

$$-32r^2\left(4\left(z^2-z+1\right)^2\left(z^2+3z+2\right)+4\left(13z^6+16z^5-42z^4-16z^3+53z^2+8z-16\right)\epsilon^4\right.$$

$$-\left(71z^6+112z^5-182z^4+32z^3+263z^2+56z-48\right)\epsilon^3$$

$$+\left(68z^6+73z^5-87z^4+171z^3+99z^2+12z+56\right)\epsilon^2$$

$$\left.-\left(29z^6+29z^5-29z^4+87z^3+28z^2-16z+48\right)\epsilon\right)$$

$$+2r\left(16\left(z^2-z+1\right)^2\left(2z^2+3z+1\right)+16\left(27z^6+4z^5-56z^4+16z^3+57z^2-8z-8\right)\epsilon^4\right.$$

$$-8\left(78z^6+11z^5-82z^4+95z^3+98z^2+20z-12\right)\epsilon^3$$

$$+\left(593z^6-216z^5+94z^4+984z^3-207z^2+432z+112\right)\epsilon^2$$

$$\left.-\left(241z^6-112z^5+110z^4+384z^3-143z^2+160z+96\right)\epsilon\right)$$

$$+z\left(-8(z+1)\left(z^2-z+1\right)^2-4\left(25z^5-50z^3+32z^2+41z-16\right)\epsilon^4\right.$$

$$+\left(137z^5+8z^4-130z^3+232z^2+89z+16\right)\epsilon^3-\left(135z^5-82z^4+66z^3+214z^2-133z+136\right)\epsilon^2$$

$$\left.\left.\left.\left.+\left(58z^5-50z^4+52z^3+70z^2-66z+64\right)\epsilon\right)\right)\right]\right\}M_{11}^{\mathrm{R}}$$

$$+\left\{-\frac{1}{z(z+1)(\epsilon-1)(2\epsilon-1)(4\epsilon-1)\left(1-4r\right)^2}\left[2r\left(z\left(\left(-48z^2+64z+304\right)r^2\right.\right.\right.\right.$$

$$\left.+4\left(9z^2-19z-44\right)r+128(z+1)r^3-3z^2+10z+23\right)$$

$$+8\epsilon^3\left(16\left(9z^3+16z^2-9z-16\right)r^2-4\left(19z^3+31z^2-18z-32\right)r+9z^3+16z^2\right.$$

$$\left.-9z-16\right)+2\epsilon^2\left(-16\left(17z^3+40z^2-81z-80\right)r^2+4\left(45z^3+59z^2-168z-160\right)r\right.$$

$$\left.+128z(z+1)r^3-17z^3-32z^2+85z+80\right)-\epsilon\left(16\left(-13z^3+4z^2+109z+32\right)r^2\right.$$

$$\left.\left.\left.\left.+4\left(43z^3-53z^2-236z-64\right)r+384z(z+1)r^3-13z^3+26z^2+121z+32\right)\right)\right]\right\}M_{12}^{\mathrm{R}}$$

$$
\begin{aligned}
&+\Bigg\{ \\
&-\frac{1}{(\epsilon-1)\epsilon(2\epsilon-1)(4\epsilon-1)\left(1-4r\right)^{2}\left(4r-z+1\right)(z-1)^{2}z^{2}} \\
&\Bigg[256(3\epsilon-2)\left(64(z-1)^{3}(z+1)\epsilon^{4}\right. \\
&-4\left(25z^{4}-50z^{3}+29z^{2}+24z-12\right)\epsilon^{3} \\
&+2\left(53z^{4}-106z^{3}+137z^{2}-56z+28\right)\epsilon^{2} \\
&+\left(-55z^{4}+110z^{3}-159z^{2}+96z-48\right)\epsilon+9z^{4}-18z^{3}+25z^{2} \\
&\left.-16z+8\right)r^{4}-32\left(48\left(13z^{5}-13z^{4}-5z^{3}+5z^{2}+16z-8\right)\epsilon^{5}\right. \\
&-4\left(407z^{5}-455z^{4}+257z^{3}-17z^{2}+464z-136\right)\epsilon^{4} \\
&+2\left(1001z^{5}-1251z^{4}+1515z^{3}-337z^{2}+784z+72\right)\epsilon^{3} \\
&-\left(1339z^{5}-1809z^{4}+2617z^{3}-579z^{2}+544z+512\right)\epsilon^{2} \\
&+\left(443z^{5}-633z^{4}+921z^{3}-187z^{2}+64z+240\right)\epsilon \\
&\left.-54z^{5}+82z^{4}-114z^{3}+22z^{2}-32\right)r^{3} \\
&+8\left(48\left(2z^{6}+24z^{5}-45z^{4}+22z^{3}+29z^{2}-24z+8\right)\epsilon^{5}\right. \\
&-4\left(100z^{6}+591z^{5}-1038z^{4}+875z^{3}+304z^{2}-216z+136\right)\epsilon^{4} \\
&+\left(614z^{6}+2207z^{5}-4121z^{4}+6373z^{3}-2865z^{2}\right. \\
&\left.+2288z-144\right)\epsilon^{3}+\left(-434z^{6}-1251z^{5}+2823z^{4}\right. \\
&\left.-5661z^{3}+4219z^{2}-3104z+512\right)\epsilon^{2} \\
&+2\left(70z^{6}+202z^{5}-547z^{4}+1116z^{3}-913z^{2}+632z-120\right)\epsilon \\
&\left.-4\left(4z^{6}+13z^{5}-40z^{4}+77z^{3}-62z^{2}+40z-8\right)\right)r^{2} \\
&-2\left(-32z^{6}+58z^{5}-70z^{4}+14z^{3}+30z^{2}-64z+\epsilon^{2}\left(-829z^{6}\right.\right. \\
&\left.+1285z^{5}-2349z^{4}+2199z^{3}-1394z^{2}-480z+512\right) \\
&+24\epsilon^{5}\left(8z^{6}+17z^{5}-33z^{4}-17z^{3}+89z^{2}-64z+16\right) \\
&+\epsilon\left(274z^{6}-485z^{5}+735z^{4}-431z^{3}+35z^{2}+416z-240\right) \\
&-2\epsilon^{4}\left(382z^{6}+7z^{5}+189z^{4}-1447z^{3}+2725z^{2}-1472z+272\right) \\
&+\epsilon^{3}\left(1159z^{6}-1204z^{5}+2710z^{4}-4124z^{3}+4595z^{2}\right. \\
&\left.\left.-1280z-144\right)+32\right)r+\left(12\epsilon^{3}-23\epsilon^{2}+13\epsilon\right. \\
&\left.-2\right)(z-1)z\left(\left(5z^{4}-5z^{3}+19z^{2}-27z+16\right)\epsilon^{2}\right. \\
&+\left(-9z^{4}+19z^{3}-43z^{2}+45z-24\right)\epsilon \\
&\left.+2\left(2z^{4}-5z^{3}+9z^{2}-8z+4\right)\right)\Bigg]\Bigg\}M_{13}^{\mathrm{R}}
\end{aligned}
$$

$$
\begin{aligned}
&+\left\{\frac{1}{(4\epsilon-1)(2r-z-1)}\left[r\left(8(z-1)\epsilon^2\left(-2\left(z^2+1\right)r+2(z+3)r^2+z^3-z^2\right.\right.\right.\right.\\
&-3z-1\Big)-(r-1)\left(\left(5z^2+2z-23\right)r+8(z-1)r^2-z^3-4z^2+z+4\right)\\
&+2\epsilon\left(-8\left(z^2+4z+3\right)r^2+4\left(3z^3-10z^2+z+6\right)r\right.\\
&\left.\left.\left.\left.+16(z+3)r^3-3\left(z^4-4z^3-2z^2+4z+1\right)\right)\right)\right]\right\}M_{14}^{\mathrm{R}}\\
&+\left\{-\frac{2(z-1)(2\epsilon+1)r^2(8r+4(z-1)\epsilon-3z+7)}{4\epsilon-1}\right\}M_{15}^{\mathrm{R}}\\
&+\left\{-\frac{1}{(4\epsilon-1)(2r-z-1)}\right.\\
&\left[2r\left(\left(-11z^2+4z+15\right)r+4\epsilon^2\left(\left(-6z^2+4z+2\right)r+4(z-1)r^2+3z^3+z^2-3z-1\right)\right.\right.\\
&+\epsilon\left(\left(62z^2-36z-58\right)r+(16-64z)r^2+32r^3-13z^3+5z^2+21z+3\right)\\
&\left.\left.\left.+(11z+1)r^2-8r^3+2\left(z^3-2z-1\right)\right)\right]\right\}M_{16}^{\mathrm{R}}.
\end{aligned}
$$

### 6.2.1 Quark Channels

For the real emission involving light quarks we consider the following channels:

$$q(p_1)+\bar{q}(p_2)\to H(p_H)+g(p_3) \tag{6.2}$$
$$q(p_1)+g(p_2)\to H(p_H)+q(p_3) \tag{6.3}$$
$$q(p_2)+g(p_1)\to H(p_H)+q(p_3) \tag{6.4}$$
$$\bar{q}(p_1)+g(p_2)\to H(p_H)+\bar{q}(p_3) \tag{6.5}$$
$$\bar{q}(p_2)+g(p_1)\to H(p_H)+\bar{q}(p_3) \tag{6.6}$$

Quarks are in the initial state in the first channel, whereas the other channels have quarks both in initial and final states. The latter channels have the same cross section since the cross section is invariant under the exchange of two initial state momenta, $p_1$ and $p_2$. Therefore we present only one result below.

Similar to the gluon channel we write the squared matrix element as

$$\sum_{\text{pols}} 2\Re\mathcal{A}^{(0)}_{q\bar{q}\to gH}\left(\mathcal{B}^{(0)}_{q\bar{q}\to gH}\right)^* = -\frac{c_H\, g_s^4 N_c\left(N_c^2-1\right)}{v}A_{q\bar{q}}, \tag{6.7}$$

where

$$
\begin{aligned}
A_{q\bar{q}} &= \left\{\frac{8(\epsilon-1)r((z-1)\epsilon+1)}{2\epsilon-3}\right\}M_4^{\mathrm{R}}+\left\{-\frac{8(\epsilon-1)r}{2\epsilon-3}\right\}M_6^{\mathrm{R}}\\
&+\left\{\frac{4(z-1)(\epsilon-1)r\,(4r+(z-1)\epsilon-z+1)}{2\epsilon-3}\right\}M_8^{\mathrm{R}},
\end{aligned}
$$

$$A_{qg} = \left\{ \frac{8((z+2)\epsilon - z)}{z(z+1)(2\epsilon-1)} \right\} M_1^{\mathrm{R}}$$

$$+ \left\{ -\frac{4r\left(z^3 - \left(3z^2+z+2\right)z\epsilon - z^2 + 2\left(z^3+3z^2-z-1\right)\epsilon^2 + 2\right) + z(z+1)(\epsilon-1)^2\left(z^2(\epsilon-1)+2z-2\right)}{z^2(z+1)(\epsilon-1)\epsilon} \right\}$$

$$M_2^{\mathrm{R}} + \left\{ \frac{4r\left(z^3(\epsilon-1)\epsilon + z^2\left(2\epsilon^2+\epsilon+1\right) + z\left(\epsilon^2-6\epsilon-3\right)+4\right)}{(z-1)z(z+1)(\epsilon-1)} \right\} M_4^{\mathrm{R}}$$

$$+ \left\{ \frac{1}{(z-1)z^2(z+1)(\epsilon-1)\epsilon(2\epsilon-1)} \Big[ 2r\left(4(z+1)\left(2\epsilon^2-3\epsilon+1\right)r\left(-z^2+(4z-2)\epsilon+2z-2\right) \right. \right.$$

$$+ z\left(z^3 - z^2 - \left(3z^3+8z^2+17z-4\right)\epsilon^3 + \left(5z^3+11z^2+10\right)\epsilon^2 - \left(5z^3-2z^2+z+8\right)\epsilon \right.$$

$$\left.\left.\left. + 2z(z+1)^2\epsilon^4 + 2\right)\right) \Big] \right\} M_5^{\mathrm{R}}$$

$$+ \left\{ \frac{1}{(z-1)z^2(z+1)(\epsilon-1)\epsilon} \Big[ 2r\left(4(z+1)(2\epsilon-1)r\left(\left(z^2+2z-2\right)\epsilon - z^2+2z-2\right) \right. \right.$$

$$\left.\left. + z\left(-z^3+z^2-\left(3z^3+5z^2+6z-4\right)\epsilon^2 + \left(3z^3+2z^2-3z+6\right)\epsilon + z(z+1)^2\epsilon^3 - 2\right)\right) \Big] \right\} M_{11}^{\mathrm{R}}$$

$$+ \left\{ \frac{8r((z+2)\epsilon - z)}{z(z+1)(\epsilon-1)(2\epsilon-1)} \right\} M_{12}^{\mathrm{R}}$$

$$+ \left\{ \frac{(3\epsilon-2)\left(4r\left(\left(z^2+2z-2\right)\epsilon - z^2+2z-2\right) + z(\epsilon-1)\left(z^2(\epsilon-1)+2z-2\right)\right)}{(z-1)z^2(\epsilon-1)\epsilon} \right\} M_{13}^{\mathrm{R}}.$$

## 7. Conclusion

In this paper we have computed the top-bottom interference contributions to the QCD NLO correction to the hadronic cross section for Higgs production via gluon fusion, i.e. where one side involves bottom-quark loop and the other side is treated in the large top-mass effective theory. We have calculated the required two-loop master integrals as an expansion in the bottom-quark mass $m_b$ using a method of expansion from differential equations. This method is general and systematic enough such that it can be applied to a variety of problems. While our results for master integrals appearing in the virtual contribution are known in the literature, we have presented the first analytic computation of the masters integrals appearing in the real contributions. We believe that our results and method will be useful for a similar $\mathrm{N^2LO}$ computation which, as a three-loop process including three mass scales ($m_b$, $m_t$ and $m_H$), is not feasible at the moment. Such a computation is required to reduce the theoretical uncertainty to the Higgs production cross section due to the bottom quark mass at levels comparable to the experiments. We believe that our method makes the completion of the calculation at $\mathrm{N^2LO}$ feasible since the appearance of elliptic functions might be circumvented by considering an expansion in small $m_b$.

## Acknowledgements

We gratefully thank Falko Dulat and Bernhard Mistlberger for providing us checks for our calculation and for useful discussions. We thank Babis Anastasiou for detailed discussions. This work was supported by ERC Starting Grant IterQCD and by the Swiss National Science Foundation through 'Theoretische Forschungen auf dem Gebiet der Elementarteilchen'.

## A. Iterative expansion

Here we comment on a modification of the algorithm presented in section 4. Our algorithm can become computationally intensive especially for big systems of dozens of master integrals. In this case the computation of the Jordan decomposition becomes intractable and it is necessary to resort to a different strategy. Although the number of master integrals can be quite big, the system is usually almost decoupled in the following sense: by permuting the order of the master integrals, it is possible to bring the system in a block-lower-triangular form where the biggest blocks are of moderate sizes, typically $2 \times 2$ or $3 \times 3$. The system can then be solved iteratively block by block, starting from the upper left corner of the matrix. Only blocks of moderate size have to be handled, avoiding the computation of Jordan decompositions of big matrices. Such an iterative procedure can improve significantly the time needed to solve the system as an expansion.

Let us consider the problem of solving

$$\partial_r \mathbf{M}(r,\epsilon) = \mathrm{A}(r,\epsilon)\, \mathbf{M}(r,\epsilon), \tag{A.1}$$

with a matrix A that has a the following block triangular form

$$\mathrm{A}(r) = \left( \begin{array}{c|c} \mathrm{B}(r) & 0 \\ \hline \mathrm{L}(r) & \mathrm{R}(r) \end{array} \right),$$

where that A is $m \times m$ matrix, L is a $m \times q$ matrix and R is a $q \times q$ matrix. Let us split the vector of master integrals accordingly as

$$\mathbf{M}^{\mathsf{T}} = (\mathbf{M}_0^{\mathsf{T}}, \mathbf{M}_1^{\mathsf{T}}),$$

where $\mathbf{M}_0$ and $\mathbf{M}_1$ have length $m$ and $q$, respectively. We assume that the subsystem of differential equations $\partial_r \mathbf{M}_0 = \mathrm{B}\mathbf{M}_0$ has already been solved and that $\mathbf{M}_0(r,\epsilon)$ is known. We then see that $\mathbf{M}_1$ must satisfy the following inhomogeneous partial differential equation

$$\partial_r \mathbf{M}_1(r) = \mathrm{R}(r)\mathbf{M}_1(r) + \mathbf{C}(r), \tag{A.2}$$

where $\mathbf{C} = \mathrm{L}\,\mathbf{M}_0$ is an inhomogeneity coming from the known part of the solution.

Let us denote the solution of the homogeneous system ($\mathbf{C} = 0$) as

$$\mathbf{M}_1(r) = \mathrm{F}(r,\epsilon)\, \mathbf{B}_1(\epsilon)$$

where F denotes the fundamental matrix of the system as constructed in section 4. It is well known that the solution of the inhomogeneous system can then be written as

$$\mathbf{M}_1(x) = \mathrm{F}(r,\epsilon) \left\{ \mathbf{B}_1(\epsilon) + \int^r \mathrm{d}r'\, \mathrm{F}^{-1}(r',\epsilon)\, \mathbf{C}(r') \right\}, \tag{A.3}$$

where the first term gives the homogeneous contribution. Note that we take the primitive rather than computing the integral, as a lower boundary of $r = 0$ in the integration would create unregulated divergences. Compared to our original algorithm this amounts to a redefinition of the boundary conditions.

The computation of the inverse of the fundamental matrix $\mathrm{F}^{-1}(r)$ as a series in $r$ can be easily obtained by the method of *series reversion.* Up to the pre-factor of $r^{\mathrm{R}}$ that can easily be inverted, the fundamental matrix has an expansion of the form

$$\mathrm{F}(r) = \sum_{n=0}^{\infty} \mathrm{F}_n\, r^n,$$

where $\mathrm{F}_n$ are matrix coefficients that do not depend on $r$. We are looking for a left-multiplicative inverse $\mathrm{G}(r)$ of $\mathrm{F}(r)$ with $\mathrm{G}(r)\mathrm{F}(r) = \mathbb{1}$: Setting $\mathrm{G}(r) = \sum_n \mathrm{G}_n r^n$, Cauchy multiplication formula directly gives

$$\begin{aligned} \mathrm{G}_0\mathrm{F}_0 &= \mathbb{1}, \\ \mathrm{G}_1\mathrm{F}_0 + \mathrm{G}_0\mathrm{F}_1 &= 0, \\ \mathrm{G}_2\mathrm{F}_0 + \mathrm{G}_1\mathrm{F}_1 + \mathrm{G}_0\mathrm{F}_2 &= 0, \\ &\vdots \end{aligned}$$

such that the matrix $\mathrm{G}(r)$ can be defined order by order as

$$\mathrm{G}_0 = \mathrm{F}_0^{-1}, \tag{A.4}$$

$$\mathrm{G}_n = -\sum_{k=0}^{n-1} \mathrm{G}_k \mathrm{F}_{n-k} \mathrm{F}_0^{-1}, \qquad (n \geq 1). \tag{A.5}$$

Equation A.3 can then be applied iteratively to all the subsystems of the differential equation under consideration to generate a solution of the complete differential equations.